\pdfoutput=1
\documentclass{article}
\usepackage{microtype}
\usepackage{graphicx}
\usepackage{booktabs}
\usepackage{array}
\usepackage{tabularx}
\usepackage{threeparttable}
\usepackage{enumitem}
\usepackage{placeins}
\usepackage{hyperref}
\usepackage[accepted]{mlsys2025}
\makeatletter\renewcommand{\Notice@String}{}\makeatother
\hypersetup{pdftitle={EpiLoRA: Fusing Many LoRA Adapters into Low-Precision GEMMs for Serving and RL}, pdfauthor={Hieu Tran, Hong Yu}}
\usepackage{amsmath}
\usepackage{amssymb}
\usepackage{mathtools}
\usepackage[capitalize,noabbrev]{cleveref}
\usepackage{tikz}
\usetikzlibrary{arrows.meta,decorations.pathreplacing}
\renewcommand{\topfraction}{0.85}%
\renewcommand{\textfraction}{0.15}
\renewcommand{\floatpagefraction}{0.8}
\graphicspath{{./}}
\mlsystitlerunning{EpiLoRA: Fusing Many LoRA Adapters into Low-Precision GEMMs for Serving and RL}

\begin{document}
\twocolumn[
\mlsystitle{EpiLoRA: Fusing Many LoRA Adapters into Low-Precision GEMMs for Serving and RL}
\begin{mlsysauthorlist}
\mlsysauthor{Hieu Tran}{amh}
\mlsysauthor{Hong Yu}{amh,low}
\end{mlsysauthorlist}
\mlsysaffiliation{amh}{University of Massachusetts Amherst, Amherst, MA, USA}
\mlsysaffiliation{low}{University of Massachusetts Lowell, Lowell, MA, USA}
\mlsyscorrespondingauthor{Hieu Tran}{hieutran@umass.edu}
\mlsyskeywords{LoRA, quantization, FP8, FP4, LLM serving, reinforcement learning}
\vskip 0.3in
\begin{abstract}
Low-rank adapters (LoRA) let one base model serve many fine-tuned variants: multi-tenant engines keep a single copy of the base and batch the requests of many \emph{tenants}, each with its own adapter.
As base models move to 8-bit (FP8) and 4-bit formats, their general matrix multiplications (GEMMs) become cheaper, but the adapters do not: engines run them as separate small GPU kernels beside the GEMM, whose latency takes a growing share of every decoding step.
An adapter projects its input down to a low rank and back up, and our key observation is that the GEMM needs the down-projection only in its \emph{epilogue}, the final stage in which it writes its output.
EpiLoRA, a plugin for the vLLM engine, therefore launches the down-projection first and the low-precision GEMM as a dependent kernel that streams its weights meanwhile and adds every tenant's up-projection in its epilogue.
On NVIDIA Hopper GPUs, EpiLoRA decodes for a single adapter with up to 24\% lower time per token than vLLM's fastest configuration, and faster than SGLang and TensorRT-LLM; with many tenants, its adapter kernels deliver more throughput than vLLM's on most GPU--model pairs we tested.
The same low-precision base also trains: a LoRA reinforcement-learning trainer that computes on the engine's FP8 weights needs no 16-bit copy, so a 27B model trains on one 48\,GB GPU, and an engine that then keeps activations in 16 bits removes most of the remaining mismatch between the model generating training data and the one trained.
\end{abstract}
]
\printAffiliationsAndNotice{}

\section{Introduction}
\label{sec:intro}

Low-rank adapters (LoRA)~\citep{hu2022lora} have become the standard way to specialize a large language model without copying it.
An adapter adds a low-rank update $BA$ to each frozen weight matrix $W$ of the model, where $A$ projects the layer's input down to a small rank $r$ (typically 16--128) and $B$ projects it back up; together they hold a fraction of a percent of the model's parameters.
One base model can therefore carry thousands of fine-tunes.
Multi-tenant serving engines exploit this by keeping a single copy of the base in GPU memory and batching the requests of many \emph{tenants}, the users or applications that each bring their own adapter~\citep{chen2024punica,sheng2023slora}.
Adapters are also what LoRA reinforcement learning (RL) trains: LoRA matches full fine-tuning at a fraction of the memory~\citep{schulman2025lorawithoutregret}, and several services now train adapters for many users on shared bases~\citep{tinker2025,mint2026,marlaas2026}.

At the same time, base models are moving to low precision.
NVIDIA's Ada (e.g., L40S) and Hopper (H100 and H200) GPUs compute in 8-bit floating point (FP8) at twice their rate in the 16-bit bf16 format~\citep{micikevicius2022fp8}, and Blackwell GPUs compute 4-bit formats such as NVFP4 at four times that rate~\citep{rouhani2023mx,nvidia2025nvfp4}.
Serving systems therefore quantize the base model's general matrix multiplications (GEMMs) to these formats~\citep{lin2024awq,lin2025qserve,frantar2025marlin}, and RL frameworks increasingly generate their training samples, called \emph{rollouts}, with FP8 inference engines~\citep{flashrl2025,jetrl2026,fp8rl2026}.
Lower-precision bases make inference cheaper, but the cost of serving LoRA adapters remains largely unchanged in current engines.

The reason is that the adapter path stays outside the low-precision GEMM.
Merging an adapter into a quantized base rounds the update onto the base's coarse grid and can erase it~\citep{pulse2026,scaleqlora2026}, so engines keep adapters separate and compute them in their own GPU kernels beside the base GEMM~\citep{kwon2023vllm,chen2024punica}.
During \emph{decoding}, when the engine generates one token per request at each step, this adapter path is a chain of small dependent kernels whose latency does not shrink with the base.
The cheaper the base, the larger the adapter's share of each step (\cref{fig:precision}).

Existing systems address other parts of the problem.
Punica and S-LoRA made batching many adapters practical~\citep{chen2024punica,sheng2023slora}, and later systems decide where adapters are stored and which tenants share a batch~\citep{wu2024dlora,li2024caraserve,chen2026infinilora}.
SVDQuant fuses one low-rank branch, shared by every token, into a 4-bit GEMM~\citep{li2025svdquant}, and LoRAFusion fuses several adapters into a bf16 GEMM for fine-tuning~\citep{zhu2026lorafusion}.
However, none computes many tenants' adapters in a low-precision GEMM for serving.

\begin{figure}[t]
\centering
\resizebox{\columnwidth}{!}{\begin{tikzpicture}[x=1cm, y=1cm, font=\scriptsize,
  box/.style={draw=black!55, line width=0.4pt, minimum height=0.34cm, inner sep=1pt, anchor=west, align=center},
  lane/.style={anchor=east, align=right, inner sep=1pt, text=black!80},
  flow/.style={-{Stealth[length=1.3mm]}, draw=black!65, line width=0.45pt}]
\definecolor{obOurs}{RGB}{0,114,178}\definecolor{obV}{RGB}{140,140,140}
\node[anchor=west, font=\scriptsize\bfseries] at (-1.55, 1.05) {(a) one stream: the adapter's kernels follow the GEMM};
\node[lane] at (-0.05, 0.55) {stream};
\node[box, fill=black!8, minimum width=0.35cm] (q) at (0.0, 0.55) {q};
\node[box, fill=black!8, minimum width=3.0cm] (g) at (0.38, 0.55) {FP8 GEMM: stream $W$};
\node[box, fill=obV!45, minimum width=0.75cm] (s) at (3.41, 0.55) {shrink};
\node[box, fill=obV!45, minimum width=0.75cm] (e) at (4.19, 0.55) {expand};
\draw[decorate, decoration={brace, amplitude=2.5pt, mirror}, black!60] (3.41, 0.33) -- (4.94, 0.33)
  node[midway, below=1.5pt, text=black!70, align=center] {the adapter's latency};
\node[anchor=west, font=\scriptsize\bfseries] at (-1.55, -0.30) {(b) two streams: the LoRA branch beside the GEMM, added after it};
\node[lane] at (-0.05, -0.75) {stream 1};
\node[box, fill=black!8, minimum width=0.35cm] (qb) at (0.0, -0.75) {q};
\node[box, fill=black!8, minimum width=3.0cm] (gb) at (0.38, -0.75) {FP8 GEMM: stream $W$};
\node[box, fill=obV!45, minimum width=0.45cm] (ab) at (3.41, -0.75) {add};
\node[lane] at (-0.05, -1.20) {stream 2};
\node[box, fill=obV!45, minimum width=0.75cm] (sb) at (0.0, -1.20) {shrink};
\node[box, fill=obV!45, minimum width=0.75cm] (eb) at (0.78, -1.20) {expand};
\draw[flow] (eb.east) .. controls (2.4, -1.20) and (3.2, -1.15) .. node[pos=0.45, below, text=black!70] {join} (ab.south);
\draw[decorate, decoration={brace, amplitude=2.5pt, mirror}, black!60] (3.41, -1.42) -- (3.89, -1.42)
  node[midway, below=1.5pt, text=black!70, align=center] {join + add};
\node[anchor=west, font=\scriptsize\bfseries, text=obOurs] at (-1.55, -2.05) {(c) EpiLoRA's decode path: $h$ is needed only in the epilogue};
\node[lane] at (-0.05, -2.50) {kernel 1};
\node[box, fill=obOurs!12, minimum width=0.35cm] (q2) at (0.0, -2.50) {q};
\node[box, fill=obOurs!25, minimum width=0.75cm] (s2) at (0.38, -2.50) {shrink};
\node[box, fill=obOurs!12, minimum width=3.0cm] (g2) at (0.38, -2.95) {K2: stream $W$, prefetch $B_\ell$};
\node[box, fill=obOurs!35, minimum width=0.45cm] (ep) at (3.41, -2.95) {+$hB$};
\node[lane] at (-0.05, -2.95) {kernel 2};
\draw[flow] (s2.east) .. controls (1.6, -2.50) and (3.2, -2.55) .. node[pos=0.55, above, text=black!70] {$h$ ready} (ep.north);
\draw[decorate, decoration={brace, amplitude=2.5pt, mirror}, black!60] (3.41, -3.17) -- (3.89, -3.17)
  node[midway, below=1.5pt, text=black!70, align=center] {the adapter's cost};
\end{tikzpicture}}
\caption{\textbf{The GEMM needs the adapter's down-projection only in its epilogue.} One projection during decoding; a CUDA stream runs its kernels (GPU functions) in order, and two streams can run at once. (a) vLLM runs the down-projection (\emph{shrink}, $h = xA^\top$) and the up-projection (\emph{expand}, $y \mathrel{+}= hB^\top$) after the FP8 GEMM; (b) its dual-stream mode and TensorRT-LLM run them on a second stream and add the result after the GEMM; (c) EpiLoRA launches the shrink first and the GEMM as a dependent kernel, which starts while the shrink runs and adds $hB^\top$ in its epilogue.}
\label{fig:decode}
\end{figure}
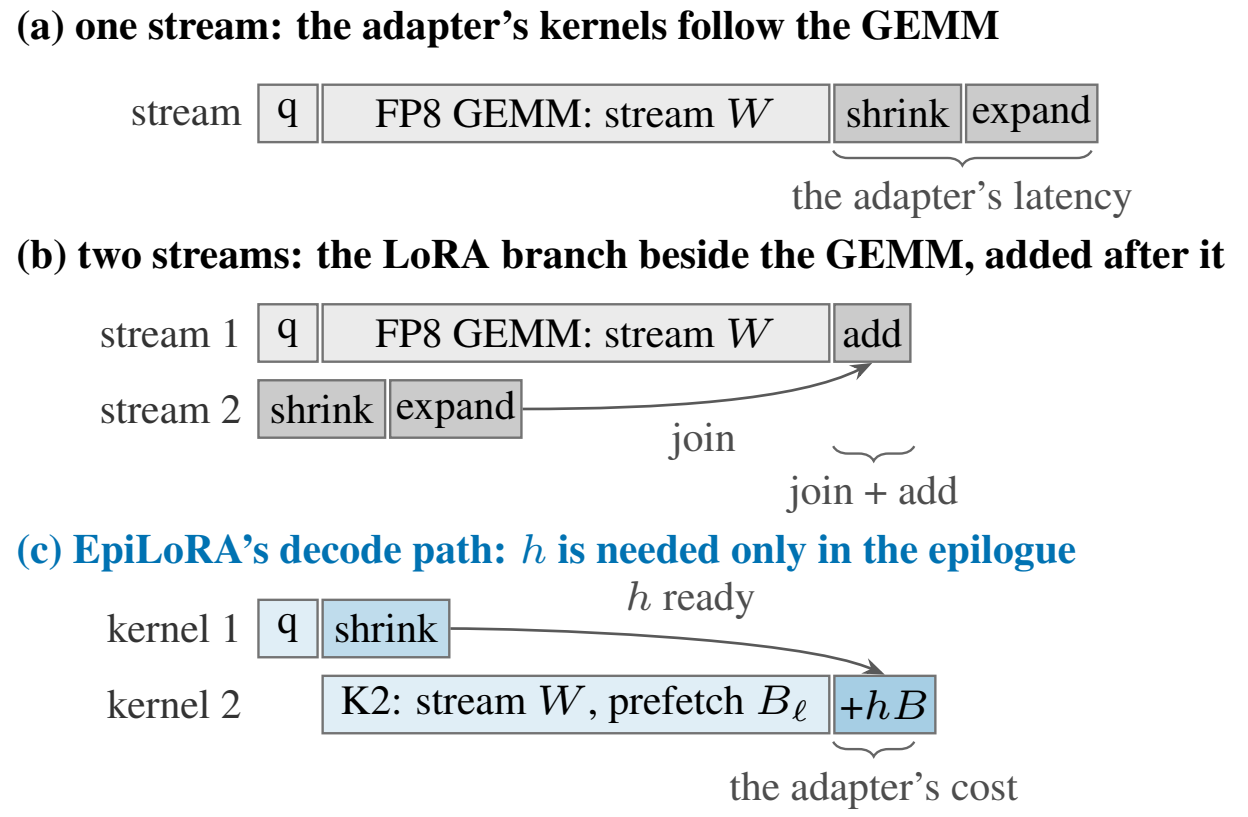

We present \textbf{EpiLoRA}, a plugin for the vLLM engine that computes every tenant's adapter together with the low-precision base GEMM.
A GEMM kernel streams the weights through its \emph{main loop} and applies scales and writes its output in its \emph{epilogue}.
Our key observation is that the GEMM needs an adapter's down-projection only in that epilogue, after the main loop has finished (\cref{fig:decode}).
During decoding, EpiLoRA therefore launches the down-projection first and its own FP8 or NVFP4 GEMM as a dependent kernel: the GEMM streams its weights while the down-projection runs, and then adds each tenant's up-projection in its epilogue.
The down-projection's latency is then mostly hidden: on full-size layers with rank-64 adapters, at most about 2\,\textmu s of it stays exposed.
On Ada GPUs, whose standard FP8 GEMM is slow during decoding, EpiLoRA instead fuses all resident adapters into two kernels of its own.
The kernels quantize activations exactly as vLLM does, so the base GEMM receives the same FP8 inputs as in vLLM.

The low-precision base also benefits RL training, where frameworks that place the inference engine and the trainer on the same GPUs keep a separate bf16 copy of the base for the trainer~\citep{hu2024openrlhf,sheng2025hybridflow,mei2025real}.
Because the FP8 engine then samples from a slightly different model than the trainer optimizes, FP8 rollout pipelines reweight each sample with an importance weight~\citep{flashrl2025}.
EpiLoRA's trainer instead computes on the engine's own FP8 weights, which saves enough memory to train a 27B model on one 48\,GB L40S, where its bf16 weights alone do not fit, and lets one copy serve both the engine and the trainer.
A \emph{matched} engine also keeps its activations in bf16, as the trainer does, which removes most of the mismatch that remains once the weights agree.

We evaluate EpiLoRA on models from 8B to 72B parameters on Ada, Hopper and Blackwell GPUs, against the strongest configuration of each baseline we could build.
Serving a single request (batch size 1), EpiLoRA takes 0.76--0.88$\times$ the time per output token of vLLM's fastest configuration for 8B--9B models on Hopper GPUs, is faster than SGLang and TensorRT-LLM, and keeps its lead on a 72B model, on one GPU and across two.
On the L40S it matches vLLM's fastest configuration.
With 32 tenants, whose batches are too large for the decode path, it delivers 1.05--1.09$\times$ vLLM's throughput on 7 of 9 GPU--model pairs, from its adapter kernels on Hopper GPUs and fused adapters on the L40S.

In summary, our contributions are:
\begin{itemize}[leftmargin=1.2em]
\item The \emph{decode path}, which removes most of an adapter's latency by adding the up-projection in the GEMM's epilogue and overlapping the down-projection with its main loop, and a simple condition that bounds when the overlap hides the down-projection (\cref{sec:kernels,sec:eval_kernels}).
\item Multi-adapter kernels for FP8 and NVFP4 bases in a drop-in vLLM plugin that works with CUDA graphs and tensor parallelism (\cref{sec:plugin,sec:eval_kernels,sec:eval_e2e}).
\item A LoRA RL system on the same low-precision base, with a decomposition of the engine--trainer mismatch showing what each design choice removes (\cref{sec:rlsys,sec:eval_rl}).
\end{itemize}
\section{Background and motivation}
\label{sec:background}

\textbf{Why engines do not merge adapters.}
Merging an adapter into the base ($W \to W+BA$) produces one model per adapter, so the engine can no longer batch tenants.
On a low-precision base, merging is also lossy: rounding $W+BA$ to its grid changes only entries whose update exceeds half a quantization step, erasing small fine-tunes.

\begin{figure}[t]
\centering
\includegraphics[width=\columnwidth]{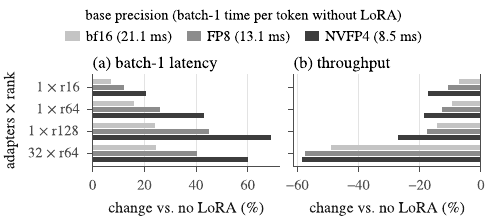}
\caption{\textbf{Adapters cost relatively more on cheaper bases.} Latency added at batch size 1 and throughput lost with vLLM's LoRA path and default kernels, relative to the engine without LoRA: Qwen3-8B with bf16, FP8 and NVFP4 bases on a partition of a Blackwell GPU, one adapter of rank 16--128 or 32 of rank 64.}
\label{fig:precision}
\end{figure}

\textbf{The cost of the separate adapter path.}
Each decoding step runs every linear layer (\emph{projection}) of the model over the tokens of the current batch.
Consider a projection over $M$ tokens, each belonging to a request that uses one of $L$ resident adapters.
On an FP8 base, vLLM~0.28 runs four kernels for it~\citep{chen2024punica}: it quantizes the activations $x$ to FP8, runs the FP8 GEMM, computes the down-projection $h=xA_\ell^\top$ for each adapter's tokens (the \emph{shrink}), and adds the up-projection $hB_\ell^\top$ to the output (the \emph{expand}).
The activations are read twice and the output is written, read and written again, and the adapter weights are read by kernels that cannot share the GEMM's work.
None of these costs shrinks with the base. Tuning the LoRA kernels' launch configurations, which we do for every vLLM baseline, reduces them, but with tuned kernels or vLLM's dual-stream mode the adapter still adds 42--55\% to a layer's time at batch size 1 on Hopper GPUs (\cref{tab:waterfall}).
On a Blackwell partition, one rank-64 adapter adds 16\% to Qwen3-8B's latency at batch size 1 with a bf16 base, 26\% with an FP8 base and 43\% with an NVFP4 base with vLLM's default kernels (\cref{fig:precision}).

\textbf{Programmatic dependent launch.}
A GPU kernel runs as many \emph{thread blocks}, which the GPU schedules onto its streaming multiprocessors (SMs).
On Hopper and Blackwell GPUs, programmatic dependent launch (PDL) lets a kernel's successor start before it finishes: the dependent's thread blocks start as soon as the producer signals that they may, and wait for its results only at an explicit wait instruction.
Inference engines use PDL to hide the launch latency between consecutive kernels, placing the wait at the start of the dependent kernel, so that only its setup overlaps the producer.
EpiLoRA moves the wait to the end of the GEMM, just before its epilogue, so that the GEMM's entire main loop overlaps the adapter's down-projection.

\textbf{LoRA RL on a low-precision engine.}
RL post-training alternates between generating rollouts with an inference engine and updating the model with a trainer, and it spends much of its time generating~\citep{sheng2025hybridflow,fu2025areal}.
An FP8 inference engine and a bf16 trainer, however, produce slightly different token distributions because they compute at different precisions.
Existing quantized rollout pipelines correct this mismatch with importance weights, the ratio of the trainer's to the engine's probability of each sampled token~\citep{flashrl2025,fp8rlstack2026,ais2026}.
Placing engine and trainer on one GPU also duplicates the base in memory: two bf16 copies of a 27B model exceed an 80\,GB GPU, and moving the trainer's copy to host memory (\emph{offloading}) copies it every step and pins it there.
As LoRA RL updates only the adapter, \cref{sec:rlsys} addresses both problems at their source.
\section{Design}
\label{sec:design}

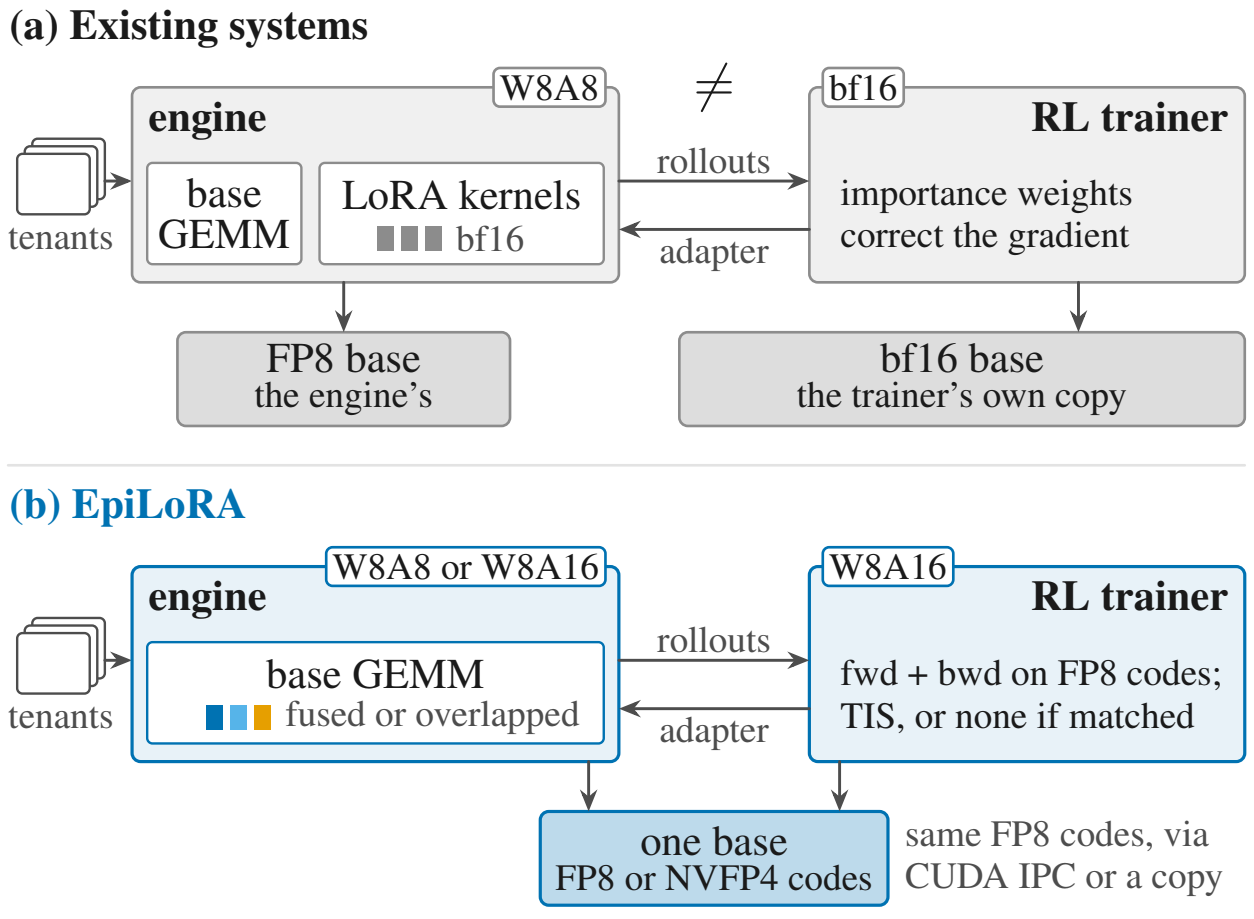
\begin{figure}[t]
\centering
\definecolor{obInk}{HTML}{1A1A1A}%
\definecolor{obInkTwo}{HTML}{4D4D4D}%
\definecolor{obGrid}{HTML}{E3E3E3}%
\definecolor{obVllm}{HTML}{8C8C8C}%
\definecolor{obOurs}{HTML}{0072B2}%
\definecolor{obFPeight}{HTML}{56B4E9}%
\definecolor{obFPfour}{HTML}{E69F00}%
\definecolor{obMatched}{HTML}{009E73}%
\def\obLabel{\fontsize{8}{9.2}\selectfont}%
\def\obNote{\fontsize{7}{8}\selectfont}%
\tikzset{
  obfig/.style={font=\obLabel, text=obInk, line width=0.5pt, line cap=round, line join=round},
  obarrow/.style={-{Stealth[length=1.45mm, width=1.15mm]}, draw=obInkTwo, line width=0.55pt},
  obnote/.style={font=\obNote, text=obInkTwo, inner sep=1pt},
  obtitle/.style={font=\obLabel\bfseries, text=obInk, anchor=base west, inner sep=0pt},
  obvllm/.style={draw=obVllm, fill=obVllm!13, rounded corners=1.6pt, line width=0.5pt},
  obours/.style={draw=obOurs, fill=obOurs!9, rounded corners=1.6pt, line width=0.6pt},
  obtensor/.style={draw=obInkTwo, fill=white, line width=0.45pt, rounded corners=0.6pt, inner sep=1.5pt, align=center},
}%
\begin{tikzpicture}[obfig, x=1cm, y=1cm]
\def\tTa{0.06}\def\tTb{0.64}     %
\def\tEa{0.82}\def\tEb{4.04}     %
\def\tRa{5.30}\def\tRb{8.18}     %
\def\tCa{0.95}\def\tCb{2.24}     %
\def\tIa{1.07}\def\tIb{1.74}     %
\def\tBh{0.62}                   %
\def\tTy{1.95}                   %
\def\tGx{4.67}                   %
\newcommand\tTenants{%
  \foreach \d in {0.10, 0.05, 0}
    \draw[draw=obInkTwo, fill=white, line width=0.45pt, rounded corners=1pt] (\tTa+\d, 1.40+\d) rectangle (\tTb-0.10+\d, 1.80+\d);
  \node[obnote, anchor=north] at ({(\tTa+\tTb)/2}, 1.36) {tenants};
  \draw[obarrow] (\tTb, 1.62) -- (\tEa, 1.62);
}
\newcommand\tTab[4]{%
  \node[draw=#3, fill=white, line width=0.5pt, rounded corners=1.2pt, inner sep=1.2pt, font=\obNote, text=obInk, anchor=#1]
    at (#2, \tCb) {#4};
}
\newcommand\tLinks{%
  \draw[obarrow] (\tEb, 1.62) -- node[obnote, above=0.2pt] {rollouts} (\tRa, 1.62);
  \draw[obarrow] (\tRa, 1.30) -- node[obnote, below=0.2pt] {adapter} (\tEb, 1.30);
}
\newcommand\tSlab[5]{%
  \draw[#3] (#1, 0) rectangle (#2, \tBh);
  \node[align=center] at ({(#1+#2)/2}, \tBh/2) {#4\\[-2.5pt]{\obNote #5}};
}
\begin{scope}
\node[obtitle] at (0, 2.56) {(a) Existing systems};
\tTenants
\draw[obvllm] (\tEa, \tCa) rectangle (\tEb, \tCb);
\node[anchor=base west, inner sep=0pt, font=\obLabel\bfseries] at (\tEa+0.1, \tTy) {engine};
\draw[draw=obVllm, fill=white, line width=0.45pt, rounded corners=1pt] (\tEa+0.1, \tIa) rectangle (\tEa+1.12, \tIb);
\node[align=center] at (\tEa+0.61, {(\tIa+\tIb)/2}) {base\\[-1.5pt]GEMM};
\draw[draw=obVllm, fill=white, line width=0.45pt, rounded corners=1pt] (\tEa+1.24, \tIa) rectangle (\tEb-0.1, \tIb);
\node at ({(\tEa+1.24+\tEb-0.1)/2}, 1.53) {LoRA kernels};
\foreach \i in {0,1,2}
  \fill[obVllm] ({(\tEa+1.24+\tEb-0.1)/2-0.56+\i*0.16}, 1.16) rectangle ++(0.11, 0.16);
\node[obnote, anchor=west] at ({(\tEa+1.24+\tEb-0.1)/2-0.08}, 1.24) {bf16};
\draw[obvllm] (\tRa, \tCa) rectangle (\tRb, \tCb);
\node[anchor=base east, inner sep=0pt, font=\obLabel\bfseries] at (\tRb-0.1, \tTy) {RL trainer};
\node[anchor=west, align=left, font=\obNote] at (\tRa+0.08, 1.38) {importance weights\\correct the gradient};
\tTab{east}{\tEb-0.08}{obVllm}{W8A8}
\tTab{west}{\tRa+0.08}{obVllm}{bf16}
\node[font=\fontsize{10}{12}\selectfont] at (\tGx, \tCb) {$\neq$};
\tLinks
\tSlab{\tEa+0.30}{\tEa+2.50}{obvllm, fill=obVllm!30}{FP8 base}{the engine's}
\tSlab{\tRb-3.74}{\tRb}{obvllm, fill=obVllm!30}{bf16 base}{the trainer's own copy}
\draw[obarrow] (\tEa+1.40, \tCa) -- (\tEa+1.40, \tBh);
\draw[obarrow] (\tRb-1.10, \tCa) -- (\tRb-1.10, \tBh);
\end{scope}
\draw[obGrid, line width=0.5pt] (0, -0.26) -- (\tRb, -0.26);
\begin{scope}[yshift=-3.17cm]
\node[obtitle] at (0, 2.56) {\color{obOurs}(b) EpiLoRA};
\tTenants
\draw[obours] (\tEa, \tCa) rectangle (\tEb, \tCb);
\node[anchor=base west, inner sep=0pt, font=\obLabel\bfseries] at (\tEa+0.1, \tTy) {engine};
\draw[draw=obOurs, fill=white, line width=0.45pt, rounded corners=1pt] (\tEa+0.1, \tIa) rectangle (\tEb-0.1, \tIb);
\node at ({(\tEa+\tEb)/2}, 1.53) {base GEMM};
\foreach \c/\i in {obOurs/0, obFPeight/1, obFPfour/2}
  \fill[\c] ({(\tEa+\tEb)/2-1.12+\i*0.16}, 1.16) rectangle ++(0.11, 0.16);
\node[obnote, anchor=west] at ({(\tEa+\tEb)/2-0.64}, 1.24) {fused or overlapped};
\draw[obours] (\tRa, \tCa) rectangle (\tRb, \tCb);
\node[anchor=base east, inner sep=0pt, font=\obLabel\bfseries] at (\tRb-0.1, \tTy) {RL trainer};
\node[anchor=west, align=left, font=\obNote] at (\tRa+0.08, 1.38) {fwd + bwd on FP8 codes;\\TIS, or none if matched};
\tTab{east}{\tEb-0.08}{obOurs}{W8A8 or W8A16}
\tTab{west}{\tRa+0.08}{obOurs}{W8A16}
\tLinks
\tSlab{\tGx-1.15}{\tGx+1.15}{obours, fill=obOurs!26}{one base}{FP8 or NVFP4 codes}
\draw[obarrow] (\tEb-0.20, \tCa) -- (\tEb-0.20, \tBh);
\draw[obarrow] (\tRa+0.20, \tCa) -- (\tRa+0.20, \tBh);
\node[obnote, anchor=west, align=left] at (\tGx+1.22, \tBh/2) {same FP8 codes, via\\CUDA IPC or a copy};
\end{scope}
\end{tikzpicture}
\caption{\textbf{One low-precision base for serving and for RL.} (a) Existing systems run adapters in kernels beside the FP8 GEMM, and an RL trainer keeps its own bf16 copy of the base, with importance weights correcting for the engine's rounding. (b) EpiLoRA computes the adapters with the base GEMM, and its trainer computes on the engine's FP8 weights, with rollouts from an FP8 engine with importance sampling or from a matched engine that needs no correction (W8A8: 8-bit weights and activations; W8A16: 16-bit activations); serving tenants from the copy RL trains works but does not pay (Appendix~\ref{app:discussion}).}
\label{fig:thesis}
\end{figure}

EpiLoRA has three parts: kernels that compute every adapter together with the low-precision base GEMM, a vLLM plugin that replaces vLLM's LoRA layers, and an RL system on the plugin; \cref{fig:thesis} compares it with existing systems.

\subsection{Multi-adapter kernels}
\label{sec:kernels}

Fusing the adapters of many tenants into a low-precision GEMM is harder than fusing one low-rank branch that every token shares~\citep{li2025svdquant}, for three reasons.
First, a GEMM divides its output into tiles, each computed by one thread block, and tokens of different adapters share a tile, so the epilogue must apply a different up-projection to each group of rows.
Second, the kernels must quantize the activations exactly as the engine's own GEMM does, so the base GEMM sees the same FP8 inputs.
Third, during decoding the adapter path is a chain of small dependent kernels whose latency, rather than its memory traffic, sets its cost.
We address the first two with two kernels, K1 and K2, that fuse the adapters' down- and up-projections into the base GEMM, and the third with the decode path, which overlaps the down-projection with the GEMM's main loop.

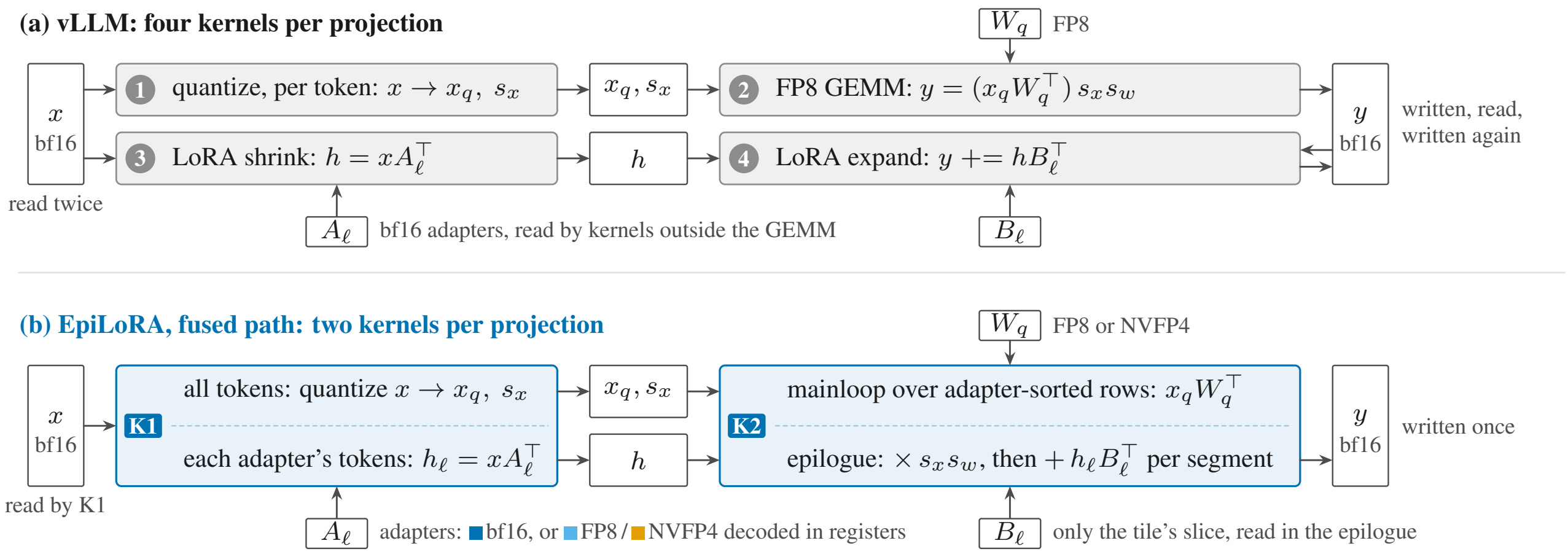
\begin{figure*}[t]
\centering
\definecolor{obInk}{HTML}{1A1A1A}%
\definecolor{obInkTwo}{HTML}{4D4D4D}%
\definecolor{obGrid}{HTML}{E3E3E3}%
\definecolor{obVllm}{HTML}{8C8C8C}%
\definecolor{obOurs}{HTML}{0072B2}%
\definecolor{obFPeight}{HTML}{56B4E9}%
\definecolor{obFPfour}{HTML}{E69F00}%
\definecolor{obMatched}{HTML}{009E73}%
\def\obLabel{\fontsize{8}{9.2}\selectfont}%
\def\obNote{\fontsize{7}{8}\selectfont}%
\tikzset{
  obfig/.style={font=\obLabel, text=obInk, line width=0.5pt, line cap=round, line join=round},
  obarrow/.style={-{Stealth[length=1.45mm, width=1.15mm]}, draw=obInkTwo, line width=0.55pt},
  obnote/.style={font=\obNote, text=obInkTwo, inner sep=1pt},
  obtitle/.style={font=\obLabel\bfseries, text=obInk, anchor=base west, inner sep=0pt},
  obvllm/.style={draw=obVllm, fill=obVllm!13, rounded corners=1.6pt, line width=0.5pt},
  obours/.style={draw=obOurs, fill=obOurs!9, rounded corners=1.6pt, line width=0.6pt},
  obtensor/.style={draw=obInkTwo, fill=white, line width=0.45pt, rounded corners=0.6pt, inner sep=1.5pt, align=center},
}%
\begin{tikzpicture}[obfig, x=1cm, y=1cm]
\def\kXa{0.10}\def\kXb{0.70}      %
\def\kCa{1.05}\def\kCb{5.80}      %
\def\kMa{6.15}\def\kMb{7.20}      %
\def\kDa{7.55}\def\kDb{13.80}     %
\def\kYa{14.15}\def\kYb{14.75}    %
\def\kRh{0.28}                   %
\def\kRu{0.37}                   %
\def\kCm{3.425}\def\kDm{10.675}   %
\newcommand\kChip[5]{\draw[obtensor] (#1, #3) rectangle (#2, #4); \node at ({(#1+#2)/2}, {(#3+#4)/2}) {#5};}
\newcommand\kTensors{%
  \kChip{\kXa}{\kXb}{-0.65}{0.65}{} \node at ({(\kXa+\kXb)/2}, 0.09) {$x$}; \node[obnote] at ({(\kXa+\kXb)/2}, -0.2) {bf16};
  \kChip{\kMa}{\kMb}{\kRu-\kRh}{\kRu+\kRh}{$x_q, s_x$}
  \kChip{\kMa}{\kMb}{-\kRu-\kRh}{-\kRu+\kRh}{$h$}
  \kChip{\kYa}{\kYb}{-0.65}{0.65}{} \node at ({(\kYa+\kYb)/2}, 0.09) {$y$}; \node[obnote] at ({(\kYa+\kYb)/2}, -0.2) {bf16};
  \kChip{\kDm-0.33}{\kDm+0.33}{0.92}{1.24}{$W_q$}
  \draw[obarrow] (\kDm, 0.92) -- (\kDm, 0.65);
  \draw[obarrow] (\kCm, -1.0) -- (\kCm, -0.65);
  \draw[obarrow] (\kDm, -1.0) -- (\kDm, -0.65);
}
\begin{scope}
\node[obtitle] at (0, 1.0) {(a) vLLM: four kernels per projection};
\draw[obvllm] (\kCa, \kRu-\kRh) rectangle (\kCb, \kRu+\kRh);
\draw[obvllm] (\kCa, -\kRu-\kRh) rectangle (\kCb, -\kRu+\kRh);
\draw[obvllm] (\kDa, \kRu-\kRh) rectangle (\kDb, \kRu+\kRh);
\draw[obvllm] (\kDa, -\kRu-\kRh) rectangle (\kDb, -\kRu+\kRh);
\kTensors
\foreach \n/\px/\py in {1/\kCa/\kRu, 3/\kCa/-\kRu, 2/\kDa/\kRu, 4/\kDa/-\kRu}
  \node[circle, fill=obVllm, text=white, inner sep=0pt, minimum size=3.2mm, font=\obNote\bfseries] at (\px+0.26, \py) {\n};
\node[anchor=west] at (\kCa+0.48, \kRu) {quantize, per token: $x \to x_q,\ s_x$};
\node[anchor=west] at (\kCa+0.48, -\kRu) {LoRA shrink: $h = x A_\ell^{\top}$};
\node[anchor=west] at (\kDa+0.48, \kRu) {FP8 GEMM: $y = (x_q W_q^{\top})\, s_x s_w$};
\node[anchor=west] at (\kDa+0.48, -\kRu) {LoRA expand: $y \mathrel{+}= h B_\ell^{\top}$};
\kChip{\kCm-0.33}{\kCm+0.33}{-1.32}{-1.0}{$A_\ell$}
\kChip{\kDm-0.33}{\kDm+0.33}{-1.32}{-1.0}{$B_\ell$}
\draw[obarrow] (\kXb, \kRu) -- (\kCa, \kRu);
\draw[obarrow] (\kXb, -\kRu) -- (\kCa, -\kRu);
\draw[obarrow] (\kCb, \kRu) -- (\kMa, \kRu);
\draw[obarrow] (\kMb, \kRu) -- (\kDa, \kRu);
\draw[obarrow] (\kCb, -\kRu) -- (\kMa, -\kRu);
\draw[obarrow] (\kMb, -\kRu) -- (\kDa, -\kRu);
\draw[obarrow] (\kDb, \kRu) -- (\kYa, \kRu);
\draw[obarrow] (\kYa, -\kRu+0.09) -- (\kDb, -\kRu+0.09);
\draw[obarrow] (\kDb, -\kRu-0.09) -- (\kYa, -\kRu-0.09);
\node[obnote, anchor=north] at ({(\kXa+\kXb)/2}, -0.72) {read twice};
\node[obnote, anchor=west, align=left] at (\kYb+0.1, 0) {written, read,\\written again};
\node[obnote, anchor=west] at (\kCm+0.42, -1.16) {bf16 adapters, read by kernels outside the GEMM};
\node[obnote, anchor=west] at (\kDm+0.42, 1.08) {FP8};
\end{scope}
\draw[obGrid, line width=0.5pt] (0, -1.6) -- (16.65, -1.6);
\begin{scope}[yshift=-3.25cm]
\node[obtitle] at (0, 1.0) {\color{obOurs}(b) EpiLoRA, fused path: two kernels per projection};
\draw[obours] (\kCa, -0.65) rectangle (\kCb, 0.65);
\draw[obours] (\kDa, -0.65) rectangle (\kDb, 0.65);
\kTensors
\draw[obOurs!40, line width=0.4pt, dash pattern=on 1.2pt off 1.2pt] (\kCa+0.6, 0) -- (\kCb-0.1, 0);
\draw[obOurs!40, line width=0.4pt, dash pattern=on 1.2pt off 1.2pt] (\kDa+0.6, 0) -- (\kDb-0.1, 0);
\foreach \k/\px in {K1/\kCa, K2/\kDa}
  \node[rounded corners=1pt, fill=obOurs, text=white, inner sep=1.3pt, font=\obNote\bfseries] at (\px+0.29, 0) {\k};
\node[anchor=west] at (\kCa+0.6, \kRu) {all tokens: quantize $x \to x_q,\ s_x$};
\node[anchor=west] at (\kCa+0.6, -\kRu) {each adapter's tokens: $h_\ell = x A_\ell^{\top}$};
\node[anchor=west] at (\kDa+0.6, \kRu) {mainloop over adapter-sorted rows: $x_q W_q^{\top}$};
\node[anchor=west] at (\kDa+0.6, -\kRu) {epilogue: ${\times}\, s_x s_w$, then ${+}\, h_\ell B_\ell^{\top}$ per segment};
\kChip{\kCm-0.33}{\kCm+0.33}{-1.32}{-1.0}{$A_\ell$}
\kChip{\kDm-0.33}{\kDm+0.33}{-1.32}{-1.0}{$B_\ell$}
\draw[obarrow] (\kXb, 0) -- (\kCa, 0);
\draw[obarrow] (\kCb, \kRu) -- (\kMa, \kRu);
\draw[obarrow] (\kMb, \kRu) -- (\kDa, \kRu);
\draw[obarrow] (\kCb, -\kRu) -- (\kMa, -\kRu);
\draw[obarrow] (\kMb, -\kRu) -- (\kDa, -\kRu);
\draw[obarrow] (\kDb, -\kRu) -- (\kYa, -\kRu);
\node[obnote, anchor=north] at ({(\kXa+\kXb)/2}, -0.72) {read by K1};
\node[obnote, anchor=west, align=left] at (\kYb+0.1, 0) {written once};
\node[obnote, anchor=west] at (\kCm+0.42, -1.16) {adapters:
  \tikz[baseline=-0.55ex]{\fill[obOurs] (0,-0.07) rectangle (0.14,0.07);}\,bf16, or
  \tikz[baseline=-0.55ex]{\fill[obFPeight] (0,-0.07) rectangle (0.14,0.07);}\,FP8\,/\,%
  \tikz[baseline=-0.55ex]{\fill[obFPfour] (0,-0.07) rectangle (0.14,0.07);}\,NVFP4 decoded in registers};
\node[obnote, anchor=west] at (\kDm+0.42, -1.16) {only the tile's slice, read in the epilogue};
\node[obnote, anchor=west] at (\kDm+0.42, 1.08) {FP8 or NVFP4};
\end{scope}
\end{tikzpicture}
\caption{\textbf{Two kernels per projection instead of four.} Boxes are tensors in GPU memory, arrows reads and writes. (a) vLLM~0.28 on an FP8 base, in launch order. (b) EpiLoRA's fused path: K1 quantizes the activations once, exactly as vLLM does, and computes each adapter's down-projection; K2 runs the FP8 or NVFP4 GEMM on tokens sorted by adapter and adds each up-projection in its epilogue.}
\label{fig:kernels}
\end{figure*}

\textbf{K1 and K2.}
K1 (\cref{fig:kernels}b) processes the tokens grouped by adapter, using the batching metadata vLLM already maintains.
It quantizes the layer's activations once for the base GEMM, with the same per-token scale and rounding as vLLM, and computes each adapter's down-projection $h_\ell = x A_\ell^\top$ from the bf16 activations.
Because a decode batch fills only a few dozen thread blocks, K1 splits the sum over the input dimension across several blocks (split-K) to use more of the GPU.
K2 then runs the low-precision GEMM over the tokens in adapter order, so that each output tile meets only a few contiguous groups of adapters.
Its epilogue applies the scales and, for each adapter group in the tile, adds $h_\ell B_\ell^\top$, reading only the tile's slice of $B_\ell$.
Adapters may also be stored in FP8 or NVFP4 and are dequantized in registers, which halves (FP8) or nearly quarters (NVFP4) the adapter bytes the epilogue reads, though 6 of 24 accuracy checks of stored adapters failed.

\textbf{When fusion pays.}
The fused path removes vLLM's separate adapter kernels but replaces the vendor's GEMM with ours.
This pays on Ada GPUs against their default FP8 GEMM, slow during decoding; on Hopper GPUs our GEMM only matches vLLM's CUTLASS GEMM (\cref{tab:waterfall}), and the fused path runs the down-projection first.

\textbf{The decode path.}
During decoding, the base GEMM streams megabytes of weights, while the adapter is a chain of dependent steps: K2's epilogue cannot add $h_\ell B_\ell^\top$ before the down-projection has finished.
On the H100 this chain costs about 13\,\textmu s per projection with vLLM's tuned kernels (\cref{tab:waterfall}), although reading $A_\ell$ and $B_\ell$ takes only about a microsecond.
For decode batches of up to 32 tokens, EpiLoRA therefore launches the down-projection first as a small kernel of its own, without PDL so that it starts only after $x$ is complete, and launches K2 as its dependent with PDL (\cref{sec:background}; \cref{fig:decode}c).
A few extra thread blocks of this kernel, or for some shapes K2 itself, quantize the activations exactly as vLLM does.
K2 starts as soon as the down-projection has begun, prefetches $B_\ell$, streams its weights once the activations are quantized, and waits for $h$ only before its epilogue.
Reductions are summed in a fixed order, so results are deterministic for a given batch.
The same design runs on Blackwell with NVFP4 bases.

\textbf{Why not a second stream?}
vLLM's dual-stream mode and TensorRT-LLM also overlap the adapter with the GEMM, by running the adapter on a second CUDA stream, a kernel queue that the GPU can run concurrently (\cref{fig:decode}b).
Because their GEMM writes its output before the adapter's result is added, the two streams must join after the GEMM, and a separate kernel (vLLM's up-projection, TensorRT-LLM's add) then reads and writes the output again.
Launching the GEMM as the down-projection's dependent avoids this join: the GEMM itself adds the up-projection in its epilogue, so the output is written once and nothing follows.

\textbf{Portability.}
The decode path needs only PDL and a GEMM epilogue that can add a low-rank product for groups of rows, both available outside Triton, the GPU language of our kernels, so a CUTLASS-based engine could adopt it for any base format whose GEMM's main loop streams its weights.

\textbf{When the overlap hides the adapter.}
With the decode path, a projection's adapter cost is approximately $t_{\mathrm{epi}} + \max(0,\, t_{\mathrm{shrink}} - t_{\mathrm{main}})$: the epilogue's extra work plus whatever part of the down-projection outlasts K2's main loop.
At decode, $t_{\mathrm{main}}$ is the time to stream the projection's weights, which falls with the GPU's memory bandwidth and when tensor parallelism splits each layer's weights across GPUs, while $t_{\mathrm{shrink}}$, the latency of a small kernel, hardly changes, so the condition is easiest to meet for large projections on GPUs of modest bandwidth; \cref{fig:exposure} measures where it holds.
Where the down-projection lasts most of the main loop, the two kernels also compete for SMs, and a few microseconds stay exposed (\cref{sec:eval_kernels}).
The condition therefore bounds the exposure from below: when the GEMM launches more thread blocks than the GPU has SMs, only its first blocks start before the down-projection ends, so less than the whole main loop can hide it.

\subsection{The vLLM plugin}
\label{sec:plugin}

EpiLoRA installs as a vLLM plugin that replaces vLLM's LoRA linear layers; any layer it cannot serve falls back to vLLM.
The kernels read vLLM's own adapter buffers and batching metadata, so adapters are never copied, and the fused call is a registered custom operator that vLLM can capture in its CUDA graphs, which record a step's kernel launches once and replay them to avoid launch overhead.

\textbf{Timed dispatch.}
When the model loads, the plugin times up to three paths for each projection shape and batch size and keeps the fastest: the fused kernels, run as the decode path on Hopper GPUs; the \emph{hybrid} path, the vendor GEMM followed by our adapter kernels; and the vendor GEMM followed by vLLM's own adapter kernels.
On Hopper GPUs it selects the decode path for decode batches of 8B--9B models. For larger batches, including \emph{prefill}, which processes a request's prompt, it selects the vendor GEMM with EpiLoRA's or vLLM's adapter kernels, whichever is faster.
On the L40S it selects the fused path (\cref{tab:waterfall}).

\textbf{Memory.}
vLLM sizes its KV cache, which stores every request's attention keys and values, from the memory left after the model loads, so the plugin shares scratch buffers across layers, reuses vLLM's weight codes on NVFP4 checkpoints and stores low-precision adapters in smaller slots; with bf16 adapters, its load-time warm-up fragments memory and leaves 0.85$\times$ vLLM's cache on the L40S (Figure~\ref{fig:pareto}).

\textbf{Implementation and correctness.}
EpiLoRA is about 6,500 lines of Python and Triton, including 48 Triton kernels and the RL trainer's kernels, and needs no changes to vLLM.
Under tensor parallelism it shards the adapters as vLLM does: projections whose outputs are split across GPUs split the up-projection's output, and projections whose inputs are split split the down-projection's input.
We test correctness at three levels: EpiLoRA's FP8 and NVFP4 quantization reproduces vLLM's codes and scales bit for bit, its kernels match vLLM's LoRA path and are bitwise reproducible, and served adapters keep vLLM's accuracy (\cref{fig:acceff}).

\subsection{LoRA RL on the same base}
\label{sec:rlsys}

The RL system places a trainer and a vLLM rollout engine on one GPU and trains with GRPO, a policy-gradient algorithm that compares groups of sampled responses to the same prompt~\citep{shao2024deepseekmath,sheng2025hybridflow}.
After each optimizer step the trainer publishes only its adapter, which the engine loads into an adapter slot.

\textbf{A trainer on the engine's FP8 weights.}
A bf16 trainer differentiates a slightly different model than the FP8 engine samples from.
Our \emph{FP8 trainer} instead computes on the engine's FP8 weights with bf16 activations, which removes the weight part of this mismatch and shrinks the frozen base weights from 16.7 to 10.2\,GiB for a 9B model (Table~\ref{tab:rlparts}).
With \emph{one copy}, the trainer maps the engine's weights directly through CUDA's inter-process memory sharing instead of loading its own, and the engine releases its KV cache while the trainer runs, so no weights move between phases.
The trainer's projections run a training GEMM that reads the FP8 weights in place and adds the LoRA update in its epilogue, as K2 does, so no 16-bit copy of the weights is ever made.

\textbf{Matched-precision rollouts.}
To first order, the engine's and the trainer's log-probabilities of a sampled token $y$ differ by three terms, from rounding the weights, rounding the activations, and differences between the two sides' kernels:
\begin{equation}
\log\pi_{\mathrm{eng}}(y) - \log\pi_{\mathrm{tr}}(y) \approx \delta_W + \delta_A + \delta_K.
\label{eq:mismatch}
\end{equation}
The FP8 trainer removes $\delta_W$, but not $\delta_A$, the engine's rounding of each GEMM's input activations.
Importance corrections compensate for this difference in the gradient~\citep{flashrl2025}; we remove it in the arithmetic instead.
In the \emph{matched} engine, every FP8 layer computes with 8-bit weights and 16-bit activations (W8A16) instead of rounding both to 8 bits (W8A8) as vLLM's FP8 engine does, using the trainer's arithmetic on the same weights: FP8 weights widened to bf16 in registers, bf16 activations and fp32 accumulation.
vLLM's weight-only FP8 path, Marlin~\citep{frantar2025marlin}, also leaves activations unrounded but repacks the weights into a layout the trainer cannot share.
\section{Evaluation}
\label{sec:eval}

We ask what the kernels save (\cref{sec:eval_kernels}), what that buys end to end (\cref{sec:eval_e2e}), and whether the same base trains as well as bf16 with less memory (\cref{sec:eval_rl}).

\textbf{Setup.}
We run vLLM~0.28 with CUDA graphs and PyTorch's compiler, kernels cached before we measure.
We serve Qwen3-8B~\citep{yang2025qwen3}, Llama-3.1-8B~\citep{grattafiori2024llama3}, Qwen3.5-9B~\citep{qwen35}, which replaces most attention layers with linear attention, and Qwen2.5-72B, with FP8 bases that vLLM quantizes when the model loads, and our NVFP4 checkpoint of Qwen3-8B; RL trains Qwen3.5-9B and Qwen3.8-27B~\citep{qwen38}.
The GPUs are the L40S (Ada), the H100, H200 and GH200 (Hopper), and a 48\,GB partition of an RTX PRO 6000 (Blackwell), which we call the Blackwell slice.
For serving, the baselines are vLLM's LoRA path on its fastest base for each GPU, with LoRA kernels we tuned for each model, SGLang~\citep{zheng2024sglang} and TensorRT-LLM~\citep{trtllm}.
For RL, they are FlashRL's FP8 engine beside a bf16 trainer with truncated importance sampling (TIS), which multiplies each token's gradient by its importance weight capped at 2~\citep{flashrl2025}, bf16 training with the trainer's weights offloaded during rollouts~\citep{sheng2025hybridflow}, and one bf16 copy shared by engine and trainer, as in Unsloth~\citep{unsloth2025grpo} but keeping its KV cache.
We measure latency as the time per output token (TPOT) of a single request, and throughput with 256 requests of 512 prompt and 256 output tokens, using random adapters of rank 16--128, since speed depends only on the adapters' shapes.
Each engine is started four times and we report the mean of the last three unless a table says otherwise; the appendices give every measurement.
Kernel timings use CUDA graphs with the L2 cache flushed; unless noted, adapters have rank 64, and bold in tables marks the best value, with ties within 2\%.

\subsection{Kernels}
\label{sec:eval_kernels}

\begin{table}[t]
\caption{\textbf{Where one adapter's batch-1 cost goes.} One Qwen3-8B decoder layer (four projections) at one decode token, in microseconds; adapter rows give each system's cost over its own base GEMM, against vLLM's fastest path on each GPU (Marlin on the L40S, CUTLASS elsewhere). L2-flushed kernel times only, so they do not add up to Table~\ref{tab:wins}'s time per token.}
\label{tab:waterfall}
\centering\small\setlength{\tabcolsep}{3.2pt}
\begin{tabular}{@{}lrrrr@{}}
\toprule
\textmu s $\downarrow$ & L40S & H100 & H200 & Blackwell \\
\midrule
Base GEMM, vLLM & \textbf{413} & 126 & \textbf{104} & 286 \\
Base GEMM, EpiLoRA & \textbf{416} & \textbf{121} & \textbf{102} & \textbf{276} \\
\addlinespace[2pt]
Adapter, vLLM & 97 & 53 & 57 & 54 \\
Adapter, EpiLoRA & \textbf{46} & \textbf{12} & \textbf{10} & \textbf{17} \\
\quad EpiLoRA's path & fused & overlap & overlap & overlap \\
\addlinespace[2pt]
Layer, EpiLoRA / vLLM $\downarrow$ & 0.91$\times$ & 0.75$\times$ & 0.70$\times$ & 0.86$\times$ \\
\bottomrule
\end{tabular}
\end{table}

\textbf{The adapter's cost falls by 2--6$\times$ on every GPU} (\cref{tab:waterfall}).
On Hopper GPUs, one rank-64 adapter costs 53--57\,\textmu s per layer with vLLM's tuned kernels and 10--12\,\textmu s with the decode path, so the layer takes 0.70--0.75$\times$ vLLM's time.
On the H100, nine tenths of the saving is in the adapter and a tenth in EpiLoRA's own GEMM.
The overlap pays because the GEMM needs the down-projection only in its epilogue, so it streams its weights while the adapter is computed; vLLM's dual-stream mode overlaps the adapter only with the GEMM as a whole (\cref{fig:decode}b).
On the L40S, fusing adapters into EpiLoRA's own GEMM halves their cost, and on Blackwell the decode path cuts it from 54 to 17\,\textmu s.

\begin{figure}[t]
\centering
\includegraphics[width=\columnwidth]{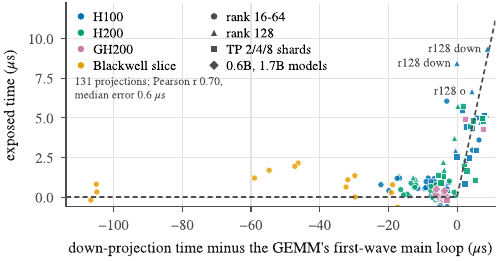}
\caption{\textbf{What stays exposed of the down-projection under the decode path.} One point per projection: Qwen3-8B layers at 1--16 decode tokens with ranks 16--128, their tensor-parallel shards and two smaller models, on four GPUs. The x axis subtracts the GEMM's first-wave main loop, a window fitted after the data, from the down-projection's time; dashed: the condition of \cref{sec:kernels}.}
\label{fig:exposure}
\end{figure}

\begin{table}[t]
\caption{\textbf{EpiLoRA against vLLM's fastest configuration.} Time per output token with one adapter at batch 1, and throughput with 32 tenants; vLLM runs its fastest base (NVFP4 on the Blackwell slice, FP8 elsewhere) with LoRA kernels tuned for each model (Qwen3.5-9B's on the H100 for all its Hopper rows). On 2$\times$H100, both systems use tiles searched for the shards; the H200 72B row comes from an earlier job than Table~\ref{tab:e2e72b}, which gives 0.94$\times$.}
\label{tab:wins}
\centering\footnotesize\setlength{\tabcolsep}{4pt}
\begin{tabular}{@{}llrrr@{}}
\toprule
GPU & Model & vLLM & EpiLoRA & Ratio \\
\midrule
\multicolumn{5}{@{}l}{\textit{One adapter at batch 1, ms per token} $\downarrow$} \\
H100 & Qwen3.5-9B & 6.26 & \textbf{5.14} & 0.82$\times$ \\
H100 & Qwen3-8B & 6.02 & \textbf{4.81} & 0.80$\times$ \\
H100 & Llama-3.1-8B & 5.59 & \textbf{4.76} & 0.85$\times$ \\
H200 & Qwen3.5-9B & 6.11 & \textbf{4.81} & 0.79$\times$ \\
H200 & Qwen3-8B & 5.94 & \textbf{4.51} & 0.76$\times$ \\
GH200 & Qwen3.5-9B & 6.70 & \textbf{5.91} & 0.88$\times$ \\
H200 & Qwen2.5-72B & 28.00 & \textbf{26.80} & 0.96$\times$ \\
2$\times$H100 & Qwen2.5-72B & 20.31 & \textbf{17.75} & 0.87$\times$ \\
L40S & Qwen3.5-9B & 15.39 & \textbf{14.97} & 0.97$\times$ \\
L40S & Qwen3-8B & \textbf{14.05} & \textbf{13.95} & 0.99$\times$ \\
Blackwell & Qwen3-8B & 10.27 & \textbf{9.46} & 0.92$\times$ \\
\addlinespace[3pt]
\multicolumn{5}{@{}l}{\textit{32 tenants, thousand tokens per second} $\uparrow$} \\
H100 & Qwen3.5-9B & 5.30 & \textbf{5.64} & 1.06$\times$ \\
H100 & Qwen3-8B & 5.89 & \textbf{6.22} & 1.05$\times$ \\
H100 & Llama-3.1-8B & \textbf{6.26} & \textbf{6.32} & 1.01$\times$ \\
H200 & Qwen3.5-9B & 5.29 & \textbf{5.78} & 1.09$\times$ \\
H200 & Qwen3-8B & 5.83 & \textbf{6.16} & 1.06$\times$ \\
GH200 & Qwen3.5-9B & 5.57 & \textbf{6.00} & 1.08$\times$ \\
L40S & Qwen3.5-9B & 1.47 & \textbf{1.59} & 1.08$\times$ \\
L40S & Qwen3-8B & 1.63 & \textbf{1.76} & 1.08$\times$ \\
Blackwell & Qwen3-8B & \textbf{2.08} & \textbf{2.11} & 1.02$\times$ \\
\bottomrule
\end{tabular}
\end{table}

\begin{table}[t]
\caption{\textbf{EpiLoRA against three serving systems on Qwen3-8B, each pair measured in one job.} Time per output token at batch 1 and throughput with 32 tenants, relative to EpiLoRA's, each system in its fastest LoRA configuration (TRT-LLM: TensorRT-LLM's PyTorch backend). $^{a}$At its fastest KV-cache share, 0.6. $^{b}$Without LoRA, 1.09--1.14$\times$ vLLM's batch-1 time and 0.95$\times$ its throughput; H200: one start. $^{c}$From NVIDIA's Nsight Systems profiler (Table~\ref{tab:profbody}); vLLM in dual-stream mode with its default kernels.}
\label{tab:systems}
\centering\footnotesize\setlength{\tabcolsep}{4pt}
\begin{tabular}{@{}lrrrr@{}}
\toprule
 & EpiLoRA & vLLM & SGLang$^{b}$ & TRT-LLM \\
\midrule
\multicolumn{5}{@{}l}{\textit{Batch 1: time per token relative to EpiLoRA's} $\downarrow$} \\
H100 & \textbf{1.00$\times$} & 1.25$\times$ & 1.65$\times$ & 1.43$\times$ \\
H200 & \textbf{1.00$\times$} & 1.32$\times$ & 1.81$\times$ & 1.52$\times$ \\
\addlinespace[3pt]
\multicolumn{5}{@{}l}{\textit{Batch 1: GPU time the adapter adds per token (ms)$^{c}$} $\downarrow$} \\
H100 & \textbf{0.40} & 1.87 & 3.01 & 2.87 \\
\addlinespace[3pt]
\multicolumn{5}{@{}l}{\textit{32 tenants: throughput relative to EpiLoRA's} $\uparrow$} \\
H100 & \textbf{1.00$\times$} & 0.95$\times$ & 0.94$\times$ & 0.97$\times$$^{a}$ \\
H200 & 1.00$\times$ & 0.95$\times$ & 0.93$\times$ & \textbf{1.05$\times$} \\
\bottomrule
\end{tabular}
\end{table}

\textbf{What stays exposed follows the overlap condition} (\cref{fig:exposure}).
The measured exposure tracks the condition with a window we fitted after the data: the down-projection stays hidden while it finishes within the main loop of the GEMM's first wave of thread blocks, when the first epilogues need it.
With a rank-64 adapter on full-size layers, the down-projection ends within the GEMM's main loop on every projection we measured, and at most 2.1\,\textmu s of it remains exposed.
At ranks 16 and 128 and on tensor-parallel shards, up to 9\,\textmu s per projection stays exposed, mostly where the GEMM launches more blocks than the GPU has SMs.

\begin{table}[t]
\caption{\textbf{From 1 to 16 decode tokens, the decode path keeps one adapter's cost to a few percent of the layer.} One Qwen3-8B decoder layer, in microseconds; cost: the time over the layer without an adapter (vLLM: the better of its tuned kernels and dual-stream mode).}
\label{tab:pdlbody}
\centering\footnotesize\setlength{\tabcolsep}{2.5pt}
\begin{tabular}{@{}r rrrrr rr@{}}
\toprule
\textmu s $\downarrow$ & No & \multicolumn{2}{c}{vLLM} & \multicolumn{2}{c}{EpiLoRA} & \multicolumn{2}{c}{Cost (\%) $\downarrow$} \\
\cmidrule(lr){3-4}\cmidrule(lr){5-6}\cmidrule(l){7-8}
Tokens & adapter & tuned & dual & serial & overlap & vLLM & EpiLoRA \\
\midrule
\multicolumn{8}{@{}l}{\textit{H100, FP8 base}} \\
1 & 126 & 179 & 180 & 163 & \textbf{134} & 42 & \textbf{6} \\
4 & 128 & 179 & 172 & 162 & \textbf{133} & 35 & \textbf{4} \\
16 & 128 & 180 & 174 & 167 & \textbf{136} & 36 & \textbf{7} \\
\addlinespace[3pt]
\multicolumn{8}{@{}l}{\textit{H200, FP8 base}} \\
1 & 104 & 161 & 167 & 144 & \textbf{112} & 55 & \textbf{8} \\
4 & 106 & 162 & 157 & 147 & \textbf{112} & 48 & \textbf{6} \\
16 & 106 & 162 & 159 & 150 & \textbf{115} & 49 & \textbf{9} \\
\bottomrule
\end{tabular}
\end{table}

\textbf{The lead holds from 1 to 16 decode tokens} (\cref{tab:pdlbody}).
With the decode path, one adapter adds 4--9\% to the layer on Hopper GPUs, against 35--55\% with vLLM's tuned kernels or dual-stream mode, because the down-projection finishes within the GEMM's main loop (\cref{fig:exposure}); run serially, the layer takes 1.13--1.31$\times$ as long, depending on the down-projection kernel (Tables~\ref{tab:floor},~\ref{tab:splitk}), so at one decode token the overlap accounts for a third to two thirds of the saving.

\begin{figure}[t]
\centering
\includegraphics[width=\columnwidth]{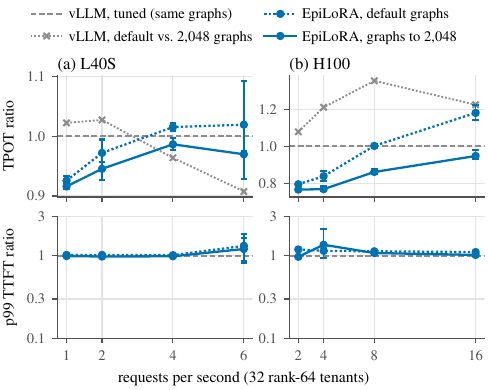}
\caption{\textbf{Online, EpiLoRA lowers the time per output token at most loads.} Median time per output token (top) and p99 time to first token (bottom, log scale) with 32 tenants, relative to vLLM with tuned kernels. Solid: CUDA graphs captured up to 2,048 tokens; dotted: vLLM's default capture, up to 512. Qwen3.5-9B, Poisson arrivals, 512 input and 256 output tokens; bars: range over three starts.}
\label{fig:online}
\end{figure}

\begin{table}[t]
\caption{\textbf{On a real trace, EpiLoRA lowers the time per output token at every load.} Per start, 1,000 of 3,383 first turns of WildChat conversations (short prompts: median 23 tokens) with gamma arrivals (burstiness 0.5), 32 tenants, Qwen3.5-9B at vLLM's default CUDA-graph settings; goodput counts requests meeting a 2\,s TTFT and a 50\,ms TPOT. $^{a}$One of vLLM's three starts took 3.7\,s (the others 0.19--0.20\,s), so neither value is bold.}
\label{tab:trace}
\centering\small\setlength{\tabcolsep}{1.9pt}
\begin{tabular}{@{}rrrrrrr@{}}
\toprule
 & \multicolumn{2}{c}{Median TPOT (ms) $\downarrow$} & \multicolumn{2}{c}{p99 TTFT (s) $\downarrow$} & \multicolumn{2}{c}{Goodput (req/s) $\uparrow$} \\
\cmidrule(lr){2-3}\cmidrule(lr){4-5}\cmidrule(l){6-7}
Req/s & vLLM & EpiLoRA & vLLM & EpiLoRA & vLLM & EpiLoRA \\
\midrule
\multicolumn{7}{@{}l}{\textit{H200}} \\
2 & 9.1 & \textbf{7.0} & \textbf{0.09} & \textbf{0.09} & \textbf{1.98} & \textbf{1.99} \\
4 & 12.8 & \textbf{8.9} & 0.10 & \textbf{0.09} & \textbf{3.88} & \textbf{3.92} \\
8 & 18.0 & \textbf{14.6} & 0.12 & \textbf{0.11} & 7.26 & \textbf{7.43} \\
16 & 30.6 & \textbf{26.4} & \textbf{0.16} & \textbf{0.16} & 12.08 & \textbf{12.70} \\
\addlinespace[2pt]
\multicolumn{7}{@{}l}{\textit{H100}} \\
2 & 9.9 & \textbf{7.6} & \textbf{0.16} & 0.18 & \textbf{1.98} & \textbf{1.99} \\
4 & 14.3 & \textbf{11.1} & \textbf{0.18} & 0.19 & \textbf{3.87} & \textbf{3.91} \\
8 & 21.8 & \textbf{19.0} & 1.37$^{a}$ & 0.22 & 7.15 & \textbf{7.35} \\
16 & 33.1 & \textbf{32.4} & \textbf{0.24} & 0.27 & 12.00 & \textbf{12.38} \\
\bottomrule\end{tabular}
\end{table}

\begin{figure}[t]
\centering
\includegraphics[width=\columnwidth]{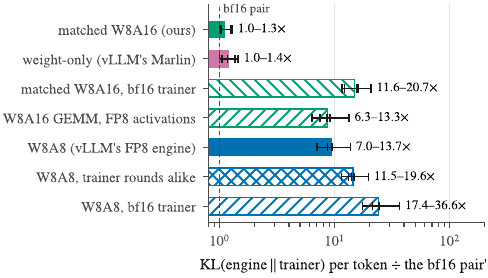}
\caption{\textbf{With the weights matched, rounded activations cause most of the remaining mismatch.} Per-token KL from engine to trainer relative to a bf16 engine and trainer, five 9B adapters on the H100 (bars: geometric mean; whiskers: range); the trainer computes on the engine's FP8 weights unless marked bf16, and hatched bars change one component of the bar above. Earlier, smaller decompositions measured up to 2.1$\times$ for the unrounded engines.}
\label{fig:kl}
\end{figure}

\begin{figure}[t]
\centering
\includegraphics[width=\columnwidth]{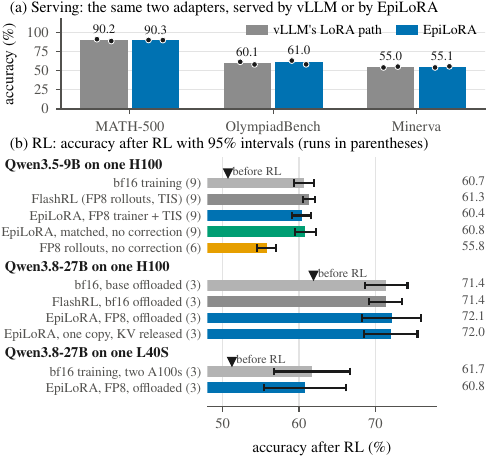}
\caption{\textbf{EpiLoRA keeps the accuracy of vLLM's LoRA path in serving and, with nine seeds on the 9B, of bf16 training in RL.} (a) Two RL adapters of Qwen3.5-9B served by vLLM and by EpiLoRA (dots: each adapter; differences $-1.0$ to $+1.6$ points, McNemar $p \ge 0.27$). (b) Accuracy after RL on OlympiadBench and Minerva (the 27B on the H100 also MATH-500), with 95\% intervals over runs; bf16 training of the 27B does not fit one L40S, so its reference uses two A100s.}
\label{fig:acceff}
\end{figure}

\subsection{End-to-end serving}
\label{sec:eval_e2e}

\textbf{At batch 1, EpiLoRA is faster than every system we ran on Hopper GPUs} (\cref{tab:wins,tab:systems}).
For 8B--9B models it takes 0.76--0.88$\times$ the time per token of vLLM's fastest configuration, including Llama-3.1-8B with both systems' kernels tuned for its shapes.
SGLang and TensorRT-LLM take 1.4--1.8$\times$ EpiLoRA's time on Qwen3-8B.
Profiles show why (\cref{tab:systems}): per token on the H100, their adapter kernels add 2.9--3.0\,ms of GPU time, vLLM's dual-stream mode 1.9 and EpiLoRA's 0.4, leaving EpiLoRA 9--14\% slower than without LoRA (Table~\ref{tab:sglang_h100}).

The lead is largest on the H200, whose faster memory shortens the base GEMM so the adapter is a larger share of each step, and smallest on the GH200, where both systems run slower than on the H100.
On the L40S, where vLLM is fastest with Marlin's weight-only GEMM, EpiLoRA's weight-only mode (bf16 activations) matches it; on the Blackwell slice's NVFP4 base, it takes 0.92$\times$ vLLM's time.

\textbf{At batch 1, the lead extends to a 72B model} (\cref{tab:wins}).
A larger model leaves less to save, as its GEMMs grow with the square of the hidden size and the adapter linearly: Qwen2.5-72B takes 0.96$\times$ the time per token of tuned vLLM on one H200 and 0.87$\times$ across two H100s.

\textbf{With many tenants, EpiLoRA keeps more of the base's throughput} (\cref{tab:wins}).
With 32 rank-64 tenants it delivers 1.05--1.09$\times$ the throughput of vLLM's best configuration on 7 of the 9 GPU--model pairs.
These batches exceed the 32 tokens the decode path covers, so the gain comes from EpiLoRA's adapter kernels after the vendor GEMM: on the H200, swapping in vLLM's turns 1.10$\times$ into 0.97$\times$ in one job (Table~\ref{tab:dispatch}). They beat vLLM's on projections with three or four outputs of unequal widths, such as attention's query, key and value, where vLLM's up-projection costs 1.7$\times$ as much per output column (Table~\ref{tab:expandkind}); at 32 tokens, the lead per layer shrinks as a batch mixes more distinct adapters (Figure~\ref{fig:tenants}).
On the L40S the gain comes from fusing adapters into EpiLoRA's GEMM; TensorRT-LLM beats EpiLoRA on the H200 (\cref{tab:systems}).

\textbf{Online, EpiLoRA lowers the time per output token at most loads} (\cref{fig:online}).
When CUDA graphs cover the prefill sizes, its median time per output token with 32 tenants is 0.77--0.95$\times$ vLLM's on the H100 and 0.92--0.99$\times$ on the L40S, and its 99th-percentile (p99) time to first token (TTFT) stays within 1.1$\times$ of vLLM's at 6 of these 8 GPU--rate points.
On a WildChat trace, EpiLoRA takes 0.70--0.86$\times$ vLLM's time per output token on the H200 and 0.77--0.98$\times$ on the H100 at vLLM's defaults, and serves at least vLLM's goodput at every rate (\cref{tab:trace}).
Its lead is largest at low load, where batches fit the decode path.

At vLLM's default graph sizes (dotted), prefills longer than 512 tokens run outside CUDA graphs, where every kernel launch costs host time: EpiLoRA then trails vLLM slightly on the L40S at 4--6 requests/s and by 1.18$\times$ on the H100 at 16 requests/s, and its p99 TTFT is 1.1--1.2$\times$ vLLM's on the H100; Tables~\ref{tab:defcap} and~\ref{tab:longfast} add the H200 and long prompts.

\subsection{LoRA RL on the same base}
\label{sec:eval_rl}

GRPO~\citep{shao2024deepseekmath} trains a rank-64 adapter on competition math (DAPO-Math~\citep{yu2025dapo}) with engine and trainer on one GPU: Qwen3.5-9B for 200 steps of 16 problems and Qwen3.8-27B for 100 steps of 8 problems, 8 samples each, with samples and answers capped at 2,048 tokens.
We report accuracy on OlympiadBench~\citep{he2024olympiadbench} and Minerva~\citep{lewkowycz2022minerva}, with 3--9 seeds and 95\% intervals (Tables~\ref{tab:arms},~\ref{tab:arms2}); RL lifts the 9B from 50.5\% to 60--61\% (\cref{fig:acceff}b).

\textbf{Matching the weights removes the largest source of mismatch, and leaving activations unrounded removes most of the rest} (\cref{fig:kl}).
We measure the mismatch as the KL divergence between the engine's and the trainer's next-token distributions.
Rounding the weights raises it the most, to 12--21$\times$ that of a bf16 engine and trainer, and the FP8 trainer removes this part by computing on the engine's own weights.
Almost all of the rest comes from rounding activations: engines that round them retain 6--14$\times$, while our matched engine and vLLM's Marlin retain 1.0--1.4$\times$, and the order holds on a dense second model (Table~\ref{tab:decomp4b}).

\textbf{On the 9B, corrected or matched FP8 RL ends where bf16 training does} (\cref{fig:acceff}b).
We call two configurations equivalent when the 95\% interval of their difference lies within $\pm$5 points, about half of RL's gain.
With nine seeds, W8A8 rollouts with TIS, FlashRL's recipe and the matched engine without any correction end within 0.6 points of bf16 training, every interval within $\pm$2.6 points (Table~\ref{tab:paired}), while uncorrected W8A8 rollouts end 4.9 points below it; on an L40S, the matched engine with one copy that keeps its KV cache ended 3.4 points below bf16 training ($[-5.7, -1.1]$), but against two copies one copy moved it by $-2.3$ and W8A8 rollouts with TIS by $+2.4$ (Welch intervals spanning zero, paired ones excluding it in opposite directions), so the one-copy effect is inconclusive (Appendix~\ref{app:discussion}).
The matched engine needs no correction because it leaves the mismatch at the bf16 pair's level (\cref{fig:kl}) and keeps it there throughout training (Table~\ref{tab:drift}).
On the 27B, EpiLoRA's configurations end at 72.0--72.1\% (with MATH-500) against 71.4\% for bf16 training and FlashRL, every interval within $\pm$4 points with three seeds (Table~\ref{tab:arms2}).

\textbf{An FP8 trainer fits the 27B on one 48\,GB GPU.}
Training on the engine's FP8 weights saves 6.5\,GiB for the 9B and 22.7\,GiB for the 27B (Table~\ref{tab:rlparts}).
So the 27B trains on one 48\,GB L40S, which its bf16 weights (50\,GiB) exceed: with the trainer's copy offloaded, EpiLoRA reaches 60.8\%, against 61.7\% for bf16 training on two A100s (\cref{fig:acceff}b).

\begin{table}[t]
\caption{\textbf{With long rollouts, one FP8 copy runs close to the fastest placement that fits, with a fraction of offloading's host memory.} Seconds per RL step and peak host memory (GiB) by placement of the trainer's base, with rollouts of up to 8,192 tokens (12-step probes; details in Table~\ref{tab:rlprobedetail}); FlashRL's placement runs in our trainer and engine; the bf16 copy keeps its KV cache, which Unsloth and EpiLoRA's FP8 copy release while training; OOM: out of GPU memory. $^{a}$Engine at 55\% of the GPU; OOM at 70\%.}
\label{tab:rlprobe}
\centering\footnotesize\setlength{\tabcolsep}{2.5pt}
\begin{tabular}{@{}lrrrr@{}}
\toprule
 & \multicolumn{2}{c}{9B, one L40S} & \multicolumn{2}{c}{27B, one H100} \\
\cmidrule(lr){2-3}\cmidrule(l){4-5}
Placement of the trainer's base & s/step $\downarrow$ & GiB $\downarrow$ & s/step $\downarrow$ & GiB $\downarrow$ \\
\midrule
One bf16 copy, KV kept$^{a}$ & 2{,}216 & \textbf{8} & \multicolumn{2}{c}{OOM} \\
Two FP8 copies & \multicolumn{2}{c}{OOM} & \multicolumn{2}{c}{OOM} \\
bf16, trainer offloaded & 1{,}314 & 48 & 659 & 160 \\
FlashRL, trainer offloaded & 1{,}076 & 41 & \textbf{421} & 126 \\
EpiLoRA, FP8 trainer offloaded & \textbf{1{,}025} & 32 & \textbf{418} & 94 \\
EpiLoRA, one FP8 copy & \textbf{1{,}012} & 9 & \textbf{422} & \textbf{9} \\
\bottomrule
\end{tabular}
\end{table}

\begin{table}[t]
\caption{\textbf{In full runs, one FP8 copy trains the 27B in 0.69$\times$ bf16 offloading's GPU-hours with a fraction of the host memory.} Single 100-step RL runs (two for one copy, which releases its KV cache) with rollouts of up to 8,192 tokens on one H100; per M tok.: GPU-hours per million trained tokens; Gen.: generation seconds per step; Host: peak host memory (GiB); Tokens: trained tokens per step. These runs compare cost: at our evaluation they change accuracy little (Table~\ref{tab:longroll}).}
\label{tab:rlong27}
\centering\footnotesize\setlength{\tabcolsep}{2.5pt}
\resizebox{\columnwidth}{!}{\begin{tabular}{@{}lrrrrr@{}}
\toprule
 & \multicolumn{2}{c}{GPU-h $\downarrow$} & & & \\
\cmidrule(lr){2-3}
Placement & run & per M tok. & Gen.\ s $\downarrow$ & Host $\downarrow$ & Tokens \\
\midrule
bf16, offloaded & 14.67 & 0.91 & 434 & 160 & 162k \\
FlashRL, offloaded & 9.84 & \textbf{0.60} & \textbf{247} & 127 & 164k \\
EpiLoRA, FP8 offloaded & \textbf{9.56} & 0.62 & \textbf{245} & 95 & 155k \\
EpiLoRA, one FP8 copy & 10.05 & \textbf{0.61} & 257 & \textbf{10} & 165k \\
\bottomrule
\end{tabular}}
\end{table}

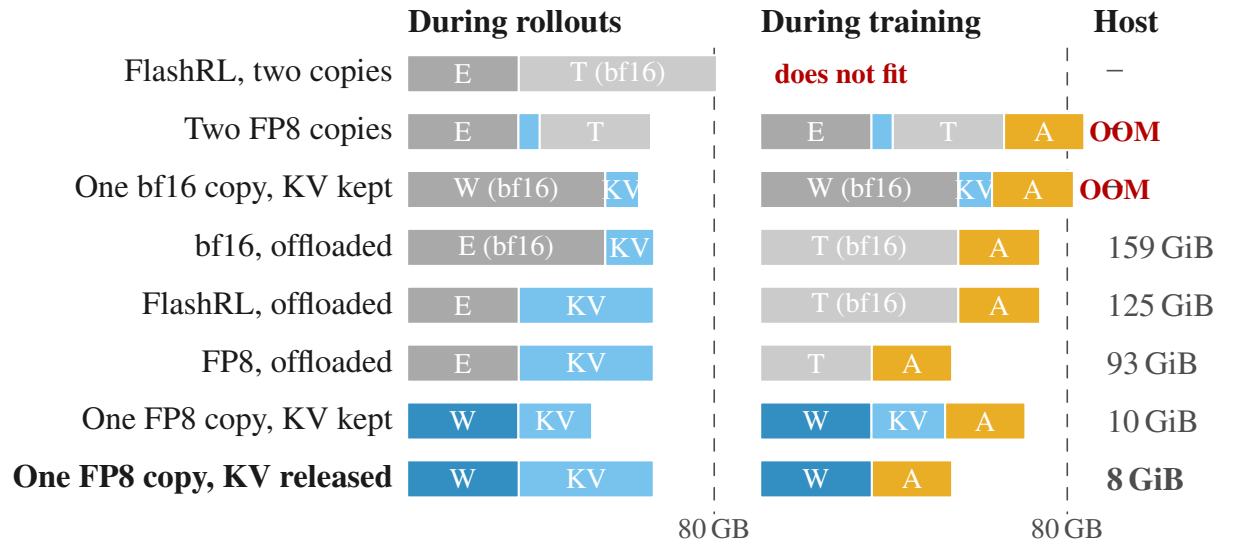
\begin{figure}[t]
\centering
\resizebox{0.97\columnwidth}{!}{\definecolor{obInk}{HTML}{1A1A1A}%
\definecolor{obInkTwo}{HTML}{4D4D4D}%
\definecolor{obGrid}{HTML}{E3E3E3}%
\definecolor{obVllm}{HTML}{8C8C8C}%
\definecolor{obOurs}{HTML}{0072B2}%
\definecolor{obFPeight}{HTML}{56B4E9}%
\definecolor{obFPfour}{HTML}{E69F00}%
\definecolor{obMatched}{HTML}{009E73}%
\def\obLabel{\fontsize{8}{9.2}\selectfont}%
\def\obNote{\fontsize{7}{8}\selectfont}%
\tikzset{
  obfig/.style={font=\obLabel, text=obInk, line width=0.5pt, line cap=round, line join=round},
  obarrow/.style={-{Stealth[length=1.45mm, width=1.15mm]}, draw=obInkTwo, line width=0.55pt},
  obnote/.style={font=\obNote, text=obInkTwo, inner sep=1pt},
  obtitle/.style={font=\obLabel\bfseries, text=obInk, anchor=base west, inner sep=0pt},
  obvllm/.style={draw=obVllm, fill=obVllm!13, rounded corners=1.6pt, line width=0.5pt},
  obours/.style={draw=obOurs, fill=obOurs!9, rounded corners=1.6pt, line width=0.6pt},
  obtensor/.style={draw=obInkTwo, fill=white, line width=0.45pt, rounded corners=0.6pt, inner sep=1.5pt, align=center},
}%
\begin{tikzpicture}[obfig, x=1cm, y=1cm]
\def\s{0.0335}%
\def\xr{0}\def\xt{3.05}%
\newcommand{\seg}[5]{%
  \fill[#4] ({#1*\s}, #3-0.16) rectangle ({(#1+#2)*\s}, #3+0.16);
  \draw[white, line width=0.5pt] ({#1*\s}, #3-0.16) rectangle ({(#1+#2)*\s}, #3+0.16);
  \node[font=\obNote, text=white, inner sep=0pt] at ({(#1+#2/2)*\s}, #3) {#5};}
\newcommand{\row}[3]{\node[anchor=east, font=\obLabel, text=obInk, align=right, inner sep=1pt] at (-0.08, #1) {#2};
  \node[anchor=west, font=\obLabel, text=obInkTwo, inner sep=1pt] at (6.0, #1) {#3};}
\node[obtitle] at (\xr, 0.36) {During rollouts};
\node[obtitle] at (\xt, 0.36) {During training};
\node[anchor=base west, font=\obLabel\bfseries, text=obInk, inner sep=0pt] at (5.92, 0.36) {Host};
\foreach \x in {\xr, \xt} {
  \draw[obInkTwo, line width=0.4pt, dashed] ({\x + 79.2*\s}, 0.22) -- ({\x + 79.2*\s}, -3.80);
  \node[font=\obNote, text=obInkTwo, anchor=north, inner sep=1pt] at ({\x + 79.2*\s}, -3.80) {80\,GB};}
\def\oom#1#2{\node[font=\obNote\bfseries, text=red!70!black, anchor=west, inner sep=1pt] at ({#1*\s}, #2) {OOM};}
\row{0}{FlashRL, two copies}{--}
\begin{scope}[shift={(\xr,0)}]\seg{0}{28.8}{0}{obVllm!75}{E}\seg{28.8}{51.1}{0}{obVllm!45}{T (bf16)}\end{scope}
\begin{scope}[shift={(\xt,0)}]\node[font=\obNote\bfseries, text=red!70!black, anchor=west] at (0, 0) {does not fit};\end{scope}
\row{-0.50}{Two FP8 copies}{--}
\begin{scope}[shift={(\xr,0)}]\seg{0}{28.7}{-0.50}{obVllm!75}{E}\seg{28.7}{5.5}{-0.50}{obFPeight!80}{}\seg{34.2}{28.7}{-0.50}{obVllm!45}{T}\end{scope}
\begin{scope}[shift={(\xt,0)}]\seg{0}{28.7}{-0.50}{obVllm!75}{E}\seg{28.7}{5.5}{-0.50}{obFPeight!80}{}\seg{34.2}{28.7}{-0.50}{obVllm!45}{T}%
  \seg{62.9}{20.7}{-0.50}{obFPfour!85}{A}\oom{83.6}{-0.50}\end{scope}
\row{-1.00}{One bf16 copy, KV kept}{--}
\begin{scope}[shift={(\xr,0)}]\seg{0}{51.1}{-1.00}{obVllm!75}{W (bf16)}\seg{51.1}{8.7}{-1.00}{obFPeight!80}{KV}\end{scope}
\begin{scope}[shift={(\xt,0)}]\seg{0}{51.1}{-1.00}{obVllm!75}{W (bf16)}\seg{51.1}{8.7}{-1.00}{obFPeight!80}{KV}\seg{59.8}{21.1}{-1.00}{obFPfour!85}{A}\oom{81.0}{-1.00}\end{scope}
\row{-1.50}{bf16, offloaded}{159\,GiB}
\begin{scope}[shift={(\xr,0)}]\seg{0}{51.1}{-1.50}{obVllm!75}{E (bf16)}\seg{51.1}{12.6}{-1.50}{obFPeight!80}{KV}\end{scope}
\begin{scope}[shift={(\xt,0)}]\seg{0}{51.1}{-1.50}{obVllm!45}{T (bf16)}\seg{51.1}{21.1}{-1.50}{obFPfour!85}{A}\end{scope}
\row{-2.00}{FlashRL, offloaded}{125\,GiB}
\begin{scope}[shift={(\xr,0)}]\seg{0}{28.8}{-2.00}{obVllm!75}{E}\seg{28.8}{34.8}{-2.00}{obFPeight!80}{KV}\end{scope}
\begin{scope}[shift={(\xt,0)}]\seg{0}{51.1}{-2.00}{obVllm!45}{T (bf16)}\seg{51.1}{21.1}{-2.00}{obFPfour!85}{A}\end{scope}
\row{-2.50}{FP8, offloaded}{93\,GiB}
\begin{scope}[shift={(\xr,0)}]\seg{0}{28.8}{-2.50}{obVllm!75}{E}\seg{28.8}{34.8}{-2.50}{obFPeight!80}{KV}\end{scope}
\begin{scope}[shift={(\xt,0)}]\seg{0}{28.8}{-2.50}{obVllm!45}{T}\seg{28.8}{20.7}{-2.50}{obFPfour!85}{A}\end{scope}
\row{-3.00}{One FP8 copy, KV kept}{10\,GiB}
\begin{scope}[shift={(\xr,0)}]\seg{0}{28.7}{-3.00}{obOurs!80}{W}\seg{28.7}{19.0}{-3.00}{obFPeight!80}{KV}\end{scope}
\begin{scope}[shift={(\xt,0)}]\seg{0}{28.7}{-3.00}{obOurs!80}{W}\seg{28.7}{19.0}{-3.00}{obFPeight!80}{KV}\seg{47.7}{20.7}{-3.00}{obFPfour!85}{A}\end{scope}
\row{-3.50}{\textbf{One FP8 copy, KV released}}{\textbf{8\,GiB}}
\begin{scope}[shift={(\xr,0)}]\seg{0}{28.8}{-3.50}{obOurs!80}{W}\seg{28.8}{34.8}{-3.50}{obFPeight!80}{KV}\end{scope}
\begin{scope}[shift={(\xt,0)}]\seg{0}{28.8}{-3.50}{obOurs!80}{W}\seg{28.8}{20.7}{-3.50}{obFPfour!85}{A}\end{scope}
\end{tikzpicture}}
\caption{\textbf{Releasing its KV cache, one FP8 copy keeps offloading's cache without pinning the trainer's copy in host memory.} GPU memory of one Qwen3.8-27B RL step on one 80\,GB H100 with rollouts of up to 8,192 tokens, to scale: E, T and W are the engine's, the trainer's and a shared copy of the base (FP8 unless marked bf16), KV the engine's cache and A the trainer's activations; right: median host memory pinned while training.}
\label{fig:memory}
\end{figure}

\textbf{With long rollouts, one FP8 copy comes close to the fastest placement with a fraction of its host memory} (\cref{tab:rlprobe,tab:rlong27}).
With rollouts of up to 8,192 tokens, two FP8 copies fit neither model, and one bf16 copy that keeps its KV cache fits only the 9B, at 2.2$\times$ the step time (1.9$\times$ that of an FP8 copy that also keeps it, Table~\ref{tab:longroll}); Unsloth releases that cache while training, a variant we did not run (Appendix~\ref{app:discussion}).
One FP8 copy then takes 0.94--1.00$\times$ the time of FlashRL's offloaded recipe and is 1.3--1.6$\times$ as fast as bf16 training with offloading, while pinning 9\,GiB of host memory instead of 32--160\,GiB; in full runs of the 27B it takes 1.02--1.05$\times$ the GPU-hours of the offloaded FP8 recipes, or 0.98--1.02$\times$ per trained token (\cref{tab:rlong27}).
The difference is in generation: FP8 weights take half the memory of bf16 weights and leave the engine a KV cache almost three times as large (\cref{fig:memory}), and with long rollouts the cache sets how many sequences decode at once; FlashRL's FP8 engine gets the same cache.
Where EpiLoRA does not pay (Appendix~\ref{app:discussion}): where two copies fit, FlashRL's faster bf16 trainer makes its recipe slightly cheaper; at 2,048 tokens a bf16 copy that keeps its KV cache ties one FP8 copy on the 9B; a trainer rounding activations like the engine trains 1.2$\times$ faster at a larger mismatch; and the matched engine takes 1.35$\times$ (vLLM's Marlin, also unrounded, 1.26$\times$) the GPU-hours of corrected W8A8 rollouts on the H100.
\section{Related work}
\label{sec:related}

\textbf{Multi-adapter serving.}
Other systems accelerate per-request and per-token adapters~\citep{wen2024flora,kong2024loraswitch}, tile adapters of mixed ranks~\citep{mi2025valora}, multiply each group of rows by its own matrix~\citep{ye2025flashinfer}, or cache, pool, compress and co-quantize adapters or share KV caches across them~\citep{xia2024loralinlaid,iliakopoulou2025chameleon,yao2023deltazip,zhang2025loquetier,bruelgabrielsson2025compress,wang2026forkkv,yu2026plora}.
All of them compute adapters beside the base GEMM (Table~\ref{tab:related}); EpiLoRA can be their kernel layer.

\textbf{Low-precision serving.}
AWQ and QServe quantize weights to 4 bits with 16- or 8-bit activations~\citep{lin2024awq,lin2025qserve}, Marlin provides fast mixed-precision GEMMs~\citep{frantar2025marlin}, and Hopper and Blackwell GPUs support FP8 and block-scaled 4-bit formats natively~\citep{micikevicius2022fp8,rouhani2023mx,nvidia2025nvfp4}; EpiLoRA keeps each engine's quantization bit for bit and changes only how adapters are computed around the GEMM.

\textbf{Fusing low-rank branches into GEMMs.}
SVDQuant's Nunchaku fuses one low-rank branch, shared by every token, into the 4-bit GEMMs of diffusion models~\citep{li2025svdquant}.
For fine-tuning, LoRAFusion fuses the up-projection into the bf16 base GEMM and routes the tiles of a mixed-adapter batch to their adapters~\citep{zhu2026lorafusion}.
K2 does the same for serving over FP8 and NVFP4 GEMMs, gathering rows sorted by adapter as mixture-of-experts GEMMs gather rows sorted by expert~\citep{gale2023megablocks}.
The decode path applies a known principle, overlapping a producer with its consumer's main loop, as systems that overlap a GEMM with communication between GPUs do tile by tile~\citep{wang2023overlap,pati2024t3,chang2024flux} and NanoFlow does across a batch's operations~\citep{zhu2025nanoflow}.
Unlike weight prefetches issued before a dependent kernel's wait, our producer is a side branch the GEMM needs only in its epilogue, so its whole main loop overlaps it; LoRAFusion needs the down-projection before its GEMM starts, and two-stream modes add the adapter's output after the GEMM ends.
Compressing adapters~\citep{yadav2023compeft,ping2024deltacome,liu2024bitdelta,mirzaei2025loraquant} and training them for low-precision merges~\citep{dettmers2023qlora,li2024loftq,guo2024lqlora,xu2024qalora,scaleqlora2026} are complementary: they change the adapters.

\textbf{Low-precision RL.}
FlashRL, our RL baseline, pairs FP8 rollouts with truncated importance sampling beside a bf16 trainer~\citep{flashrl2025}.
Quantized rollouts also speed up full-weight RL, which must re-quantize the weights at every step~\citep{sparserlsync2026,pulse2026}; other systems train and generate at one low precision, rounding activations on both sides, which leaves a mismatch unless both sides' kernels agree bit for bit~\citep{jetrl2026,fp8rlstack2026,fp8rl2026,qarl2026}, correct the rounding of 4-bit rollouts~\citep{quads2026,trace2026}, or sample from 4-bit weight-only kernels~\citep{qerl2025}.
Adaptive importance sampling tunes the correction~\citep{ais2026} and score centering removes the drift a mismatched sampler adds to the policy gradient~\citep{scorecentering2026}; the matched engine instead removes most of the mismatch itself.
FlexLLM co-serves inference and fine-tuning on one bf16 base~\citep{miao2024flexllm}, batch-invariant kernels make engine and trainer agree bit for bit~\citep{he2025nondeterminism,vllm2025bitwise}, Unsloth shares one 4-bit or FP8 copy of the base between engine and trainer and releases its KV cache while training~\citep{unsloth2025grpo,unsloth2025fp8,unsloth2025standby}, and a TRL patch shares FP8 weights with vLLM~\citep{trlfp8share2026}.
QLoRA first trained adapters on a frozen quantized base~\citep{dettmers2023qlora}.
EpiLoRA trains on the engine's FP8 codes with bf16 activations, and its matched engine keeps them unrounded, as Marlin and QeRL do, in the trainer's layout.
\section{Conclusion}
\label{sec:discussion}

Adapters have stayed outside the low-precision GEMM, so a cheaper base makes them relatively costlier to serve.
EpiLoRA, a drop-in vLLM plugin, brings them back into the GEMM: on Ada GPUs it fuses every tenant's adapter into its own low-precision GEMM, and on Hopper and Blackwell GPUs it launches the down-projection first and lets the GEMM stream its weights while the adapter is computed, waiting for it only in the epilogue.
This change makes one adapter at batch 1 faster than vLLM, SGLang and TensorRT-LLM on Hopper GPUs, and with many tenants its adapter kernels keep more of the base's throughput than vLLM's on most GPU--model pairs we tested.
The same low-precision weights also serve RL: a trainer that computes on them needs no bf16 copy of the model, so a 27B model trains on one 48\,GB GPU, and an engine that leaves activations unrounded needs no importance correction.
Served adapters keep the accuracy of vLLM's LoRA path, and on the 9B, corrected or matched FP8 RL ends where bf16 training does.
Sharing one FP8 copy between engine and trainer saves most of a second copy's memory, which matters most for models that do not fit twice on a GPU; with long rollouts it nears the fastest placement with a fraction of its host memory.
More broadly, a kernel can wait for its producer where it first needs the data, not where it starts.

\bibliography{refs}

\begin{thebibliography}{77}
\providecommand{\natexlab}[1]{#1}
\providecommand{\url}[1]{\texttt{#1}}
\expandafter\ifx\csname urlstyle\endcsname\relax
  \providecommand{\doi}[1]{doi: #1}\else
  \providecommand{\doi}{doi: \begingroup \urlstyle{rm}\Url}\fi

\bibitem[Br{\"u}el-Gabrielsson et~al.(2025)Br{\"u}el-Gabrielsson, Zhu,
  Bhardwaj, Choshen, Greenewald, Yurochkin, and
  Solomon]{bruelgabrielsson2025compress}
Br{\"u}el-Gabrielsson, R., Zhu, J., Bhardwaj, O., Choshen, L., Greenewald, K.,
  Yurochkin, M., and Solomon, J.
\newblock Compress then serve: Serving thousands of {LoRA} adapters with little
  overhead.
\newblock In \emph{Proceedings of the 42nd International Conference on Machine
  Learning (ICML)}, pp.\  18062--18095, 2025.
\newblock arXiv:2407.00066.

\bibitem[Chang et~al.(2024)Chang, Bao, Hou, Jiang, Zheng, Zhong, Zhang, Song,
  Yao, Jiang, Lin, Jin, and Liu]{chang2024flux}
Chang, L.-W., Bao, W., Hou, Q., Jiang, C., Zheng, N., Zhong, Y., Zhang, X.,
  Song, Z., Yao, C., Jiang, Z., Lin, H., Jin, X., and Liu, X.
\newblock {FLUX}: Fast software-based communication overlap on {GPUs} through
  kernel fusion.
\newblock \emph{arXiv preprint arXiv:2406.06858}, 2024.

\bibitem[Chen et~al.(2026{\natexlab{a}})Chen, Jiang, Wei, Zhong, Li, Zhang,
  Lin, and Venkataraman]{fp8rl2026}
Chen, F., Jiang, Z., Wei, Z., Zhong, Z., Li, D., Zhang, C., Lin, H., and
  Venkataraman, S.
\newblock Towards full pipeline {FP8} reinforcement learning for {LLMs}.
\newblock \emph{arXiv preprint arXiv:2609.22870}, 2026{\natexlab{a}}.

\bibitem[Chen et~al.(2026{\natexlab{b}})Chen, Ruan, Xu, Li, Chen, Leng, He,
  Guo, and Sun]{chen2026infinilora}
Chen, H., Ruan, L., Xu, Z., Li, Y., Chen, X., Leng, J., He, B., Guo, M., and
  Sun, S.
\newblock {InfiniLoRA}: Disaggregated multi-{LoRA} serving for large language
  models.
\newblock \emph{arXiv preprint arXiv:2604.07173}, 2026{\natexlab{b}}.

\bibitem[Chen et~al.(2024)Chen, Ye, Wu, Zhuo, Ceze, and
  Krishnamurthy]{chen2024punica}
Chen, L., Ye, Z., Wu, Y., Zhuo, D., Ceze, L., and Krishnamurthy, A.
\newblock Punica: Multi-tenant {LoRA} serving.
\newblock In \emph{Proceedings of Machine Learning and Systems (MLSys)}, pp.\
  1--13, 2024.
\newblock arXiv:2310.18547.

\bibitem[Dettmers et~al.(2023)Dettmers, Pagnoni, Holtzman, and
  Zettlemoyer]{dettmers2023qlora}
Dettmers, T., Pagnoni, A., Holtzman, A., and Zettlemoyer, L.
\newblock {QLoRA}: Efficient finetuning of quantized {LLMs}.
\newblock In \emph{Advances in Neural Information Processing Systems}, pp.\
  10088--10115, 2023.

\bibitem[Frantar et~al.(2025)Frantar, Castro, Chen, Hoefler, and
  Alistarh]{frantar2025marlin}
Frantar, E., Castro, R.~L., Chen, J., Hoefler, T., and Alistarh, D.
\newblock {MARLIN}: Mixed-precision auto-regressive parallel inference on large
  language models.
\newblock In \emph{Proceedings of the 30th ACM SIGPLAN Annual Symposium on
  Principles and Practice of Parallel Programming (PPoPP)}, pp.\  239--251,
  2025.

\bibitem[Fu et~al.(2025)Fu, Gao, Shen, Zhu, Mei, He, Xu, Wei, Mei, Wang, Yang,
  Yuan, and Wu]{fu2025areal}
Fu, W., Gao, J., Shen, X., Zhu, C., Mei, Z., He, C., Xu, S., Wei, G., Mei, J.,
  Wang, J., Yang, T., Yuan, B., and Wu, Y.
\newblock {AReaL}: A large-scale asynchronous reinforcement learning system for
  language reasoning.
\newblock \emph{arXiv preprint arXiv:2505.24298}, 2025.

\bibitem[Gale et~al.(2023)Gale, Narayanan, Young, and
  Zaharia]{gale2023megablocks}
Gale, T., Narayanan, D., Young, C., and Zaharia, M.
\newblock {MegaBlocks}: Efficient sparse training with mixture-of-experts.
\newblock In \emph{Proceedings of Machine Learning and Systems (MLSys)}, pp.\
  288--304, 2023.

\bibitem[Gu et~al.(2026)Gu, Wang, Liu, Li, Zhu, Liu, Xu, Wang, Yang, Lin, Han,
  and Guo]{qarl2026}
Gu, H., Wang, H., Liu, J., Li, L., Zhu, Q., Liu, B., Xu, B., Wang, L., Yang,
  X., Lin, S., Han, S., and Guo, Y.
\newblock {QaRL}: Rollout-aligned quantization-aware {RL} for fast and stable
  training under training--inference mismatch.
\newblock \emph{arXiv preprint arXiv:2604.07853}, 2026.

\bibitem[Guo et~al.(2024)Guo, Greengard, Xing, and Kim]{guo2024lqlora}
Guo, H., Greengard, P., Xing, E.~P., and Kim, Y.
\newblock {LQ-LoRA}: Low-rank plus quantized matrix decomposition for efficient
  language model finetuning.
\newblock In \emph{International Conference on Learning Representations
  (ICLR)}, 2024.

\bibitem[He et~al.(2024)He, Luo, Bai, Hu, Thai, Shen, Hu, Han, Huang, Zhang,
  Liu, Qi, Liu, and Sun]{he2024olympiadbench}
He, C., Luo, R., Bai, Y., Hu, S., Thai, Z.~L., Shen, J., Hu, J., Han, X.,
  Huang, Y., Zhang, Y., Liu, J., Qi, L., Liu, Z., and Sun, M.
\newblock {OlympiadBench}: A challenging benchmark for promoting {AGI} with
  olympiad-level bilingual multimodal scientific problems.
\newblock In \emph{ACL}, pp.\  3828--3850, 2024.

\bibitem[He \& {Thinking Machines Lab}(2025)He and {Thinking Machines
  Lab}]{he2025nondeterminism}
He, H. and {Thinking Machines Lab}.
\newblock Defeating nondeterminism in {LLM} inference.
\newblock Thinking Machines Lab: Connectionism, 2025.

\bibitem[Hu et~al.(2022)Hu, Shen, Wallis, Allen-Zhu, Li, Wang, Wang, and
  Chen]{hu2022lora}
Hu, E.~J., Shen, Y., Wallis, P., Allen-Zhu, Z., Li, Y., Wang, S., Wang, L., and
  Chen, W.
\newblock {LoRA}: Low-rank adaptation of large language models.
\newblock In \emph{International Conference on Learning Representations}, 2022.

\bibitem[Hu et~al.(2024)Hu, Wu, Zhu, Xianyu, Wang, Zhang, and
  Cao]{hu2024openrlhf}
Hu, J., Wu, X., Zhu, Z., Xianyu, Wang, W., Zhang, D., and Cao, Y.
\newblock {OpenRLHF}: An easy-to-use, scalable and high-performance {RLHF}
  framework.
\newblock \emph{arXiv preprint arXiv:2405.11143}, 2024.

\bibitem[Hu et~al.(2026)Hu, Zhao, Zhu, Zhang, Zhang, Yin, and
  Zhao]{sparserlsync2026}
Hu, L., Zhao, R., Zhu, I., Zhang, Z., Zhang, H., Yin, H., and Zhao, J.
\newblock {SparseRL-Sync}: Lossless weight synchronization with $\sim$100x less
  communication.
\newblock \emph{arXiv preprint arXiv:2605.07330}, 2026.

\bibitem[Huang et~al.(2025)Huang, Ge, Yang, Xiao, Mao, Lin, Ye, Liu, Cheung,
  Yin, Lu, Qi, Han, and Chen]{qerl2025}
Huang, W., Ge, Y., Yang, S., Xiao, Y., Mao, H., Lin, Y., Ye, H., Liu, S.,
  Cheung, K.~C., Yin, H., Lu, Y., Qi, X., Han, S., and Chen, Y.
\newblock {QeRL}: Beyond efficiency -- quantization-enhanced reinforcement
  learning for {LLMs}.
\newblock \emph{arXiv preprint arXiv:2510.11696}, 2025.

\bibitem[Iliakopoulou et~al.(2025)Iliakopoulou, Stojkovic, Alverti, Xu, Franke,
  and Torrellas]{iliakopoulou2025chameleon}
Iliakopoulou, N., Stojkovic, J., Alverti, C., Xu, T., Franke, H., and
  Torrellas, J.
\newblock Chameleon: Adaptive caching and scheduling for many-adapter {LLM}
  inference environments.
\newblock In \emph{Proceedings of the 58th IEEE/ACM International Symposium on
  Microarchitecture (MICRO)}, pp.\  217--231, 2025.

\bibitem[{ingambe}(2026)]{trlfp8share2026}
{ingambe}.
\newblock {FP8} training for {GRPO}: replay harness, {FP8} recipes, {LoRA} over
  an {FP8} base and {FP8} weights shared with {vLLM}.
\newblock GitHub pull request \#119 to a fork of {TRL},
  \url{https://github.com/ingambe/trl/pull/119}, 2026.

\bibitem[Kong et~al.(2024)Kong, Li, Fang, Feng, He, Dong, Wang, Li, Kong, and
  Liu]{kong2024loraswitch}
Kong, R., Li, Q., Fang, X., Feng, Q., He, Q., Dong, Y., Wang, W., Li, Y., Kong,
  L., and Liu, Y.
\newblock {LoRA-Switch}: Boosting the efficiency of dynamic {LLM} adapters via
  system-algorithm co-design.
\newblock \emph{arXiv preprint arXiv:2405.17741}, 2024.

\bibitem[Kwon et~al.(2023)Kwon, Li, Zhuang, Sheng, Zheng, Yu, Gonzalez, Zhang,
  and Stoica]{kwon2023vllm}
Kwon, W., Li, Z., Zhuang, S., Sheng, Y., Zheng, L., Yu, C.~H., Gonzalez, J.~E.,
  Zhang, H., and Stoica, I.
\newblock Efficient memory management for large language model serving with
  {PagedAttention}.
\newblock In \emph{Proceedings of the 29th Symposium on Operating Systems
  Principles (SOSP)}, pp.\  611--626, 2023.

\bibitem[Lewkowycz et~al.(2022)Lewkowycz, Andreassen, Dohan, Dyer, Michalewski,
  Ramasesh, Slone, Anil, Schlag, Gutman-Solo, Wu, Neyshabur, Gur-Ari, and
  Misra]{lewkowycz2022minerva}
Lewkowycz, A., Andreassen, A., Dohan, D., Dyer, E., Michalewski, H., Ramasesh,
  V., Slone, A., Anil, C., Schlag, I., Gutman-Solo, T., Wu, Y., Neyshabur, B.,
  Gur-Ari, G., and Misra, V.
\newblock Solving quantitative reasoning problems with language models.
\newblock In \emph{NeurIPS}, pp.\  3843--3857, 2022.

\bibitem[Li et~al.(2025)Li, Lin, Zhang, Cai, Li, Guo, Xie, Meng, Zhu, and
  Han]{li2025svdquant}
Li, M., Lin, Y., Zhang, Z., Cai, T., Li, X., Guo, J., Xie, E., Meng, C., Zhu,
  J.-Y., and Han, S.
\newblock {SVDQuant}: Absorbing outliers by low-rank components for 4-bit
  diffusion models.
\newblock In \emph{International Conference on Learning Representations
  (ICLR)}, 2025.
\newblock arXiv:2411.05007.

\bibitem[Li et~al.(2024{\natexlab{a}})Li, Lu, Wu, Yu, Weng, Chen, Shan, Yuan,
  and Wang]{li2024caraserve}
Li, S., Lu, H., Wu, T., Yu, M., Weng, Q., Chen, X., Shan, Y., Yuan, B., and
  Wang, W.
\newblock {CaraServe}: {CPU}-assisted and rank-aware {LoRA} serving for
  generative {LLM} inference.
\newblock \emph{arXiv preprint arXiv:2401.11240}, 2024{\natexlab{a}}.

\bibitem[Li et~al.(2026)Li, Huang, Wang, and Gotei]{scaleqlora2026}
Li, T.-L., Huang, J., Wang, L.-C., and Gotei, J.~R.
\newblock {Scale-QLoRA}: Code-invariant adapter merging for native 4-bit
  microscaling {LLMs}.
\newblock \emph{arXiv preprint arXiv:2609.04526}, 2026.

\bibitem[Li et~al.(2024{\natexlab{b}})Li, Yu, Liang, He, Karampatziakis, Chen,
  and Zhao]{li2024loftq}
Li, Y., Yu, Y., Liang, C., He, P., Karampatziakis, N., Chen, W., and Zhao, T.
\newblock {LoftQ}: {LoRA}-fine-tuning-aware quantization for large language
  models.
\newblock In \emph{International Conference on Learning Representations
  (ICLR)}, 2024{\natexlab{b}}.

\bibitem[Lin et~al.(2024)Lin, Tang, Tang, Yang, Chen, Wang, Xiao, Dang, Gan,
  and Han]{lin2024awq}
Lin, J., Tang, J., Tang, H., Yang, S., Chen, W.-M., Wang, W.-C., Xiao, G.,
  Dang, X., Gan, C., and Han, S.
\newblock {AWQ}: Activation-aware weight quantization for on-device {LLM}
  compression and acceleration.
\newblock In \emph{Proceedings of Machine Learning and Systems}, pp.\  87--100,
  2024.

\bibitem[Lin et~al.(2025)Lin, Tang, Yang, Zhang, Xiao, Gan, and
  Han]{lin2025qserve}
Lin, Y., Tang, H., Yang, S., Zhang, Z., Xiao, G., Gan, C., and Han, S.
\newblock {QServe}: {W4A8KV4} quantization and system co-design for efficient
  {LLM} serving.
\newblock In \emph{Proceedings of Machine Learning and Systems}, 2025.

\bibitem[Liu et~al.(2024)Liu, Xiao, Li, Lee, Han, Dao, and
  Cai]{liu2024bitdelta}
Liu, J., Xiao, G., Li, K., Lee, J.~D., Han, S., Dao, T., and Cai, T.
\newblock {BitDelta}: Your fine-tune may only be worth one bit.
\newblock In \emph{Advances in Neural Information Processing Systems}, pp.\
  13579--13600, 2024.

\bibitem[Liu et~al.(2025)Liu, Yao, Zhang, Dong, Shang, and Gao]{flashrl2025}
Liu, L., Yao, F., Zhang, D., Dong, C., Shang, J., and Gao, J.
\newblock {FlashRL}: 8-bit rollouts, full power {RL}.
\newblock \url{https://fengyao.notion.site/flash-rl}, 2025.

\bibitem[{Llama Team, AI @ Meta}(2024)]{grattafiori2024llama3}
{Llama Team, AI @ Meta}.
\newblock The {Llama} 3 herd of models.
\newblock \emph{arXiv preprint arXiv:2407.21783}, 2024.

\bibitem[Marek \& Ryabinin(2026)Marek and Ryabinin]{scorecentering2026}
Marek, M. and Ryabinin, M.
\newblock Score centering stabilizes off-policy reinforcement learning.
\newblock \emph{arXiv preprint arXiv:2609.20807}, 2026.

\bibitem[Mei et~al.(2025)Mei, Fu, Li, Wang, Zhang, and Wu]{mei2025real}
Mei, Z., Fu, W., Li, K., Wang, G., Zhang, H., and Wu, Y.
\newblock {ReaL}: Efficient {RLHF} training of large language models with
  parameter reallocation.
\newblock In \emph{Proceedings of Machine Learning and Systems (MLSys)}, 2025.

\bibitem[Mi et~al.(2025)Mi, Wang, Tu, He, Kong, Fang, Dong, Zhang, Li, Li, Dai,
  Chen, and Liu]{mi2025valora}
Mi, L., Wang, W., Tu, W., He, Q., Kong, R., Fang, X., Dong, Y., Zhang, Y., Li,
  Y., Li, M., Dai, H., Chen, G., and Liu, Y.
\newblock Empower vision applications with {LoRA} {LMM}.
\newblock In \emph{Proceedings of the European Conference on Computer Systems
  (EuroSys)}, pp.\  261--277, 2025.
\newblock arXiv:2411.00915.

\bibitem[Miahi \& Belilovsky(2026)Miahi and Belilovsky]{pulse2026}
Miahi, E. and Belilovsky, E.
\newblock Understanding and exploiting weight update sparsity for
  communication-efficient distributed {RL}.
\newblock \emph{arXiv preprint arXiv:2602.03839}, 2026.

\bibitem[Miao et~al.(2024)Miao, Oliaro, Cheng, Wu, Unger, and
  Jia]{miao2024flexllm}
Miao, X., Oliaro, G., Cheng, X., Wu, M., Unger, C., and Jia, Z.
\newblock {FlexLLM}: A system for co-serving large language model inference and
  parameter-efficient finetuning.
\newblock \emph{arXiv preprint arXiv:2402.18789}, 2024.

\bibitem[Micikevicius et~al.(2022)Micikevicius, Stosic, Burgess, Cornea, Dubey,
  Grisenthwaite, Ha, Heinecke, Judd, Kamalu, Mellempudi, Oberman, Shoeybi, Siu,
  and Wu]{micikevicius2022fp8}
Micikevicius, P., Stosic, D., Burgess, N., Cornea, M., Dubey, P.,
  Grisenthwaite, R., Ha, S., Heinecke, A., Judd, P., Kamalu, J., Mellempudi,
  N., Oberman, S., Shoeybi, M., Siu, M., and Wu, H.
\newblock {FP8} formats for deep learning.
\newblock \emph{arXiv preprint arXiv:2209.05433}, 2022.

\bibitem[{Mind Lab}(2026)]{mint2026}
{Mind Lab}.
\newblock {MinT}: Managed infrastructure for training and serving millions of
  {LLMs}.
\newblock \emph{arXiv preprint arXiv:2605.13779}, 2026.

\bibitem[Mirzaei et~al.(2025)Mirzaei, Wen, Cao, and Mou]{mirzaei2025loraquant}
Mirzaei, A.~R., Wen, Y., Cao, Y., and Mou, L.
\newblock {LoRAQuant}: Mixed-precision quantization of {LoRA} to ultra-low
  bits.
\newblock \emph{arXiv preprint arXiv:2510.26690}, 2025.

\bibitem[{NVIDIA}(2025)]{nvidia2025nvfp4}
{NVIDIA}.
\newblock Pretraining large language models with {NVFP4}.
\newblock \emph{arXiv preprint arXiv:2509.25149}, 2025.

\bibitem[{NVIDIA}(2026)]{trtllm}
{NVIDIA}.
\newblock {TensorRT-LLM}.
\newblock \url{https://github.com/NVIDIA/TensorRT-LLM}, 2026.

\bibitem[Pati et~al.(2024)Pati, Aga, Islam, Jayasena, and Sinclair]{pati2024t3}
Pati, S., Aga, S., Islam, M., Jayasena, N., and Sinclair, M.~D.
\newblock {T3}: Transparent tracking \& triggering for fine-grained overlap of
  compute \& collectives.
\newblock In \emph{Proceedings of the ACM International Conference on
  Architectural Support for Programming Languages and Operating Systems
  (ASPLOS)}, pp.\  1146--1164, 2024.

\bibitem[Ping et~al.(2024)Ping, Wang, Wang, Han, Xu, Yan, Chen, Chang, Liu, and
  Sun]{ping2024deltacome}
Ping, B., Wang, S., Wang, H., Han, X., Xu, Y., Yan, Y., Chen, Y., Chang, B.,
  Liu, Z., and Sun, M.
\newblock {Delta-CoMe}: Training-free delta-compression with mixed-precision
  for large language models.
\newblock In \emph{Advances in Neural Information Processing Systems}, pp.\
  31056--31077, 2024.

\bibitem[Qiu et~al.(2026)Qiu, Yu, Zhang, Zhang, Huang, Yang, and
  Lai]{fp8rlstack2026}
Qiu, Z., Yu, S., Zhang, J., Zhang, S., Huang, X., Yang, J., and Lai, J.
\newblock {FP8-RL}: A practical and stable low-precision stack for {LLM}
  reinforcement learning.
\newblock \emph{arXiv preprint arXiv:2601.18150}, 2026.

\bibitem[{Qwen Team}(2025)]{yang2025qwen3}
{Qwen Team}.
\newblock Qwen3 technical report.
\newblock \emph{arXiv preprint arXiv:2505.09388}, 2025.

\bibitem[{Qwen Team}(2026{\natexlab{a}})]{qwen35}
{Qwen Team}.
\newblock {Qwen3.5}.
\newblock \url{https://huggingface.co/Qwen/Qwen3.5-9B}, 2026{\natexlab{a}}.

\bibitem[{Qwen Team}(2026{\natexlab{b}})]{qwen38}
{Qwen Team}.
\newblock {Qwen3.8}.
\newblock \url{https://huggingface.co/Qwen/Qwen3.8-27B}, 2026{\natexlab{b}}.

\bibitem[Rouhani et~al.(2023)Rouhani, Zhao, More, Hall, Khodamoradi, Deng,
  Choudhary, Cornea, Dellinger, Denolf, Dusan, Elango, Golub, Heinecke,
  James-Roxby, Jani, Kolhe, Langhammer, Li, Melnick, Mesmakhosroshahi,
  Rodriguez, Schulte, Shafipour, Shao, Siu, Dubey, Micikevicius, Naumov,
  Verrilli, Wittig, Burger, and Chung]{rouhani2023mx}
Rouhani, B.~D., Zhao, R., More, A., Hall, M., Khodamoradi, A., Deng, S.,
  Choudhary, D., Cornea, M., Dellinger, E., Denolf, K., Dusan, S., Elango, V.,
  Golub, M., Heinecke, A., James-Roxby, P., Jani, D., Kolhe, G., Langhammer,
  M., Li, A., Melnick, L., Mesmakhosroshahi, M., Rodriguez, A., Schulte, M.,
  Shafipour, R., Shao, L., Siu, M., Dubey, P., Micikevicius, P., Naumov, M.,
  Verrilli, C., Wittig, R., Burger, D., and Chung, E.
\newblock Microscaling data formats for deep learning.
\newblock \emph{arXiv preprint arXiv:2310.10537}, 2023.

\bibitem[Schulman \& {Thinking Machines Lab}(2025)Schulman and {Thinking
  Machines Lab}]{schulman2025lorawithoutregret}
Schulman, J. and {Thinking Machines Lab}.
\newblock {LoRA} without regret.
\newblock Thinking Machines Lab: Connectionism, 2025.

\bibitem[Shao et~al.(2024)Shao, Wang, Zhu, Xu, Song, Bi, Zhang, Zhang, Li, Wu,
  and Guo]{shao2024deepseekmath}
Shao, Z., Wang, P., Zhu, Q., Xu, R., Song, J., Bi, X., Zhang, H., Zhang, M.,
  Li, Y.~K., Wu, Y., and Guo, D.
\newblock {DeepSeekMath}: Pushing the limits of mathematical reasoning in open
  language models.
\newblock \emph{arXiv preprint arXiv:2402.03300}, 2024.

\bibitem[Sheng et~al.(2025)Sheng, Zhang, Ye, Wu, Zhang, Zhang, Peng, Lin, and
  Wu]{sheng2025hybridflow}
Sheng, G., Zhang, C., Ye, Z., Wu, X., Zhang, W., Zhang, R., Peng, Y., Lin, H.,
  and Wu, C.
\newblock {HybridFlow}: A flexible and efficient {RLHF} framework.
\newblock In \emph{Proceedings of the European Conference on Computer Systems
  (EuroSys)}, pp.\  1279--1297, 2025.

\bibitem[Sheng et~al.(2024)Sheng, Cao, Li, Hooper, Lee, Yang, Chou, Zhu, Zheng,
  Keutzer, Gonzalez, and Stoica]{sheng2023slora}
Sheng, Y., Cao, S., Li, D., Hooper, C., Lee, N., Yang, S., Chou, C., Zhu, B.,
  Zheng, L., Keutzer, K., Gonzalez, J.~E., and Stoica, I.
\newblock {SLoRA}: Scalable serving of thousands of {LoRA} adapters.
\newblock In \emph{Proceedings of Machine Learning and Systems (MLSys)}, pp.\
  296--311, 2024.

\bibitem[{Thinking Machines Lab}(2025)]{tinker2025}
{Thinking Machines Lab}.
\newblock Tinker: a training {API} for fine-tuning open-weight models.
\newblock \url{https://thinkingmachines.ai/tinker/}, 2025.

\bibitem[{Unsloth AI}(2025{\natexlab{a}})]{unsloth2025fp8}
{Unsloth AI}.
\newblock {FP8} reinforcement learning.
\newblock
  \url{https://unsloth.ai/docs/get-started/reinforcement-learning-rl-guide/fp8-reinforcement-learning},
  2025{\natexlab{a}}.

\bibitem[{Unsloth AI}(2025{\natexlab{b}})]{unsloth2025grpo}
{Unsloth AI}.
\newblock Reinforcement learning ({RL}) guide.
\newblock Unsloth documentation,
  \url{https://unsloth.ai/docs/basics/reinforcement-learning-rl-guide},
  2025{\natexlab{b}}.
\newblock Accessed 2026-10-26.

\bibitem[{Unsloth AI}(2025{\natexlab{c}})]{unsloth2025standby}
{Unsloth AI}.
\newblock Memory efficient {RL}: Unsloth standby.
\newblock \url{https://docs.unsloth.ai/new/memory-efficient-rl},
  2025{\natexlab{c}}.
\newblock Accessed 2026-10-24.

\bibitem[{vLLM Team}(2025)]{vllm2025bitwise}
{vLLM Team}.
\newblock Bitwise-consistent on-policy reinforcement learning with {vLLM} and
  {TorchTitan}.
\newblock vLLM blog, 2025.

\bibitem[Wang et~al.(2023)Wang, Wei, Sabne, Davis, Ilbeyi, Hechtman, Chen,
  Murthy, Maggioni, Zhang, Kumar, Guo, Xu, and Zhou]{wang2023overlap}
Wang, S., Wei, J., Sabne, A., Davis, A., Ilbeyi, B., Hechtman, B., Chen, D.,
  Murthy, K.~S., Maggioni, M., Zhang, Q., Kumar, S., Guo, T., Xu, Y., and Zhou,
  Z.
\newblock Overlap communication with dependent computation via decomposition in
  large deep learning models.
\newblock In \emph{Proceedings of the ACM International Conference on
  Architectural Support for Programming Languages and Operating Systems
  (ASPLOS)}, pp.\  93--106, 2023.

\bibitem[Wang et~al.(2026{\natexlab{a}})Wang, Ren, and Gui]{wang2026forkkv}
Wang, S., Ren, R., and Gui, L.
\newblock {ForkKV}: Scaling multi-{LoRA} agent serving via copy-on-write
  disaggregated {KV} cache.
\newblock \emph{arXiv preprint arXiv:2604.06370}, 2026{\natexlab{a}}.

\bibitem[Wang et~al.(2026{\natexlab{b}})Wang, Yu, Zhuge, Mao, Li, Feng, Luo,
  Zhang, Cao, Zhang, Liu, and Zhang]{trace2026}
Wang, X., Yu, H., Zhuge, Z., Mao, B., Li, Z., Feng, J., Luo, Y., Zhang, Y.,
  Cao, Y., Zhang, M., Liu, D., and Zhang, J.
\newblock {TRACE}: Rollout-guided quantization-aware training for {FP4}
  reinforcement learning of {MoE} language models.
\newblock \emph{arXiv preprint arXiv:2610.07767}, 2026{\natexlab{b}}.

\bibitem[Wen \& Chaudhuri(2024)Wen and Chaudhuri]{wen2024flora}
Wen, Y. and Chaudhuri, S.
\newblock Batched low-rank adaptation of foundation models.
\newblock In \emph{International Conference on Learning Representations
  (ICLR)}, 2024.

\bibitem[Wu et~al.(2024)Wu, Zhu, Zhang, Sun, Liu, and Jin]{wu2024dlora}
Wu, B., Zhu, R., Zhang, Z., Sun, P., Liu, X., and Jin, X.
\newblock {dLoRA}: Dynamically orchestrating requests and adapters for {LoRA}
  {LLM} serving.
\newblock In \emph{18th USENIX Symposium on Operating Systems Design and
  Implementation (OSDI)}, pp.\  911--927, 2024.

\bibitem[Xi et~al.(2026)Xi, Ruan, Liao, Lin, Cai, Zhao, Yang, Keutzer, Han, and
  Zhu]{jetrl2026}
Xi, H., Ruan, C., Liao, P., Lin, Y., Cai, H., Zhao, Y., Yang, S., Keutzer, K.,
  Han, S., and Zhu, L.
\newblock {Jet-RL}: Enabling on-policy {FP8} reinforcement learning with
  unified training and rollout precision flow.
\newblock \emph{arXiv preprint arXiv:2601.14243}, 2026.

\bibitem[Xia et~al.(2024)Xia, Fu, Zhang, Jiang, and Cui]{xia2024loralinlaid}
Xia, Y., Fu, F., Zhang, W., Jiang, J., and Cui, B.
\newblock Efficient multi-task {LLM} quantization and serving for multiple
  {LoRA} adapters.
\newblock In \emph{Advances in Neural Information Processing Systems
  (NeurIPS)}, pp.\  63686--63714, 2024.

\bibitem[Xu et~al.(2024)Xu, Xie, Gu, Chen, Chang, Zhang, Chen, Zhang, and
  Tian]{xu2024qalora}
Xu, Y., Xie, L., Gu, X., Chen, X., Chang, H., Zhang, H., Chen, Z., Zhang, X.,
  and Tian, Q.
\newblock {QA-LoRA}: Quantization-aware low-rank adaptation of large language
  models.
\newblock In \emph{International Conference on Learning Representations}, 2024.

\bibitem[Yadav et~al.(2023)Yadav, Choshen, Raffel, and
  Bansal]{yadav2023compeft}
Yadav, P., Choshen, L., Raffel, C., and Bansal, M.
\newblock {ComPEFT}: Compression for communicating parameter efficient updates
  via sparsification and quantization.
\newblock \emph{arXiv preprint arXiv:2311.13171}, 2023.

\bibitem[Yao et~al.(2025)Yao, Hu, and Klimovic]{yao2023deltazip}
Yao, X., Hu, Q., and Klimovic, A.
\newblock {DeltaZip}: Efficient serving of multiple full-model-tuned {LLMs}.
\newblock In \emph{Proceedings of the European Conference on Computer Systems
  (EuroSys)}, pp.\  110--127, 2025.

\bibitem[Ye et~al.(2025)Ye, Chen, Lai, Lin, Zhang, Wang, Chen, Kasikci, Grover,
  Krishnamurthy, and Ceze]{ye2025flashinfer}
Ye, Z., Chen, L., Lai, R., Lin, W., Zhang, Y., Wang, S., Chen, T., Kasikci, B.,
  Grover, V., Krishnamurthy, A., and Ceze, L.
\newblock {FlashInfer}: Efficient and customizable attention engine for {LLM}
  inference serving.
\newblock In \emph{Proceedings of Machine Learning and Systems (MLSys)}, 2025.

\bibitem[Yu et~al.(2025)Yu, Zhang, Zhu, Yuan, Zuo, Yue, Dai, Fan, Liu, Liu,
  Liu, Lin, Lin, Ma, Sheng, Tong, Zhang, Zhang, Zhang, Zhu, Zhu, Chen, Chen,
  Wang, Yu, Song, Wei, Zhou, Liu, Ma, Zhang, Yan, Qiao, Wu, and
  Wang]{yu2025dapo}
Yu, Q., Zhang, Z., Zhu, R., Yuan, Y., Zuo, X., Yue, Y., Dai, W., Fan, T., Liu,
  G., Liu, L., Liu, X., Lin, H., Lin, Z., Ma, B., Sheng, G., Tong, Y., Zhang,
  C., Zhang, M., Zhang, W., Zhu, H., Zhu, J., Chen, J., Chen, J., Wang, C., Yu,
  H., Song, Y., Wei, X., Zhou, H., Liu, J., Ma, W.-Y., Zhang, Y.-Q., Yan, L.,
  Qiao, M., Wu, Y., and Wang, M.
\newblock {DAPO}: An open-source {LLM} reinforcement learning system at scale.
\newblock \emph{arXiv preprint arXiv:2503.14476}, 2025.

\bibitem[Yu et~al.(2026{\natexlab{a}})Yu, Singh, Shi, Sadri, Zhang, and
  Fan]{marlaas2026}
Yu, T. T.~L., Singh, G., Shi, G., Sadri, H., Zhang, Y., and Fan, Z.
\newblock {MARLaaS}: Multi-tenant asynchronous reinforcement learning as a
  service.
\newblock \emph{arXiv preprint arXiv:2605.08527}, 2026{\natexlab{a}}.

\bibitem[Yu et~al.(2026{\natexlab{b}})Yu, Venkatachalam, Wang, Li, Lin, Yu,
  Wang, Liu, Tang, Pei, Kang, and Ding]{yu2026plora}
Yu, Z., Venkatachalam, O.~R., Wang, Z., Li, Y., Lin, Y., Yu, Z., Wang, Y., Liu,
  L., Tang, X., Pei, S., Kang, Y., and Ding, Y.
\newblock {PLoRA}: An {NDP}-enhanced pooled-memory system for cost-efficient
  multi-{LoRA} serving.
\newblock \emph{arXiv preprint arXiv:2608.05483}, 2026{\natexlab{b}}.

\bibitem[Zhang et~al.(2025)Zhang, Du, Cao, and Xu]{zhang2025loquetier}
Zhang, Y., Du, H., Cao, C., and Xu, J.
\newblock Loquetier: A virtualized multi-{LoRA} framework for unified {LLM}
  fine-tuning and serving.
\newblock \emph{arXiv preprint arXiv:2511.00101}, 2025.

\bibitem[Zheng et~al.(2024)Zheng, Yin, Xie, Sun, Huang, Yu, Cao, Kozyrakis,
  Stoica, Gonzalez, Barrett, and Sheng]{zheng2024sglang}
Zheng, L., Yin, L., Xie, Z., Sun, C., Huang, J., Yu, C.~H., Cao, S., Kozyrakis,
  C., Stoica, I., Gonzalez, J.~E., Barrett, C., and Sheng, Y.
\newblock {SGLang}: Efficient execution of structured language model programs.
\newblock In \emph{Advances in Neural Information Processing Systems}, pp.\
  62557--62583, 2024.

\bibitem[Zhou et~al.(2026)Zhou, Shao, Zheng, Fan, and Wong]{ais2026}
Zhou, J., Shao, W., Zheng, L., Fan, Y., and Wong, N.
\newblock {AIS}: Adaptive importance sampling for quantized {RL}.
\newblock \emph{arXiv preprint arXiv:2605.13907}, 2026.

\bibitem[Zhu et~al.(2024)Zhu, Gao, Zhao, Zhao, Zuo, Gu, Xie, Tang, Xu, Ye,
  Kamahori, Lin, Wang, Wang, Krishnamurthy, and Kasikci]{zhu2025nanoflow}
Zhu, K., Gao, Y., Zhao, Y., Zhao, L., Zuo, G., Gu, Y., Xie, D., Tang, T., Xu,
  Q., Ye, Z., Kamahori, K., Lin, C.-Y., Wang, Z., Wang, S., Krishnamurthy, A.,
  and Kasikci, B.
\newblock {NanoFlow}: Towards optimal large language model serving throughput.
\newblock \emph{arXiv preprint arXiv:2408.12757}, 2024.

\bibitem[Zhu et~al.(2026)Zhu, Su, Ding, Song, Wang, and
  Pekhimenko]{zhu2026lorafusion}
Zhu, Z., Su, Q., Ding, Y., Song, K., Wang, S., and Pekhimenko, G.
\newblock {LoRAFusion}: Efficient {LoRA} fine-tuning for {LLMs}.
\newblock In \emph{Proceedings of the European Conference on Computer Systems
  (EuroSys)}, pp.\  586--604, 2026.

\bibitem[Zhuge et~al.(2026)Zhuge, Yu, Wang, Li, Cao, Liu, and Zhang]{quads2026}
Zhuge, Z., Yu, H., Wang, X., Li, Z., Cao, Y., Liu, D., and Zhang, J.
\newblock {QUADS}: Stabilizing {NVFP4} reinforcement learning for {MoE} via
  {QU}antization-error alignment across dual sides.
\newblock \emph{arXiv preprint arXiv:2607.15810}, 2026.

\end{thebibliography}
\bibliographystyle{mlsys2025}
\clearpage
\appendix
\setcounter{dbltopnumber}{4}\setcounter{topnumber}{4}\setcounter{totalnumber}{8}
\renewcommand{\dbltopfraction}{0.95}\renewcommand{\topfraction}{0.95}\renewcommand{\textfraction}{0.05}
\renewcommand{\floatpagefraction}{0.8}\renewcommand{\dblfloatpagefraction}{0.8}
\makeatletter
\setlength{\@fptop}{0pt}\setlength{\@fpbot}{0pt plus 1fil}\setlength{\@dblfptop}{0pt}\setlength{\@dblfpbot}{0pt plus 1fil}
\makeatother
\noindent These appendices hold the measurements behind the paper's claims. Appendix~\ref{app:discussion} discusses where EpiLoRA helps and its limitations, and Appendix~\ref{app:rlbuys} what each part of the RL system buys; Appendices~\ref{app:related} to~\ref{app:rl_math} give the closest systems and the kernel, serving and RL measurements; Appendix~\ref{app:repro} gives the code and the software.

\section{Discussion and limitations}
\label{app:discussion}

\textbf{Where a single low-precision base helps.}
EpiLoRA helps where the adapter's cost or the base's memory is the bottleneck.
At small decode batches, vLLM's LoRA kernels add 35--55\% to a Hopper layer, and the decode path removes most of it.
With many tenants, EpiLoRA's adapter kernels keep more of the base's throughput, and adapters stored in low precision let more of them stay resident.
In RL, an FP8 trainer removes the bf16 copy of the base, which lets larger models train on smaller GPUs.
Other engines could adopt the decode path by adding the up-projection in a CUTLASS epilogue that waits for a down-projection launched first.
Computing the down-projection inside the normalization kernel that already reads the input would hide it for the query/key/value and gate/up projections, whose inputs come from a normalization, but not for the output and down projections. Its gain is bounded by what stays exposed today: at one decode token, the exposed down-projection and the second kernel's launch together take 1.4--2.2\,\textmu s of the 10--12\,\textmu s a rank-64 adapter still adds to a Hopper layer (\cref{tab:epilogue}); the rest is the epilogue's extra work, so a leaner epilogue is the next step.

\textbf{Where it does not.}
Where neither memory nor latency is the bottleneck, simpler designs are as good or better.
Where two copies fit, FlashRL's recipe is the cheapest: on the 9B at 2{,}048-token rollouts on an H100 it takes 2.96 GPU-hours per run, against 3.10 for our FP8 trainer with TIS (the difference is in training) and 3.22 for bf16 training. At 2{,}048 tokens one copy saves memory, not time, on the 9B: one bf16 copy that keeps its KV cache trains as fast as one FP8 copy on the L40S (174 against 173\,s per step), while on the 27B the FP8 copy is 1.33$\times$ faster (\cref{tab:rlshbf16}). With rollouts of up to 8{,}192 tokens, the bf16 copy that keeps its KV cache takes 2.2$\times$ EpiLoRA's step time on the 9B, or 1.9$\times$ that of an FP8 copy that also keeps it (\cref{tab:longroll}); Unsloth's bf16 mode, which releases the cache while training and which we did not run, holds bf16 offloading's memory without its weight transfers, so it would take at most bf16 offloading's 1.3$\times$.
The matched engine samples more slowly than a W8A8 engine: it takes 1.35$\times$ the GPU-hours of corrected W8A8 rollouts on the H100, almost all in generation, and 0.95--1.03$\times$ on the L40S (\cref{tab:gensplit,tab:rlparts}).
vLLM's Marlin engine, which also keeps activations in 16 bits, reaches the same accuracy at 3.9 GPU-hours, so the matched engine's advantage, a layout the trainer can share, is unproven; on the L40S one copy ended 3.4 points below bf16 training ($[-5.7, -1.1]$).
With long rollouts on the 27B, FlashRL's steps trained 19\% more tokens than ours in the same time in the 12-step probes (\cref{tab:rlprobedetail}), and in the full 100-step runs its recipe needed 2--3\% fewer GPU-hours per trained token than ours (Table~\ref{tab:rlong27}), since its bf16 trainer is faster; at 16{,}384 tokens the bf16 trainers run out of memory, and FlashRL's recipe fits only with chunked log-probabilities and then takes 0.86$\times$ our step time while training fewer tokens per step (\cref{tab:longroll}).
A trainer that rounds its activations as the engine does (our proxy for Unsloth's FP8 trainer; S8RA in \cref{tab:arms}) trains 1.2$\times$ as fast as ours on the L40S (35--39 against 45--46\,s of training per step, from the runs' logs), at 1.4--1.6$\times$ our trainer's mismatch with the engine.
With 32 tenants, TensorRT-LLM delivers 1.05$\times$ EpiLoRA's throughput on the H200 in its own job, SGLang~0.5.9 delivers 1{,}803 tokens/s on Qwen3-8B on the L40S against EpiLoRA's 1{,}761 in Table~\ref{tab:wins}'s job (1.02$\times$, in separate jobs; SGLang's engine without LoRA delivers 3{,}179), and vLLM's kernels tuned for Llama-3.1-8B match ours.

\textbf{Limitations.}
\begin{enumerate}[label=(\roman*),leftmargin=1.8em,itemsep=2pt,topsep=2pt,parsep=0pt]
\item \emph{Hardware.} Datacenter Blackwell GPUs (sm\_100) are untested; our Blackwell results come from a MIG slice of an RTX PRO 6000 (sm\_120). sm\_100's shorter main loops would leave more of the down-projection exposed. The first-wave window used in Figure~\ref{fig:exposure} is an empirical fit chosen after the data, and it overestimates the exposure beyond two waves. Our Triton GEMM trails CUTLASS on the H100 above 32 tokens, and at one token by up to 28\% on narrow tensor-parallel shards and on small models (Table~\ref{tab:k2cutlass}).
\item \emph{Scope of serving.} The tensor-parallel path is opt-in and was tested only at two-way parallelism on one 72B model, where its batch-1 lead requires tiles searched for the shards and where it ties vLLM with 8 tenants. Mixture-of-experts LoRA and block-scaled FP8 bases are untested, and speculative decoding was tested only with n-gram drafts on one GPU, where the decode path serves verification steps in 0.81--0.82$\times$ vLLM's time per token except in runs whose drafts stopped matching after a first token that run-to-run nondeterminism flipped, as in some of vLLM's own runs. TensorRT-LLM ran on one model, with its PyTorch backend and a single-process worker.
\item \emph{Workloads.} Serving speed uses random adapters and, except for one WildChat trace, synthetic requests; storage accuracy uses only RL-trained adapters; and every model except Llama-3.1-8B is a Qwen model.
\item \emph{Default CUDA-graph sizes.} At vLLM's default capture sizes, EpiLoRA's operator runs eagerly for prefill steps beyond 512 tokens. It still leads online at every rate on the H200, on the 72B with tensor parallelism and on the H100 with long prompts, but with 512-token prompts it trails vLLM on the H100 at 16 requests/s (1.18$\times$ its time per output token, p99 TTFT 1.11--1.20$\times$) and trails it slightly on the L40S at 4--6 requests/s (1.02$\times$, p99 TTFT up to 1.32$\times$; \cref{tab:defcap}). The kernel-level account (\cref{tab:kdiff}): the timed dispatch, which times its candidates on CUDA-graph replays, mixes vLLM's adapter kernels into some projections and so pays both systems' launch costs on the H100's host CPU, and EpiLoRA's own down-projection took twice vLLM's GPU time at these batch sizes. Handing every batch above 32 tokens to vLLM's kernels widens the gap (1.23$\times$ and 1.26$\times$ at 8 and 16 requests/s). Two opt-in settings chosen after these diagnostics, EpiLoRA's own kernels above the capture bound and tiles searched for large batches, narrow it to 1.07$\times$ at 16 requests/s with p99 TTFT within 1.1$\times$. Timing the dispatch's candidates as they run outside the graphs, host time included, would let it choose these settings where they help.
\item \emph{Co-serving.} Serving tenants from the copy that RL trains on works but does not pay off: on one L40S it saves only 5--8\% of the GPU-hours of two separate deployments (0.92--0.95$\times$), RL's steps take 1.75$\times$ as long, and the tenants' p99 time to first token rises from 0.26 to 2.87\,s.
\item \emph{Batch-dependent numerics.} A request's tokens are computed by the decode path in decode batches of up to 32 tokens and by the vendor GEMM with adapter kernels above, so greedy outputs can change with the batch's composition, as they do between vLLM's own kernel configurations: on 16 prompts, every alternative configuration, vLLM's tuned and dual-stream kernels included, diverges from vLLM's default path on 5--9 in the output check of our H100 end-to-end test, where EpiLoRA's decode path diverges on at most one more; without adapters, the two engines' outputs are identical. Deployments that need batch-invariant outputs need batch-invariant kernels~\citep{he2025nondeterminism}.
\end{enumerate}

\section{What each part of the RL system buys}
\label{app:rlbuys}

\Cref{tab:rlbuys} summarizes, for each part of the RL system of Section~\ref{sec:rlsys}, what it buys and what it costs, measured on Qwen3.5-9B on one L40S unless noted; \cref{tab:rlparts} gives the full per-configuration accounting.
Running the matched engine, one copy and the KV release together in a 30-step probe takes 153\,s per step, against 155\,s with the KV cache kept, with twice the KV cache (23 against 12\,GiB).

\begin{table*}[h]
\caption{\textbf{What each part of the RL system buys and costs.} Qwen3.5-9B on one L40S unless noted; GPU-hours per 200-step run; accuracy differences with 95\% confidence intervals.}
\label{tab:rlbuys}
\centering\footnotesize\setlength{\tabcolsep}{5pt}
\begin{tabular}{@{}>{\raggedright\arraybackslash}p{0.16\textwidth}>{\raggedright\arraybackslash}p{0.36\textwidth}>{\raggedright\arraybackslash}p{0.40\textwidth}@{}}
\toprule
Part & Benefit & Cost \\
\midrule
FP8 trainer weights & 6.5\,GiB less GPU memory for the 9B and 22\,GiB for the 27B; the 27B trains on one 48\,GB L40S & None measurable: 7.2 GPU-hours, the same as FlashRL's bf16 trainer \\
\addlinespace[2pt]
One shared copy & Pinned host memory falls from 48--159\,GiB to 7--11\,GiB & None measurable: 7.1 against 7.2 GPU-hours with offloading \\
\addlinespace[2pt]
KV-cache release & A 21\,GiB instead of a 10\,GiB KV cache during rollouts & None measurable at 2{,}048 tokens; at 8{,}192 tokens the 27B's KV-releasing copy took 1.05$\times$ FP8 offloading's GPU-hours \\
\addlinespace[2pt]
Unrounded (W8A16) rollouts, ours or Marlin's & No correction needed: uncorrected runs end where W8A8 with TIS does & On the H100, 4.2 GPU-hours (Marlin 3.9) against 3.1 for W8A8 with TIS; on the L40S, 8.8 against 9.3 resident and 7.2 offloaded; with one copy at engine share 0.55, $-3.4$ points $[-5.7, -1.1]$ against bf16 \\
\bottomrule
\end{tabular}
\end{table*}

\FloatBarrier
\section{The closest systems}
\label{app:related}
\Cref{tab:related} compares EpiLoRA's capabilities with those of the closest serving and RL systems. Services that train and serve many users' adapters on shared bases (Tinker, MinT, MARLaaS) do not document their base precision or kernels, so they are not in the table.

\begin{table}[t]
\caption{\textbf{EpiLoRA brings many tenants' adapters and an RL trainer onto one low-precision base.} Capabilities of the closest systems, not their speed. \emph{Fused}: adapters computed inside the low-precision base GEMM. \emph{One copy}: the trainer computes on the engine's copy of the base. \emph{Matched}: engine and trainer compute on the same weights with activations unrounded, so RL needs no off-policy correction.}
\label{tab:related}
\centering\footnotesize\setlength{\tabcolsep}{3.5pt}
\newcommand{\hd}[2]{\begin{tabular}[b]{@{}c@{}}#1\\#2\end{tabular}}
\begin{tabular}{@{}lccc@{}}
\toprule
Serving system & \hd{Many}{tenants} & Fused & \hd{Low-precision}{adapters} \\
\midrule
Punica, S-LoRA & \checkmark & $\times$ & $\times$ \\
vLLM, SGLang & \checkmark & $\times$ & $\times$ \\
TensorRT-LLM & \checkmark & $\times$ & \checkmark \\
SVDQuant & $\times$ & \checkmark & $\times$ \\
LoRAFusion$^{\mathrm{a}}$ & \checkmark & \checkmark & $\times$ \\
FlexLLM$^{\mathrm{b}}$ & \checkmark & $\times$ & $\times$ \\
\textbf{EpiLoRA} & \checkmark & \checkmark & \checkmark \\
\midrule
RL system & \hd{Low-precision}{rollouts} & \hd{One}{copy} & Matched \\
\midrule
FlashRL & \checkmark & $\times$ & $\times$ \\
Jet-RL$^{\mathrm{c}}$ & \checkmark & $\times$ & $\times$ \\
QeRL & \checkmark & $\times$ & \checkmark \\
Unsloth$^{\mathrm{d}}$ & \checkmark & \checkmark & \checkmark \\
Unsloth, FP8$^{\mathrm{c}}$ & \checkmark & \checkmark & $\times$ \\
\textbf{EpiLoRA} & \checkmark & \checkmark & \checkmark \\
\bottomrule
\end{tabular}
\par\vspace{2pt}\raggedright\scriptsize $^{\mathrm{a}}$Fine-tuning: mixed-adapter batches in a bf16 GEMM. $^{\mathrm{b}}$Co-serves inference and supervised fine-tuning, not RL. $^{\mathrm{c}}$Rounds activations alike on both sides, which leaves a mismatch (Section~\ref{sec:rlsys}). $^{\mathrm{d}}$A bf16 or 4-bit copy, weight-only on both sides.
\end{table}

\section{Kernel details}
\label{app:kernels}

\begin{figure}[tp]
\centering
\resizebox{0.92\columnwidth}{!}{\begin{tikzpicture}[x=1cm, y=0.22cm, font=\scriptsize,
  lab/.style={align=center, inner sep=1pt},
  note/.style={align=left, inner sep=1pt, text=black!70},
  flow/.style={-{Stealth[length=1.4mm]}, draw=black!60, line width=0.45pt}]
\definecolor{sA}{RGB}{66,110,180}\definecolor{sB}{RGB}{214,120,40}\definecolor{sC}{RGB}{70,150,90}
\foreach \y/\c in {0/sA,1/sA,2/sA,3/sA,4/sA,5/sB,6/sB,7/sB,8/sB,9/sB,10/sB,11/sB,12/sC,13/sC,14/sC,15/sC}
  \fill[\c!35] (0, -\y) rectangle (0.9, -\y-1);
\draw[black!60] (0,0) rectangle (0.9,-16);
\node[lab, above] at (0.45, 0.2) {$x_q$ rows\\(sorted)};
\node[lab, left] at (0, -2.5) {$\ell_1$};
\node[lab, left] at (0, -8.5) {$\ell_2$};
\node[lab, left] at (0, -14) {$\ell_3$};
\fill[black!6] (1.6,0) rectangle (3.6,-16); \draw[black!60] (1.6,0) rectangle (3.6,-16);
\node[lab] at (2.6, -7) {mainloop\\[1pt]$\mathrm{acc} \mathrel{+}=$\\$x_q W_q^\top$\\[2pt]FP8 or NVFP4\\MMA over $K$};
\node[lab, above] at (2.6, 0.2) {$B_M{\times}B_N$ tile};
\draw[flow] (0.95, -8) -- (1.55, -8);
\foreach \y/\c in {0/sA,1/sA,2/sA,3/sA,4/sA,5/sB,6/sB,7/sB,8/sB,9/sB,10/sB,11/sB,12/sC,13/sC,14/sC,15/sC}
  \fill[\c!60] (4.3, -\y) rectangle (4.75, -\y-1);
\draw[black!60] (4.3,0) rectangle (4.75,-16);
\node[lab, above] at (4.52, 0.2) {$h$\\(rank $r$)};
\fill[sA!60] (5.15, -0.5) rectangle (6.35, -3.5); \node[lab] at (5.75, -2) {$B_{\ell_1}$ slice};
\fill[sB!60] (5.15, -6.5) rectangle (6.35, -9.5); \node[lab] at (5.75, -8) {$B_{\ell_2}$ slice};
\fill[sC!60] (5.15, -12.5) rectangle (6.35, -15.5); \node[lab] at (5.75, -14) {$B_{\ell_3}$ slice};
\node[lab, above] at (5.75, 0.2) {$r{\times}B_N$};
\fill[blue!7] (7.0,0) rectangle (8.9,-16); \draw[blue!50!black] (7.0,0) rectangle (8.9,-16);
\node[lab] at (7.95, -8) {epilogue\\[1pt]$\mathrm{acc} \cdot s_x s_w$\\[2pt]per segment $\ell$\\in the tile:\\[1pt]$\mathrm{acc}[\ell] \mathrel{+}=$\\$h_\ell B_\ell^\top$\\(masked\\rank-$r$ dot)\\[2pt]scatter rows\\to $y$};
\draw[flow] (3.45, 5.2) -| (7.95, 0.05); \draw[black!60, line width=0.45pt] (3.45, 0.05) -- (3.45, 5.2); \node[lab, above] at (5.7, 5.2) {acc};
\draw[flow] (4.8, -2) -- (5.1, -2); \draw[flow] (4.8, -8) -- (5.1, -8); \draw[flow] (4.8, -14) -- (5.1, -14);
\draw[flow] (6.4, -2) -- (6.95, -4); \draw[flow] (6.4, -8) -- (6.95, -8); \draw[flow] (6.4, -14) -- (6.95, -12);
\end{tikzpicture}}
\caption{\textbf{One output tile of K2.} Rows are tokens sorted by adapter, forming contiguous groups (here three). After the low-precision main loop, the epilogue adds each group's down-projection $h$ multiplied by its adapter's slice of $B$, then writes the rows back in token order.}
\label{fig:k2tile}
\end{figure}
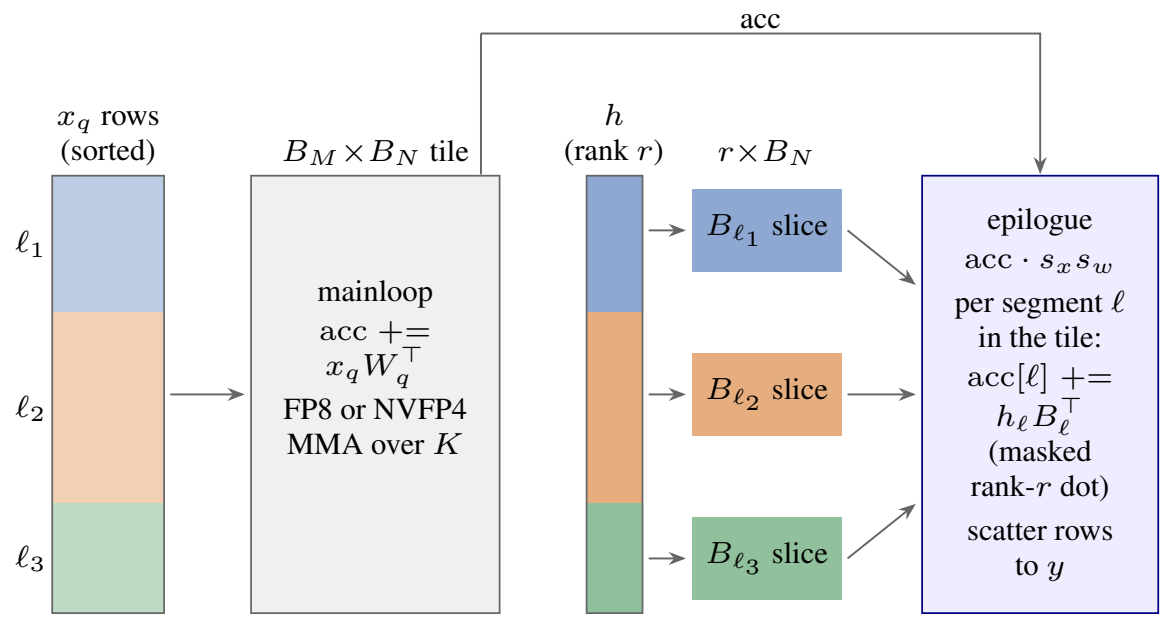

\textbf{One output tile of K2} (\cref{fig:k2tile}). K2 runs the low-precision GEMM over tokens sorted by adapter, so each output tile meets a few contiguous groups of rows, and its epilogue adds each group's up-projection.

\textbf{Table~\ref{tab:waterfall}'s L40S column} comes from the weight-only kernel benchmark: Marlin's W8A16 GEMM with vLLM's tuned LoRA kernels against EpiLoRA's fused weight-only kernel, one Qwen3-8B layer at one decode token with one rank-64 adapter, CUDA graphs with the L2 cache flushed; its other columns come from the rows of \cref{tab:pdl}.

\textbf{The decode path across decode batch sizes} (\cref{tab:pdl}). From one to sixteen tokens, one rank-64 adapter adds 4--9\% to a Qwen3-8B layer on Hopper GPUs and 3--7\% on the Blackwell slice with the decode path, against 34--61\% and 13--20\% for vLLM's tuned or dual-stream kernels.

\begin{table*}[tp]
\caption{\textbf{On the H100, the H200 and the Blackwell slice, the decode path brings one adapter's cost to a few percent of the layer.} One Qwen3-8B decoder layer (the four projections summed; FP8 base on the Hopper GPUs, our NVFP4 checkpoint on the Blackwell slice), one rank-64 adapter, CUDA graphs with the L2 cache flushed, microseconds (lower is better); each path at its best tile configuration. Previous: the plugin's best path before the decode path (Hopper: hybrid or swap-AB; Blackwell: fused or hybrid). Bold: the fastest path with the adapter. Last column: the PDL path's time over CUTLASS alone.}
\label{tab:pdl}
\centering\footnotesize\setlength{\tabcolsep}{6pt}
\begin{tabular}{@{}r rrrr rrr r@{}}
\toprule
\textmu s $\downarrow$ & & \multicolumn{3}{c}{vLLM's LoRA path} & \multicolumn{3}{c}{Ours} & PDL over \\
\cmidrule(lr){3-5}\cmidrule(lr){6-8}
Tokens & CUTLASS alone & default & tuned & dual-stream & previous & no PDL & PDL & CUTLASS (\%) \\
\midrule
\multicolumn{9}{@{}l}{\textit{H100, FP8 base}} \\
1 & 126 & 205 & 179 & 180 & 188 & 163 & \textbf{134} & +6 \\
4 & 128 & 195 & 179 & 172 & 192 & 162 & \textbf{133} & +4 \\
8 & 128 & 195 & 179 & 172 & 191 & 157 & \textbf{133} & +4 \\
16 & 128 & 196 & 180 & 174 & 181 & 167 & \textbf{136} & +7 \\
\addlinespace[3pt]
\multicolumn{9}{@{}l}{\textit{H200, FP8 base}} \\
1 & 104 & 190 & 161 & 167 & 171 & 144 & \textbf{112} & +8 \\
4 & 106 & 180 & 162 & 157 & 174 & 147 & \textbf{112} & +6 \\
8 & 106 & 180 & 162 & 156 & 175 & 144 & \textbf{113} & +7 \\
16 & 106 & 180 & 162 & 159 & 164 & 150 & \textbf{115} & +9 \\
\addlinespace[3pt]
\multicolumn{9}{@{}l}{\textit{Blackwell slice, NVFP4 base}} \\
1 & 286 & 388 & 342 & 340 & 316 & 330 & \textbf{293} & +3 \\
4 & 285 & 360 & 339 & 325 & 317 & 339 & \textbf{298} & +5 \\
8 & 288 & 359 & 343 & 326 & 320 & 342 & \textbf{298} & +3 \\
16 & 291 & 359 & 348 & 334 & 324 & 345 & \textbf{312} & +7 \\
\bottomrule
\end{tabular}
\end{table*}

\textbf{Controls for the overlap} (\cref{tab:floor,tab:splitk}). Serializing a library-fast down-projection (cuBLAS's GEMM or GEMV) before the same GEMM leaves the layer 1.17--1.25$\times$ slower than the decode path, and our own down-projection at its fastest split-K 1.13--1.17$\times$; only a serialized kernel that does no work comes within 2\%.

\begin{table*}[tp]
\caption{\textbf{The overlap beats serializing even a library-fast down-projection, and comes within 2\% of a serialized kernel that does no work (0.99--1.02$\times$).} One Qwen3-8B decoder layer's four projections at one decode token with one adapter, microseconds summed over the projections (CUDA graphs, L2 cache flushed), on one H200 and one H100. \emph{Serialized}: a kernel run before the same GEMM without overlap, which does no work (empty kernel), uses cuBLAS's GEMM or GEMV, or is our own down-projection. \emph{GEMM, $h$ given}: the GEMM with the down-projection's result already in memory, a lower bound for any overlap. \emph{Shrink alone}: the down-projection by itself.}
\label{tab:floor}
\centering\footnotesize\setlength{\tabcolsep}{4pt}
\begin{tabular}{@{}llrrrrrrrrr@{}}
\toprule
\textmu s $\downarrow$ & & & \multicolumn{4}{c}{Serialized} & GEMM, & \multicolumn{2}{c}{Shrink alone} & Empty \\
\cmidrule(lr){4-7}\cmidrule(lr){9-10}
GPU & Rank & Decode path & empty kernel & cuBLAS GEMM & cuBLAS GEMV & ours & $h$ given & cuBLAS & ours & kernel \\
\midrule
H200& 16 & \textbf{108.1} & 110.2 & 135.0 & 134.6 & 138.9 & 106.0 & \textbf{37.6} & 48.6 & 12.8 \\
 & 64 & \textbf{112.4} & 114.0 & 140.1 & 139.6 & 144.6 & 110.4 & \textbf{38.6} & 50.3 & 13.3 \\
 & 128 & \textbf{117.9} & 120.7 & 146.1 & 146.3 & 150.8 & 117.3 & \textbf{40.5} & 51.7 & 12.9 \\
\addlinespace[2pt]
H100& 16 & \textbf{129.3} & 132.0 & 154.9 & 154.5 & 156.8 & 128.7 & \textbf{42.0} & 49.4 & 20.3 \\
 & 64 & \textbf{133.4} & 134.8 & 158.0 & 158.5 & 163.0 & 130.9 & \textbf{42.1} & 53.4 & 19.6 \\
 & 128 & 139.2 & \textbf{138.2} & 163.5 & 163.0 & 164.8 & 135.3 & \textbf{44.6} & 52.2 & 19.9 \\
\bottomrule
\end{tabular}
\end{table*}

\begin{table*}[tp]
\caption{\textbf{Our own down-projection at its fastest split-K, run before the same GEMM without overlap, still trails the decode path.} \Cref{tab:floor}'s cases: one Qwen3-8B layer's four projections at one decode token with one adapter, the down-projection at every split-K from 1 to 32, serialized before the GEMM; per projection the fastest split, microseconds summed over the projections (CUDA graphs). Ratio: serialized over the decode path, for the layer and per projection. Shrink alone: the down-projection at the decode path's split and at the fastest.}
\label{tab:splitk}
\centering\footnotesize\setlength{\tabcolsep}{6pt}
\begin{tabular}{@{}llrrrrrr@{}}
\toprule
GPU & rank & decode path & serialized & ratio & per projection & \multicolumn{2}{c}{shrink alone} \\
\cmidrule(l){7-8}
 & & \textmu s $\downarrow$ & \textmu s $\downarrow$ & & & its split & fastest \\
\midrule
H200 & 16 & \textbf{109.6} & 118.3 & 1.08$\times$ & 1.05--1.12$\times$ & 11.1--14.6 & 6.1--8.6 \\
H200 & 64 & \textbf{112.1} & 130.8 & 1.17$\times$ & 1.11--1.24$\times$ & 9.0--17.8 & 6.6--9.0 \\
H200 & 128 & \textbf{117.9} & 146.1 & 1.24$\times$ & 1.17--1.33$\times$ & 9.7--18.2 & 6.9--9.7 \\
H100 & 16 & \textbf{128.8} & 140.3 & 1.09$\times$ & 1.06--1.15$\times$ & 9.3--15.2 & 7.4--9.7 \\
H100 & 64 & \textbf{133.1} & 149.9 & 1.13$\times$ & 1.09--1.19$\times$ & 10.1--18.1 & 7.5--9.9 \\
H100 & 128 & \textbf{137.9} & 162.0 & 1.17$\times$ & 1.11--1.26$\times$ & 10.6--18.2 & 7.9--10.5 \\
\bottomrule
\end{tabular}
\end{table*}

\textbf{Where the adapter's remaining cost goes} (\cref{tab:epilogue}). Under the decode path, most of what the adapter still adds to a layer is the epilogue's extra work; the exposed down-projection and its launch take 1--3\,\textmu s per layer, and EpiLoRA's GEMM without an adapter is up to 6\,\textmu s faster than CUTLASS with vLLM's separate quantization.

\begin{table}[tp]
\caption{\textbf{Under the decode path, the epilogue, not the down-projection, is most of the adapter's remaining cost.} One Qwen3-8B decoder layer (its four projections) with one rank-64 adapter at 1--16 decode tokens; CUDA graphs, L2 cache flushed. \emph{GEMM vs.\ CUTLASS}: EpiLoRA's GEMM without an adapter minus vLLM's quantization and CUTLASS GEMM; \emph{epilogue}: the GEMM with the adapter, the down-projection's result given, minus the GEMM without it; \emph{exposed + launch}: the decode path minus the former; \emph{adapter total}: the decode path over EpiLoRA's own GEMM.}
\label{tab:epilogue}
\centering\footnotesize\setlength{\tabcolsep}{3pt}
\resizebox{\columnwidth}{!}{\begin{tabular}{@{}lrrrrr@{}}
\toprule
 & & \multicolumn{4}{c}{Microseconds per layer $\downarrow$} \\
\cmidrule(l){3-6}
GPU & Tokens & GEMM vs.\ CUTLASS & Epilogue & Exposed + launch & Adapter total \\
\midrule
H100 & 1 & $-5.0$ & 10.2 & 2.2 & 12.5 \\
 & 4 & $-5.9$ & 9.6 & 1.8 & 11.4 \\
 & 8 & $-5.9$ & 9.0 & 1.8 & 10.8 \\
 & 16 & $-2.3$ & 7.7 & 3.1 & 10.8 \\
\addlinespace[2pt]
H200 & 1 & $-1.9$ & 8.8 & 1.4 & 10.2 \\
 & 4 & $-3.3$ & 7.6 & 1.6 & 9.1 \\
 & 8 & $-2.4$ & 7.8 & 1.5 & 9.3 \\
 & 16 & $+0.1$ & 6.7 & 2.5 & 9.2 \\
\bottomrule
\end{tabular}}
\end{table}

\textbf{Which path the dispatch picks} (\cref{tab:dispatch}). At engine start, the plugin times every candidate path per projection shape and batch-size bucket, on CUDA-graph replays with one adapter, and keeps the fastest. On the H100 with 32 rank-64 adapter slots it picks the decode path for every projection up to 32 tokens; above, it picks EpiLoRA's adapter kernels after the vendor GEMM for the projections that combine three or four outputs, where the projection, vendor GEMM included, takes 0.53--0.85$\times$ its time with vLLM's kernels, and vLLM's own adapter kernels for the others.

\textbf{Why EpiLoRA's adapter kernels win on these projections} (\cref{tab:expandkind}). In the profiled step of \cref{tab:kdiff}, EpiLoRA's GPU time on the H100 is 5.0\,ms below vLLM's: 3.4\,ms in the up-projection (expand) and 1.9\,ms in activation quantization and norms, which EpiLoRA's base path computes with vLLM's CUDA quantization kernel rather than the kernels vLLM's compiler generates (as much with EpiLoRA's own adapter kernels throughout, so not from them), while its down-projection takes 0.2\,ms more. We attribute the up-projection's difference to how the kernels index a projection's outputs. vLLM's expand kernel treats them as slices and launches one program per adapter, slice and tile of the widest slice, so the programs of narrower slices past their width exit without work. With several slices it reads each slice's starting column in the output from memory at run time, and when their widths differ, as for attention's query, key and value under grouped-query attention (8{,}192, 1{,}024 and 1{,}024 columns in Qwen3.5-9B) and for the linear-attention input (2{,}048, 2{,}048, 4{,}096 and 4{,}096), also each slice's strides, which the compiler then cannot specialize. EpiLoRA's expand-add tiles the concatenated output instead, each tile inside one slice, and receives the slice boundaries as kernel arguments. The costs per output column are consistent with this account: with tiles tuned for these shapes, vLLM's expand takes 7.9\,ns on the projections with one output, 13.4\,ns on gate/up's two equal ones and 21.6\,ns on the three or four unequal ones, while EpiLoRA's takes 12.1--13.1\,ns on every kind, so on a single output vLLM's kernel is the faster and the dispatch keeps it there. On the H200, vLLM's expand takes 17.4\,ns per output column on the projections with two to four outputs against EpiLoRA's 11.7, and 8.8 against 13.1\,ns on those with one. We did not time vLLM's kernel launched once per slice or with the slice offsets as constants, either of which might close the gap, and these costs come from one model's 2{,}048-token step outside CUDA graphs; at the 64--512-token batches of the many-tenant runs, \cref{tab:dispatch} shows the same ordering by kind of projection. The gain also depends on the model: on Llama-3.1-8B, whose query, key and value have Qwen3-8B's widths, vLLM's tuned kernels tie EpiLoRA with 32 tenants (Table~\ref{tab:wins}). EpiLoRA's kernels read vLLM's own adapter buffers, so vLLM could adopt them.

\begin{table*}[tp]
\caption{\textbf{Above 32 tokens, EpiLoRA's adapter kernels win on the projections that combine three or four outputs, and the dispatch keeps vLLM's kernels for the rest.} The timed dispatch's choice per projection of Qwen3.5-9B on the H100 with 32 rank-64 adapter slots (FP8 base, CUDA-graph replays, one adapter), and the projection's time with vLLM's adapter kernels over its time with EpiLoRA's at the given batch size, the same vendor GEMM included in both (above 1: EpiLoRA's are faster; our call of vLLM's kernels zero-fills their buffer, at most 0.5\,MB here, once more than vLLM does). The default-capture serving configuration's dispatch picks the same paths, and so does the dispatch on an H200, where this ratio is 1.19--2.00 on the projections with three or four outputs and the gate/up projection switches to them from 256 tokens. With 32 rank-64 tenants on the H200, EpiLoRA delivers 5{,}813 tokens/s against vLLM's 5{,}286 (1.10$\times$); with vLLM's adapter kernels above 32 tokens instead of its own, it delivers 5{,}148 (0.97$\times$), so the gain comes from EpiLoRA's kernels above 32 tokens (one job, three passes each).}
\label{tab:dispatch}
\centering\footnotesize\setlength{\tabcolsep}{4pt}
\resizebox{\textwidth}{!}{\begin{tabular}{@{}lllrrrr@{}}
\toprule
 & \multicolumn{2}{c}{Path picked at} & \multicolumn{4}{c}{vLLM's kernels' time over EpiLoRA's at ($\uparrow$)} \\
\cmidrule(lr){2-3}\cmidrule(l){4-7}
Projection & 1--32 tokens & 64--2{,}048 tokens & 64 & 128 & 256 & 512 \\
\midrule
linear attention, input (4 outputs) & decode path & EpiLoRA's kernels & \textbf{1.31} & \textbf{1.17} & \textbf{1.62} & \textbf{1.57} \\
attention, query/key/value (3 outputs) & decode path & EpiLoRA's kernels & \textbf{1.45} & \textbf{1.25} & \textbf{1.78} & \textbf{1.89} \\
MLP, gate/up (2 outputs) & decode path & vLLM's kernels; EpiLoRA's kernels at 256 & 0.87 & 0.89 & 1.00 & 0.99 \\
attention output & decode path & vLLM's kernels & 0.81 & 0.72 & 0.84 & 0.70 \\
MLP, down & decode path & vLLM's kernels & 0.84 & 0.74 & 0.89 & 0.78 \\
\bottomrule
\end{tabular}}
\end{table*}

\begin{table}[tp]
\caption{\textbf{vLLM's up-projection costs more per output column as a projection's outputs multiply; EpiLoRA's costs about the same on every kind.} The up-projection's GPU time per kind of LoRA projection of Qwen3.5-9B in the profiled H100 step of \cref{tab:kdiff}, with vLLM's kernel tuned for these shapes and with EpiLoRA's expand-add. Each kind's time is a difference between profiles whose timed dispatches send it to different kernels: vLLM sends all 152 projections to its kernel (20.91\,ms); EpiLoRA sends the 32 with unequal outputs to its own (4.71\,ms, with 12.77 left on vLLM's), with the new tiles gate/up as well (14.23; 2.23 left; a profile of the same step that \cref{tab:kdiff} does not show), and with its own kernels all but the linear attention's 24 two-column projections (17.65; 0.15 left).}
\label{tab:expandkind}
\centering\footnotesize\setlength{\tabcolsep}{3pt}
\resizebox{\columnwidth}{!}{\begin{tabular}{@{}lrrrrr@{}}
\toprule
 & & \multicolumn{2}{c}{ms per step $\downarrow$} & \multicolumn{2}{c}{ns per column $\downarrow$} \\
\cmidrule(lr){3-4}\cmidrule(l){5-6}
Outputs (projections) & Columns & vLLM & EpiLoRA & vLLM & EpiLoRA \\
\midrule
3--4 unequal (32) & 376{,}832 & 8.14 & \textbf{4.71} & 21.6 & \textbf{12.5} \\
gate/up, 2 equal (32) & 786{,}432 & 10.54 & \textbf{9.52} & 13.4 & \textbf{12.1} \\
one output (64) & 262{,}144 & \textbf{2.08} & 3.42 & \textbf{7.9} & 13.1 \\
\bottomrule
\end{tabular}}
\end{table}

\textbf{When K2 is slower than CUTLASS} (\cref{tab:k2cutlass}). Without an adapter, K2 matches CUTLASS on the 8B's and the 72B's shapes and trails it on small and narrow ones, which lengthens its main loop and so widens the window that hides the down-projection.

\begin{table}[tp]
\caption{\textbf{Without LoRA, the decode path's GEMM matches CUTLASS on the 8B's and the 72B's shapes, and trails it on small and narrow ones.} K2 without an adapter (with its own per-token quantization) over vLLM's no-adapter path (per-token quantization, then CUTLASS) per projection on the H100 ($M$ tokens, $n$ cases; CUDA graphs; CUTLASS with vLLM's quantization), from the kernel benchmark's rows (Tables~\ref{tab:pdl} and~\ref{tab:exposure}; the 72B's shards from their own run, whose tiles the job of \cref{tab:tp2} uses); K2's tiles fully searched except for Qwen3-8B's tensor-parallel shards (a coarse sweep). Where K2 is slower, its longer main loop also widens the window that hides the shrink.}
\label{tab:k2cutlass}
\centering\footnotesize\setlength{\tabcolsep}{3pt}
\begin{tabular}{@{}llrrrr@{}}
\toprule
Shapes & $M$ & $n$ & \multicolumn{3}{c}{K2 / CUTLASS $\downarrow$} \\
\cmidrule(l){4-6}
 & & & min & median & max \\
\midrule
Qwen3-8B & 1--16 & 16 & 0.93 & 0.96 & 1.00 \\
Qwen3-1.7B & 1 & 8 & 0.92 & 1.08 & 1.12 \\
Qwen3-0.6B & 1 & 8 & 1.15 & 1.16 & 1.24 \\
Qwen3-8B, TP2 & 1 & 4 & 0.98 & 1.15 & 1.28 \\
Qwen3-8B, TP4 & 1 & 4 & 0.97 & 1.06 & 1.23 \\
Qwen3-8B, TP8 & 1 & 4 & 0.99 & 1.19 & 1.22 \\
Qwen2.5-72B, TP2 & 1--32 & 20 & 0.93 & 0.99 & 1.07 \\
\bottomrule
\end{tabular}
\end{table}

\textbf{The exposure model} (\cref{tab:exposure}). On every measured projection, the exposed time against the down-projection's time and the GEMM's main loop. The window of the GEMM's first wave of thread blocks, assuming one block per SM, was fitted after the data; with the resident blocks per SM that the CUDA occupancy API reports (two in most Hopper cases), the same window tracks the exposure less well.

\begin{table*}[tp]
\caption{\textbf{What stays exposed of the down-projection, on every measured projection} (Figure~\ref{fig:exposure}). Per GPU and shape family, with one adapter at 1--16 decode tokens (tensor-parallel shards: Qwen3-8B's per-GPU shapes at one token): the down-projection alone ($t_{\mathrm{shrink}}$), the GEMM's main loop ($t_{\mathrm{main}}$), the exposed time (the decode path's time minus the same GEMM with the down-projection already computed), and the median error of the bound $\max(0,\, t_{\mathrm{shrink}} - t_{\mathrm{window}})$ with the whole main loop as the window and with the main loop of the GEMM's first wave of thread blocks, assuming one block per SM (a window fitted after the data). Where the CUDA occupancy API reports two resident blocks per SM, as it does in most Hopper cases, the first-wave window tracks the exposure less well (Pearson 0.50 against 0.71 for the one-block window on the same 115 projections; 0.70 over all projections, as in Figure~\ref{fig:exposure}).}
\label{tab:exposure}
\centering\footnotesize\setlength{\tabcolsep}{1.7pt}
\begin{tabular}{@{}llrrrrrrr@{}}
\toprule
 & & & & & \multicolumn{2}{c}{Exposed ($\mu$s) $\downarrow$} & \multicolumn{2}{c}{$|$error$|$, median ($\mu$s) $\downarrow$} \\
\cmidrule(lr){6-7}\cmidrule(l){8-9}
GPU & Shapes & Cases & $t_{\mathrm{shrink}}$ ($\mu$s) & $t_{\mathrm{main}}$ ($\mu$s) & median & max & whole loop & first wave  \\
\midrule
H100 & rank 64 & 24 & 10.4--18.3 & 16.7--50.5 & 0.6 & 1.3 & \textbf{0.6} & \textbf{0.6} \\
H100 & rank 16 & 8 & 11.8--15.5 & 18.7--50.8 & 2.6 & 6.1 & 2.6 & \textbf{0.7} \\
H100 & rank 128 & 8 & 14.0--18.5 & 18.7--50.8 & 5.0 & 9.3 & 5.0 & \textbf{2.3} \\
H100 & TP 2 shards & 4 & 10.0--14.5 & 15.0--33.3 & 3.7 & 5.1 & 3.7 & \textbf{2.1} \\
H100 & TP 4 shards & 4 & 10.7--14.2 & 14.4--20.6 & 2.5 & 4.3 & 2.5 & \textbf{1.1} \\
H100 & TP 8 shards & 4 & 10.4--14.3 & 12.6--19.3 & 3.1 & 5.4 & 2.5 & \textbf{1.9} \\
H200 & rank 64 & 16 & 9.2--17.8 & 14.9--41.5 & 0.4 & 1.2 & \textbf{0.4} & \textbf{0.4} \\
H200 & rank 16 & 8 & 8.4--14.7 & 15.3--41.4 & 0.2 & 1.0 & \textbf{0.2} & \textbf{0.2} \\
H200 & rank 128 & 8 & 9.8--18.2 & 15.3--43.0 & 1.5 & 5.7 & \textbf{1.5} & \textbf{1.5} \\
H200 & TP 2 shards & 4 & 9.5--13.7 & 14.2--28.2 & 3.2 & 4.8 & 3.2 & \textbf{1.7} \\
H200 & TP 4 shards & 4 & 9.9--13.7 & 13.1--17.6 & 1.5 & 4.8 & 1.5 & \textbf{1.1} \\
H200 & TP 8 shards & 4 & 9.7--13.5 & 11.7--17.4 & 3.2 & 5.7 & 2.7 & \textbf{1.9} \\
GH200 & TP 8 shards & 3 & 9.4--13.8 & 11.4--17.0 & 4.3 & 4.9 & \textbf{2.5} & \textbf{2.5} \\
GH200 & Qwen3-0.6B, -1.7B & 16 & 7.8--11.5 & 12.3--18.0 & 0.0 & 0.5 & \textbf{0.1} & \textbf{0.1} \\
Blackwell slice & rank 64 & 16 & 15.0--29.7 & 31.6--126.0 & 0.7 & 2.1 & \textbf{0.9} & \textbf{0.9} \\
\bottomrule
\end{tabular}
\end{table*}

\begin{figure}[tp]
\centering
\includegraphics[width=\columnwidth]{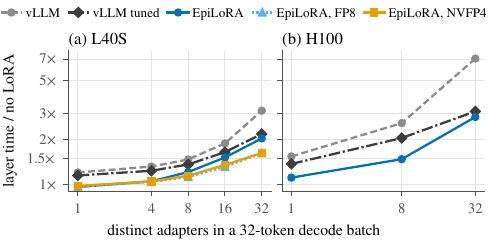}
\caption{\textbf{EpiLoRA's advantage is largest with few distinct adapters.} Time of one Qwen3-8B layer with rank-64 adapters at 32 decode tokens, relative to vLLM's FP8 GEMM without adapters, as the number of distinct adapters in the batch grows. (a) L40S, fused path, with adapters in bf16, FP8 or NVFP4; (b) H100, decode path. vLLM tuned: LoRA kernels tuned for the GPU.}
\label{fig:tenants}
\end{figure}

\textbf{Many distinct adapters} (\cref{fig:tenants}). With one adapter in a 32-token batch, EpiLoRA adds 11\% to the H100 layer where vLLM's tuned kernels add 38\%; with 32 distinct adapters the adapters' own weights dominate the step, and storing them in FP8 or NVFP4 nearly halves the overhead on the L40S.

\section{End-to-end serving}
\label{app:e2e}

\begin{figure}[tp]
\centering
\includegraphics[width=\columnwidth]{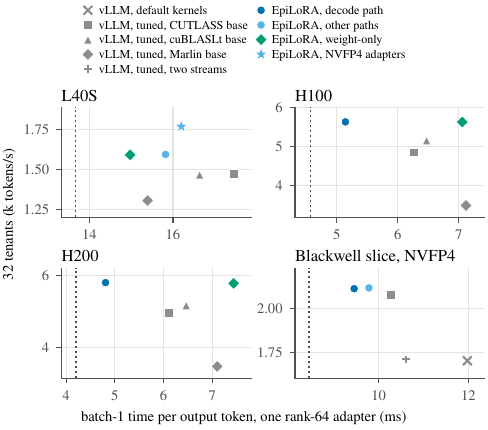}
\caption{\textbf{EpiLoRA is faster at batch 1 than every vLLM configuration and leads or ties with 32 tenants.} Each point is one configuration of one job: batch-1 time per output token with one rank-64 adapter (lower is better) against throughput with 32 rank-64 tenants (higher is better), for Qwen3.5-9B (Qwen3-8B with an NVFP4 base on the Blackwell slice). Dotted lines mark the same engine without LoRA, which is also the speed of a merged adapter. On the L40S, EpiLoRA with adapters stored in NVFP4 delivers 1{,}767 tokens/s with 32 tenants, against 1{,}475 for vLLM's best configuration in the same job (1.20$\times$). The KV cache with 32 rank-64 slots on the L40S holds 422k tokens with vLLM, 357k with EpiLoRA's bf16 adapters and 630k with NVFP4-stored ones; with bf16 adapters, the plugin's load-time warm-up fragments the memory vLLM later allocates for its adapter buffers, and an opt-in deferred warm-up restores 410k against vLLM's 422k in a job of its own. Table~\ref{tab:wins}'s Qwen3-8B row on the L40S comes from one job of its own: at batch 1, 14.05\,ms for vLLM's Marlin path and 13.95\,ms for EpiLoRA's weight-only mode; with 32 tenants, 1{,}628 tokens/s for vLLM's tuned path and 1{,}761 for EpiLoRA.}
\label{fig:pareto}
\end{figure}

\textbf{Batch 1 against throughput} (\cref{fig:pareto}). Every vLLM configuration and every EpiLoRA path on each GPU, one point per configuration.

\textbf{SGLang and TensorRT-LLM} (\cref{tab:sglang_h100,tab:sglang_h200,tab:trtllm}). Each system ran in its own job beside EpiLoRA, which gives the same-job ratios of Table~\ref{tab:systems}; \cref{tab:profbody} gives the GPU profiles behind its Nsight row.

\begin{table*}[tp]
\caption{\textbf{On the H100, EpiLoRA has 0.61$\times$ SGLang's batch-1 latency and 1.06$\times$ its throughput with 32 tenants.} Qwen3-8B with an FP8 base, rank-64 adapters; each engine's W1 (time per output token at batch 1) and W2 (throughput) without LoRA, and each LoRA path's change against its own engine; means of three starts. SGLang ran with CUDA graphs up to batch 256 and as many LoRA slots as adapters; its LoRA path adds about the same time per step with one adapter or 32 (3.2\,ms at batch 1).}
\label{tab:sglang_h100}
\centering\footnotesize\setlength{\tabcolsep}{4pt}
\begin{tabular}{@{}l rr rr rr@{}}
\toprule
 & \multicolumn{2}{c}{Without LoRA} & \multicolumn{2}{c}{1 adapter (\%)} & \multicolumn{2}{c}{32 adapters (\%)} \\
\cmidrule(lr){2-3}\cmidrule(lr){4-5}\cmidrule(l){6-7}
Engine, LoRA path & W1 (ms) $\downarrow$ & W2 (tok/s) $\uparrow$ & W1 $\downarrow$ & W2 $\uparrow$ & W1 $\downarrow$ & W2 $\uparrow$ \\
\midrule
SGLang 0.5.9, csgmv & 4.91 & 10{,}125 & +64.9 & $-$43.8 & +64.8 & $-$43.7 \\
SGLang 0.5.9, triton & 4.91 & 10{,}125 & +67.4 & $-$77.2 & +66.6 & $-$77.2 \\
vLLM 0.28, default kernels & \textbf{4.52} & \textbf{10{,}663} & +52.1 & \textbf{$-$27.5} & +83.5 & $-$54.3 \\
vLLM 0.28, dual-stream & \textbf{4.52} & \textbf{10{,}663} & +39.3 & $-$30.6 & +67.8 & $-$54.4 \\
ours, decode path & \textbf{4.52} & \textbf{10{,}663} & \textbf{+8.6} & $-$28.1 & \textbf{+8.1} & \textbf{$-$43.4} \\
\bottomrule
\end{tabular}
\end{table*}

\begin{table*}[tp]
\caption{\textbf{On the H200, EpiLoRA has 0.55$\times$ SGLang's batch-1 latency and 1.07$\times$ its throughput with 32 tenants.} The protocol of \cref{tab:sglang_h100} on one H200 NVL; one start per configuration. SGLang's LoRA path again adds about the same time per step with one adapter or 32 (3.5\,ms at batch 1).}
\label{tab:sglang_h200}
\centering\footnotesize\setlength{\tabcolsep}{4pt}
\begin{tabular}{@{}l rr rr rr@{}}
\toprule
 & \multicolumn{2}{c}{Without LoRA} & \multicolumn{2}{c}{1 adapter (\%)} & \multicolumn{2}{c}{32 adapters (\%)} \\
\cmidrule(lr){2-3}\cmidrule(lr){4-5}\cmidrule(l){6-7}
Engine, LoRA path & W1 (ms) $\downarrow$ & W2 (tok/s) $\uparrow$ & W1 $\downarrow$ & W2 $\uparrow$ & W1 $\downarrow$ & W2 $\uparrow$ \\
\midrule
SGLang 0.5.9, csgmv & 4.69 & 10{,}192 & +74.9 & $-$45.3 & +74.9 & $-$45.3 \\
SGLang 0.5.9, triton & 4.69 & 10{,}192 & +78.4 & $-$79.1 & +78.1 & $-$79.0 \\
vLLM 0.28, default kernels & \textbf{4.11} & \textbf{10{,}742} & +67.9 & $-$28.8 & +106.9 & $-$56.6 \\
vLLM 0.28, dual-stream & \textbf{4.11} & \textbf{10{,}742} & +52.1 & $-$30.8 & +90.7 & $-$55.9 \\
ours, decode path & \textbf{4.11} & \textbf{10{,}742} & \textbf{+10.4} & \textbf{$-$28.6} & \textbf{+12.5} & \textbf{$-$44.4} \\
\bottomrule
\end{tabular}
\end{table*}

\begin{table}[t]
\caption{\textbf{EpiLoRA against TensorRT-LLM, each pair measured in one job.} Qwen3-8B with an FP8 base and rank-64 adapters, CUDA graphs; for each metric, TensorRT-LLM's fastest configuration we found: $^{\mathrm{a}}$its block-FP8 checkpoint with the LoRA path overlapped with the base; $^{\mathrm{b}}$the ModelOpt FP8 checkpoint, LoRA overlapped; $^{\mathrm{c}}$the same at a KV-cache share of 0.6, found after the first run, with EpiLoRA measured again in that job. Without LoRA, EpiLoRA is vLLM's engine.}
\label{tab:trtllm}
\centering\footnotesize\setlength{\tabcolsep}{3pt}
\resizebox{\columnwidth}{!}{\begin{tabular}{@{}lrrrr@{}}
\toprule
 & \multicolumn{2}{c}{H100} & \multicolumn{2}{c}{H200} \\
\cmidrule(lr){2-3}\cmidrule(l){4-5}
 & TRT-LLM & EpiLoRA & TRT-LLM & EpiLoRA \\
\midrule
Time per token, no LoRA (ms) $\downarrow$ & 4.59$^{\mathrm{b}}$ & \textbf{4.52} & \textbf{4.08}$^{\mathrm{b}}$ & 4.11 \\
Time per token, one adapter (ms) $\downarrow$ & 6.93$^{\mathrm{a}}$ & \textbf{4.85} & 6.89$^{\mathrm{a}}$ & \textbf{4.53} \\
Throughput, one adapter (tok/s) $\uparrow$ & 6{,}858$^{\mathrm{b}}$ & \textbf{7{,}703} & 7{,}256$^{\mathrm{b}}$ & \textbf{7{,}654} \\
Throughput, 32 adapters (tok/s) $\uparrow$ & 5{,}901$^{\mathrm{c}}$ & \textbf{6{,}072}$^{\mathrm{c}}$ & \textbf{6{,}269}$^{\mathrm{b}}$ & 5{,}985 \\
\bottomrule
\end{tabular}}
\end{table}

\begin{table}[tp]
\caption{\textbf{Where the batch-1 time goes.} Time per output token for Qwen3-8B with an FP8 base on one H100, without and with one rank-64 adapter, and the GPU time the adapter adds per token, measured with NVIDIA's Nsight Systems profiler. Each system uses CUDA graphs; vLLM runs its dual-stream mode with its default LoRA kernels (6.02\,ms with tuned ones, Table~\ref{tab:wins}) and TensorRT-LLM its overlapped mode.}
\label{tab:profbody}
\centering\footnotesize\setlength{\tabcolsep}{4pt}
\begin{tabular}{@{}lrrr@{}}
\toprule
 & \multicolumn{2}{c}{Time per token (ms) $\downarrow$} & Added GPU \\
\cmidrule(lr){2-3}
System & no adapter & one adapter & time (ms) $\downarrow$ \\
\midrule
vLLM, two streams & \textbf{4.39} & 6.52 & 1.87 \\
SGLang & 5.10 & 8.23 & 3.01 \\
TensorRT-LLM & 4.60 & 7.50 & 2.87 \\
EpiLoRA & \textbf{4.39} & \textbf{4.92} & \textbf{0.40} \\
\bottomrule
\end{tabular}
\end{table}

\textbf{Served adapters' accuracy} (\cref{tab:servedacc}). The test behind Figure~\ref{fig:acceff}.

\begin{table}[tp]
\caption{\textbf{EpiLoRA serves adapters at the accuracy of vLLM's LoRA path.} Two RL adapters of Qwen3.5-9B on an FP8 base, greedy decoding, on MATH-500 (500 problems), OlympiadBench (675) and Minerva (272): accuracy with each system, the problems only one of them solves (vLLM/EpiLoRA), the difference (EpiLoRA's accuracy minus vLLM's, in points) $\pm$ the half-width of its paired 95\% interval, and McNemar's exact $p$.}
\label{tab:servedacc}
\centering\footnotesize\setlength{\tabcolsep}{4pt}
\begin{tabular}{@{}lrrrrr@{}}
\toprule
 & \multicolumn{2}{c}{Accuracy (\%) $\uparrow$} & One & Difference & \\
\cmidrule(lr){2-3}
Benchmark & vLLM & EpiLoRA & only & (95\%) & $p$ \\
\midrule
\multicolumn{6}{@{}l}{\textit{bf16-trained adapter}} \\
MATH-500 & \textbf{91.4} & 90.4 & 14/9 & $-1.0 \pm 1.9$ & 0.40 \\
Olympiad & 61.8 & \textbf{63.4} & 36/47 & $+1.6 \pm 2.6$ & 0.27 \\
Minerva & \textbf{54.4} & \textbf{54.4} & 7/7 & $+0.0 \pm 2.7$ & 1.00 \\
\addlinespace[2pt]
\multicolumn{6}{@{}l}{\textit{W8A8-trained adapter}} \\
MATH-500 & 89.0 & \textbf{90.2} & 13/19 & $+1.2 \pm 2.2$ & 0.38 \\
Olympiad & 58.4 & \textbf{58.5} & 40/41 & $+0.1 \pm 2.6$ & 1.00 \\
Minerva & 55.5 & \textbf{55.9} & 8/9 & $+0.4 \pm 3.0$ & 1.00 \\
\bottomrule
\end{tabular}
\end{table}

\textbf{With CUDA graphs over the prefill sizes} (\cref{tab:fullgraph}). Every point of Figure~\ref{fig:online}'s solid series, and the same test with rank-16 tenants.

\begin{table}[tp]
\caption{\textbf{Online serving with CUDA graphs captured up to 2{,}048 tokens for both systems.} Qwen3.5-9B with an FP8 base, 32 tenants, Poisson arrivals at the given rates, 512-token prompts and 256 output tokens; EpiLoRA's median time per output token and p99 time to first token over those of vLLM with tuned LoRA kernels, means of three server starts. On the H100 at 4 requests/s, one of EpiLoRA's three starts has an outlying p99 TTFT (335 against 150--160\,ms); at 6 requests/s the L40S is past saturation, where vLLM's p99 TTFT grows to seconds.}
\label{tab:fullgraph}
\centering\footnotesize\setlength{\tabcolsep}{3pt}
\resizebox{\columnwidth}{!}{\begin{tabular}{@{}lrrrrrrrr@{}}
\toprule
 & \multicolumn{4}{c}{Median TPOT / vLLM's $\downarrow$} & \multicolumn{4}{c}{p99 TTFT / vLLM's $\downarrow$} \\
\cmidrule(lr){2-5}\cmidrule(l){6-9}
\textit{H100}, requests/s & 2 & 4 & 8 & 16 & 2 & 4 & 8 & 16 \\
\midrule
\quad rank 64 & \textbf{0.77} & \textbf{0.77} & \textbf{0.86} & 0.95 & \textbf{0.98} & 1.37 & \textbf{1.08} & \textbf{1.03} \\
\quad rank 16 & 0.90 & 0.95 & 1.01 & \textbf{0.93} & 1.07 & \textbf{1.06} & 1.13 & 1.13 \\
\addlinespace[3pt]
\textit{L40S}, requests/s & 1 & 2 & 4 & 6 & 1 & 2 & 4 & 6 \\
\cmidrule(r){1-1}
\quad rank 64 & 0.92 & 0.95 & 0.99 & 0.97 & \textbf{1.00} & 0.98 & 0.99 & 1.21 \\
\quad rank 64, NVFP4-stored & \textbf{0.91} & \textbf{0.88} & \textbf{0.86} & \textbf{0.86} & \textbf{1.00} & 0.98 & 0.92 & \textbf{0.10} \\
\quad rank 16 & 0.94 & 0.94 & 0.95 & 0.87 & 1.02 & \textbf{0.96} & \textbf{0.90} & 0.83 \\
\bottomrule
\end{tabular}}
\end{table}

\textbf{The WildChat trace} (Table~\ref{tab:trace}). Of 4{,}000 first turns, vLLM's benchmark keeps the 3{,}383 with at most 1{,}024 prompt and 2{,}048 total tokens: prompts have a median of 23 tokens (90th percentile 247), replies 253 (759), and each request generates its recorded reply length. Each start draws 1{,}000 of them, a different sample per start and the same for both systems, with 85--102 prompt and 334--345 reply tokens on average. vLLM's p99 TTFT of 1.37\,s on the H100 at 8 requests/s comes from one of its three starts (3.7\,s; the other two 0.19--0.20\,s); EpiLoRA's p99 TTFT there is 1.06--1.13$\times$ vLLM's at the other three rates.

\textbf{At vLLM's default CUDA-graph sizes} (\cref{tab:defcap}). Steps that prefill more than 512 tokens run outside CUDA graphs; with 512-token prompts and 32 tenants, EpiLoRA lowers the time per output token at every rate on the H200 and at low rates on the H100 and the L40S.

\begin{table}[tp]
\caption{\textbf{Online serving at vLLM's default CUDA-graph sizes.} Qwen3.5-9B with an FP8 base, 32 rank-64 tenants, Poisson arrivals at the given rate (requests per second), 512-token prompts and 256 output tokens, 1{,}000 requests per rate; vLLM with LoRA kernels tuned for this model (on the H200, the H100's tuning) against EpiLoRA (with direct launches on the H100), both at vLLM's default capture up to 512 tokens. \emph{New tiles}: down-projection tiles searched for batches above 512 tokens; \emph{own kernels}: no vLLM adapter kernels in steps above the capture bound; both opt-in, each in a job of its own with vLLM measured again. Means of three server starts; ratios are EpiLoRA's over vLLM's.}
\label{tab:defcap}
\centering\footnotesize\setlength{\tabcolsep}{3pt}
\resizebox{\columnwidth}{!}{\begin{tabular}{@{}lrrrrrrr@{}}
\toprule
 &  & \multicolumn{3}{c}{Median TPOT (ms) $\downarrow$} & \multicolumn{3}{c}{p99 TTFT (ms) $\downarrow$} \\
\cmidrule(lr){3-5}\cmidrule(l){6-8}
GPU & Rate & vLLM & EpiLoRA & Ratio & vLLM & EpiLoRA & Ratio \\
\midrule
H100 & 2 & 9.9 & \textbf{7.9} & 0.79 & \textbf{213} & 254 & 1.20 \\
 & 4 & 15.5 & \textbf{13.0} & 0.84 & \textbf{238} & 274 & 1.15 \\
 & 8 & \textbf{28.9} & 29.0 & 1.00 & \textbf{268} & 305 & 1.14 \\
 & 16 & \textbf{54.8} & 64.7 & 1.18 & \textbf{360} & 400 & 1.11 \\
\addlinespace[2pt]
H100, new tiles & 2 & 9.9 & \textbf{7.7} & 0.78 & \textbf{214} & 242 & 1.13 \\
 & 4 & 15.5 & \textbf{12.5} & 0.81 & \textbf{243} & 262 & 1.08 \\
 & 8 & 29.0 & \textbf{27.7} & 0.95 & \textbf{269} & 289 & 1.08 \\
 & 16 & \textbf{54.9} & 61.1 & 1.11 & \textbf{361} & 386 & 1.07 \\
\addlinespace[2pt]
H100, new tiles, own kernels & 2 & 9.9 & \textbf{7.5} & 0.76 & 214 & \textbf{213} & 1.00 \\
 & 4 & 15.5 & \textbf{11.8} & 0.76 & 243 & \textbf{240} & 0.99 \\
 & 8 & 29.0 & \textbf{24.9} & 0.86 & 269 & \textbf{266} & 0.99 \\
 & 16 & \textbf{54.9} & 58.6 & 1.07 & \textbf{361} & 386 & 1.07 \\
\addlinespace[2pt]
H200 & 2 & 8.8 & \textbf{6.5} & 0.74 & \textbf{96} & 112 & 1.17 \\
 & 4 & 12.0 & \textbf{8.9} & 0.74 & \textbf{120} & 138 & 1.14 \\
 & 8 & 21.2 & \textbf{17.6} & 0.83 & 166 & \textbf{164} & 0.99 \\
 & 16 & 46.6 & \textbf{43.1} & 0.93 & 270 & \textbf{243} & 0.90 \\
\addlinespace[2pt]
L40S & 1 & 23.7 & \textbf{21.9} & 0.93 & \textbf{235} & 241 & 1.03 \\
 & 2 & 32.0 & \textbf{31.1} & 0.97 & \textbf{255} & 262 & 1.03 \\
 & 4 & \textbf{51.4} & 52.2 & 1.02 & \textbf{332} & 339 & 1.02 \\
 & 6 & \textbf{86.9} & 88.5 & 1.02 & \textbf{3{,}062} & 4{,}054 & 1.32 \\
\bottomrule
\end{tabular}}
\end{table}

\textbf{Long prompts, a 72B model and host time} (\cref{tab:longfast,tab:hostovh}). At vLLM's default CUDA-graph sizes, with direct launches, EpiLoRA lowers the time per output token with 2{,}048-token prompts on both Hopper GPUs and on a 72B model; \cref{tab:hostovh} gives each path's host and GPU time per layer for the steps that run outside CUDA graphs.

\begin{table*}[tp]
\caption{\textbf{At vLLM's default CUDA-graph sizes, EpiLoRA lowers the time per output token with 2{,}048-token prompts and on a 72B model.} Top: Qwen3.5-9B with an FP8 base, 32 rank-64 tenants, Poisson arrivals, 2{,}048-token prompts and 256 output tokens. Middle: Qwen2.5-72B-Instruct on one H200 with 8 tenants and 512-token prompts. Bottom: the 72B across two H100s with tensor parallelism and 8 tenants, against vLLM's kernels tuned for the shards. vLLM with tuned LoRA kernels against EpiLoRA with direct launches; means of three server starts, whose time-per-token ratios agree within 0.03.}
\label{tab:longfast}
\centering\footnotesize\setlength{\tabcolsep}{3pt}
\begin{tabular}{@{}llrrrrrr@{}}
\toprule
 &  & \multicolumn{3}{c}{Median TPOT (ms) $\downarrow$} & \multicolumn{3}{c}{p99 TTFT (ms) $\downarrow$} \\
\cmidrule(lr){3-5}\cmidrule(l){6-8}
GPU & Requests/s & vLLM & ours & ratio & vLLM & ours & ratio \\
\midrule
\multicolumn{8}{@{}l}{\emph{Qwen3.5-9B, 32 tenants, 2{,}048-token prompts}} \\
H200 & 1 & 8.00 & \textbf{5.88} & 0.73 & 175 & \textbf{173} & 0.99 \\
 & 2 & 9.79 & \textbf{7.08} & 0.72 & 235 & \textbf{210} & 0.90 \\
 & 4 & 16.70 & \textbf{11.09} & 0.66 & 338 & \textbf{277} & 0.82 \\
H100 & 1 & 8.08 & \textbf{6.41} & 0.79 & \textbf{242} & 250 & 1.03 \\
 & 2 & 10.02 & \textbf{7.84} & 0.78 & \textbf{295} & 297 & 1.01 \\
 & 4 & 16.89 & \textbf{13.04} & 0.77 & 402 & \textbf{384} & 0.95 \\
\addlinespace
\multicolumn{8}{@{}l}{\emph{Qwen2.5-72B, 8 tenants, 512-token prompts}} \\
H200 & 0.5 & 35.71 & \textbf{34.75} & 0.97 & 422 & \textbf{416} & 0.99 \\
 & 1 & 41.84 & \textbf{39.70} & 0.95 & 541 & \textbf{505} & 0.93 \\
 & 2 & 59.30 & \textbf{54.57} & 0.92 & 703 & \textbf{694} & 0.99 \\
\addlinespace
\multicolumn{8}{@{}l}{\emph{Qwen2.5-72B at TP2 on two H100s, vLLM tuned for the shards}} \\
H100 & 0.5 & 23.91 & \textbf{20.78} & 0.87 & \textbf{219} & 249 & 1.14 \\
 & 1 & 26.14 & \textbf{23.26} & 0.89 & \textbf{272} & 280 & 1.03 \\
 & 2 & 31.66 & \textbf{28.57} & 0.90 & 641 & \textbf{368} & 0.57 \\
\bottomrule
\end{tabular}
\end{table*}

\begin{table*}[tp]
\caption{\textbf{Host and GPU time per layer for the steps that run outside CUDA graphs.} One layer of Qwen3.5-9B (its four projections) with 32 adapter slots at the batch sizes of vLLM's eager steps, microseconds: the host time to launch each path (calls queued behind a GPU sleep, so the GPU never throttles them) and its GPU time (CUDA-graph replays). The H100's host CPU (Intel Xeon Gold 6448Y) pays about twice the host time per call of the H200's (AMD EPYC 9335). Direct launches halve EpiLoRA's host time and give bit-identical outputs.}
\label{tab:hostovh}
\centering\footnotesize\setlength{\tabcolsep}{3pt}
\begin{tabular}{@{}lrrrrrrrrrrrr@{}}
\toprule
 & \multicolumn{4}{c}{H100} & \multicolumn{4}{c}{H200} & \multicolumn{4}{c}{L40S} \\
\cmidrule(lr){2-5}\cmidrule(lr){6-9}\cmidrule(lr){10-13}
 & \multicolumn{2}{c}{512 tokens} & \multicolumn{2}{c}{2,048} & \multicolumn{2}{c}{512 tokens} & \multicolumn{2}{c}{2,048} & \multicolumn{2}{c}{512 tokens} & \multicolumn{2}{c}{2,048} \\
Path, \textmu s $\downarrow$ & host & GPU & host & GPU & host & GPU & host & GPU & host & GPU & host & GPU \\
\midrule
vLLM's quantize, GEMM and LoRA kernels & 648 & 916 & 639 & 2211 & 299 & 982 & 287 & 2332 & 460 & 1579 & 454 & 4320 \\
the vendor GEMM alone & 136 & 208 & 138 & 732 & 68 & 215 & 69 & 730 & 101 & 713 & 100 & 2952 \\
ours, hybrid & 541 & \textbf{870} & 532 & 1769 & 260 & \textbf{875} & 258 & \textbf{1756} & 405 & 1689 & 402 & 4211 \\
ours, hybrid, direct launches & 245 & 873 & 243 & \textbf{1768} & 125 & 876 & 125 & \textbf{1756} & 180 & 1686 & 181 & 4354 \\
ours, fused & 350 & 1699 & 345 & 3921 & 174 & 1804 & 171 & 4164 & 265 & \textbf{1554} & 259 & 3490 \\
ours, fused, direct launches & \textbf{106} & 1700 & \textbf{106} & 3919 & \textbf{59} & 1802 & \textbf{59} & 4142 & \textbf{80} & 1559 & \textbf{81} & \textbf{3474} \\
\bottomrule
\end{tabular}
\end{table*}

\textbf{Where a step outside CUDA graphs spends its time} (\cref{tab:kdiff}). In a step that prefills 2{,}048 tokens, EpiLoRA's default configuration takes less GPU time than vLLM but more host time, because its timed dispatch picks vLLM's adapter kernels for some projections and pays both systems' launch costs, which the H100's host CPU makes larger. Restricted to its own kernels, EpiLoRA matches vLLM's host time but its down-projection takes about twice vLLM's GPU time at this batch size. Tiles searched for large batches remove most of that difference, bringing the step to 1.01$\times$ vLLM's time on the H100 and 0.98$\times$ on the H200. Online on the H100, the two settings together narrow the gap at 16 requests/s from 1.18$\times$ to 1.07$\times$ and bring the p99 time to first token within 1.1$\times$ at every rate (\cref{tab:defcap}). On the H200, whose host CPU is faster, EpiLoRA's default dispatch, which mixes in vLLM's kernels, is faster still (0.94$\times$ vLLM's step time), so the settings stay opt-in.

\begin{table*}[tp]
\caption{\textbf{Where the time of a step outside CUDA graphs goes.} One four-tenant step of Qwen3.5-9B that prefills 2{,}048 tokens (FP8 base, 32 rank-64 adapter slots), above vLLM's default CUDA-graph sizes, so it runs outside the graphs; median step time over 20 steps, and the host (CPU) time and GPU time of one profiled step by kernel class, milliseconds. \emph{EpiLoRA}: its default configuration, whose timed dispatch may pick vLLM's adapter kernels inside its operator; \emph{own kernels}: EpiLoRA restricted to its own adapter kernels above 512 tokens; \emph{new tiles}: the same with the down-projection's tiles searched for large batches.}
\label{tab:kdiff}
\centering\footnotesize\setlength{\tabcolsep}{4pt}
\resizebox{\textwidth}{!}{\begin{tabular}{@{}lrrrrrrrr@{}}
\toprule
ms $\downarrow$ & \multicolumn{4}{c}{H100} & \multicolumn{4}{c}{H200} \\
\cmidrule(lr){2-5}\cmidrule(l){6-9}
 & vLLM & EpiLoRA & own kernels & own, new tiles & vLLM & EpiLoRA & own kernels & own, new tiles \\
\midrule
Step time (ms) & \textbf{66.7} & 71.0 & 71.7 & 67.4 & 65.5 & \textbf{61.4} & 68.6 & 64.5 \\
Host time, profiled (ms) & 75.3 & 88.2 & 74.3 & \textbf{71.9} & 65.7 & \textbf{59.2} & 66.2 & 62.8 \\
GPU time (ms) & 65.6 & \textbf{60.6} & 67.3 & 62.9 & 68.0 & \textbf{59.0} & 66.5 & 61.1 \\
\addlinespace[2pt]
\quad CUTLASS FP8 GEMM & \textbf{22.6} & 22.7 & 22.7 & 22.7 & 22.4 & \textbf{22.3} & 22.4 & 22.4 \\
\quad down-projection (shrink) & \textbf{5.0} & 5.2 & 11.5 & 6.8 & \textbf{5.6} & 6.8 & 13.0 & 7.3 \\
\quad up-projection (expand) & 20.9 & \textbf{17.5} & 17.8 & 18.1 & 22.7 & \textbf{16.0} & 17.2 & 17.4 \\
\quad activation quantization and norms & 11.5 & \textbf{9.6} & 9.8 & 9.9 & 11.9 & \textbf{8.6} & 8.7 & 8.8 \\
\quad everything else & 5.6 & 5.7 & \textbf{5.5} & \textbf{5.5} & 5.3 & 5.3 & \textbf{5.2} & \textbf{5.2} \\
\bottomrule
\end{tabular}}
\end{table*}

\textbf{The 72B model} (\cref{tab:tp2,tab:e2e72b}). Across two H100s with tensor parallelism, against vLLM's LoRA kernels tuned for the shard shapes, and on one H200 with the timed dispatch and with the decode path forced on every projection.

\begin{table*}[tp]
\caption{\textbf{Across two H100s, EpiLoRA leads vLLM's kernels tuned for the tensor-parallel shards at batch 1.} Qwen2.5-72B-Instruct with vLLM's online FP8 at two-way tensor parallelism, one random rank-64 adapter; W1: time per output token at batch 1 (mean of three starts and their range) and its increase over the engine without LoRA (16.39\,ms); W2: throughput with one adapter (without LoRA 2{,}877\,tokens/s). EpiLoRA runs the decode path on tiles searched for the shards; with 8 tenants vLLM delivers 1{,}567 tokens/s against EpiLoRA's 1{,}549.}
\label{tab:tp2}
\centering\footnotesize\setlength{\tabcolsep}{3pt}
\begin{tabular}{@{}lrrr@{}}
\toprule
Path & W1 (ms) $\downarrow$ & over no LoRA $\downarrow$ & W2 (tok/s) $\uparrow$ \\
\midrule
vLLM, tuned & 20.40 [20.31, 20.54] & +24\% & \textbf{2{,}141} \\
vLLM, dual, tuned & 20.31 [20.25, 20.34] & +24\% & 2{,}017 \\
EpiLoRA, searched & \textbf{17.75} [17.70, 17.77] & \textbf{+8\%} & 2{,}124 \\
\bottomrule
\end{tabular}
\end{table*}

\begin{table*}[tp]
\caption{\textbf{On one H200, EpiLoRA's timed dispatch serves the 72B fastest at batch 1.} Qwen2.5-72B-Instruct with vLLM's online FP8, random rank-64 adapters, vLLM's LoRA kernels tuned for this model on this GPU; W1 as in \cref{tab:tp2} (without LoRA 23.35\,ms), W2 with one adapter and with 8 (without LoRA 1{,}320\,tokens/s). Forcing the decode path on every projection is 1.03$\times$ slower than the timed dispatch, which serves a third of the projections with vLLM's kernels. An earlier job with the same configurations measured 28.00 against 26.80\,ms at batch 1 (0.96$\times$, the value in Table~\ref{tab:wins}) and, with 8 tenants, 985 against 951 tokens/s.}
\label{tab:e2e72b}
\centering\footnotesize\setlength{\tabcolsep}{3pt}
\begin{tabular}{@{}lrrrr@{}}
\toprule
Path & W1 (ms) $\downarrow$ & over no LoRA $\downarrow$ & W2, one (tok/s) $\uparrow$ & W2, 8 (tok/s) $\uparrow$ \\
\midrule
vLLM, tuned for the 72B & 28.20 [28.06, 28.27] & +21\% & \textbf{1{,}042} & \textbf{980} \\
vLLM, tuned, two streams & 29.18 [29.17, 29.19] & +25\% & 994 & 926 \\
EpiLoRA, timed dispatch (searched tiles) & \textbf{26.50} [26.48, 26.51] & \textbf{+13\%} & 1{,}038 & 974 \\
EpiLoRA, decode path forced & 27.32 [27.31, 27.32] & +17\% & 1{,}041 & 975 \\
\bottomrule
\end{tabular}
\end{table*}

\section{LoRA RL}
\label{app:rl_math}

\begin{figure}[tp]
\centering
\includegraphics[width=\columnwidth]{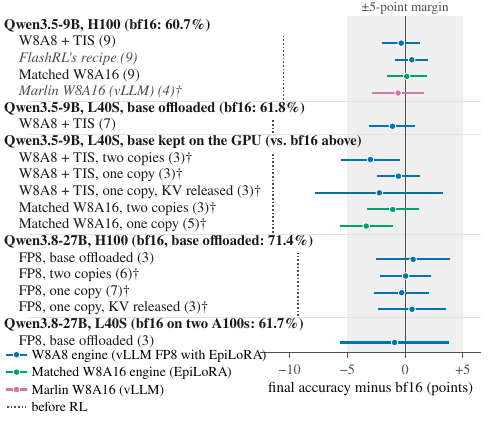}
\caption{\textbf{FP8 rollouts end close to bf16 training when they are corrected or matched.} Final accuracy minus that of bf16 training on the same GPU (in each heading), with 95\% confidence intervals and the number of runs; shaded: $\pm$5 points; dotted: before RL. Gray: other systems' recipes; KV freed: the engine releases its KV cache during training; $^\dagger$compared against this reference after the data; the 27B's two- and one-copy rows pool every completed run, as in \cref{tab:rlparts}. Accuracy on OlympiadBench and Minerva pooled, except the 27B on the H100, which also includes MATH-500 (\cref{tab:arms2}).}
\label{fig:rlacc}
\end{figure}

\textbf{Every configuration} (\cref{fig:rlacc}, \cref{tab:arms,tab:arms2}). Final accuracy of each RL configuration against bf16 training on the same GPU, with its cost; \cref{tab:paired} tests each comparison for equivalence within $\pm$5 points, also paired by seed and on one common engine.

\begin{table*}[tp]
\caption{\textbf{Every LoRA-RL configuration on Qwen3.5-9B: its configuration, cost and accuracy against its reference} (continued in \cref{tab:arms2}). Arms grouped by setting, each group named with its reference. Accuracy: OlympiadBench and Minerva pooled, greedy, each policy on its own engine. Engines: \emph{W8A8}, vLLM's FP8 engine with EpiLoRA; \emph{W8A16}, vLLM's weight-only FP8 engine (Marlin) or EpiLoRA's matched engine; \emph{asleep}, sleep mode during training. Trainers: bf16, or on the engine's FP8 weights, dequantized on every call (\emph{deq.}) or through our kernels (\emph{ours}, with flash-linear-attention, FLA). $\Delta$: final accuracy minus the reference's, Welch 95\% interval; $\le$ pre-RL: runs that ended at or below their accuracy before RL.}
\label{tab:arms}
\centering\footnotesize\setlength{\tabcolsep}{1.8pt}
\begin{threeparttable}
\begin{tabular}{@{}llllrrrrlr@{}}
\toprule
Code & Engine & Trainer, correction & GPU & Runs & GPU-h $\downarrow$ & \multicolumn{2}{c}{Accuracy (\%)} & $\Delta$ (points) $\uparrow$ & $\le$ \\
\cmidrule(lr){7-8}
 & & & & & & pre-RL & final & [95\% CI] & pre-RL \\
\midrule
\multicolumn{10}{@{}l}{\textit{Qwen3.5-9B math; against S0G}} \\
S0G & bf16 & bf16 & H100$^{\mathrm{a}}$ & 5 & 6.9 & 49.8--50.5 & 57.4--61.8 & \textit{reference} & 0/5 \\
S0T & bf16, asleep & bf16 & H100 & 3 & 5.6 & 50.5 & 61.0--61.7 & \textbf{$+1.5$} [$-0.4$, $+3.5$] & 0/3 \\
S8G & W8A8 & FP8 (deq.) & H100 & 5 & 5.6 & 50.6--51.5 & 2.3--60.7 & $-12.9$ [$-43.9$, $+18.0$] & 1/5 \\
S4F & W8A8 & bf16 & H100 & 3 & 5.3 & 49.7--51.0 & 4.3--52.4 & $-32.7$ [$-92.4$, $+27.0$] & 2/3 \\
S8T & W8A8 & FP8 (deq.), TIS & H100 & 4 & 5.7 & 50.6--51.0 & 57.6--61.0 & $-0.4$ [$-2.8$, $+2.1$] & 0/4 \\
S0TF & bf16, asleep & bf16 + FLA & H100 & 9 & \textbf{3.2} & 49.8--50.5 & 58.2--63.5 & $+0.9$ [$-1.2$, $+2.9$] & 0/9 \\
S8C & W8A8 & FP8 (ours), centering & H100 & 4 & 4.0 & 49.9--50.6 & 57.1--62.8 & $+0.7$ [$-2.9$, $+4.4$] & 0/4 \\
S8W & W8A16, Marlin & FP8 (ours) & H100 & 4 & 3.9 & 49.5--50.4 & 58.9--62.1 & $+0.3$ [$-2.2$, $+2.7$] & 0/4 \\
S8TL & W8A8 & FP8 (deq.), TIS & L40S & 3 & 13.4 & 49.9--50.5 & 60.5--60.9 & $+1.0$ [$-1.0$, $+2.9$] & 0/3 \\
S8ML & W8A16, matched & FP8 (ours) & L40S & 3 & 8.8 & 50.2--50.6 & 60.0--62.0 & $+0.9$ [$-1.5$, $+3.3$] & 0/3 \\
S8WA & W8A16, Marlin & FP8 (deq.) + FLA & A100 & 8 & 6.1 & 49.2--50.6 & 58.8--63.4 & $+0.8$ [$-1.3$, $+2.8$] & 0/8 \\
\addlinespace[3pt]
\multicolumn{10}{@{}l}{\textit{Qwen3.5-9B math, three jobs on one engine and trainer base (S9P pipelined, S9B batched); against S0G}} \\
S9P & W8A8 & FP8 (deq.) & H100 & 3 & \textbf{5.7} & 50.7--51.5 & 56.2--58.5 & \textbf{$-2.3$} [$-4.8$, $+0.1$] & 0/3 \\
S9B & W8A8 & FP8 (deq.) & H100 & 3 & 5.8 & 49.6--50.2 & 47.7--59.7 & $-6.3$ [$-20.4$, $+7.8$] & 1/3 \\
\addlinespace[3pt]
\multicolumn{10}{@{}l}{\textit{Qwen3.5-9B math, fast trainers; against S0TF}} \\
S8TF & W8A8 & FP8 (ours), TIS & H100 & 9 & 3.1 & 49.4--51.0 & 57.7--62.1 & $-0.3$ [$-2.0$, $+1.3$] & 0/9 \\
S4TF & W8A8 & bf16 + FLA, TIS & H100 & 9 & \textbf{3.0} & 50.2--51.0 & 60.1--62.8 & \textbf{$+0.6$} [$-0.9$, $+2.0$] & 0/9 \\
S8MH & W8A16, matched & FP8 (ours) & H100 & 9 & 4.2 & 50.7--50.8 & 58.2--64.4 & $+0.1$ [$-1.6$, $+1.9$] & 0/9 \\
S8UF & W8A8 & FP8 (ours) & H100 & 6 & 3.1 & 49.9--51.5 & 54.2--57.2 & $-4.9$ [$-6.5$, $-3.3$] & 0/6 \\
\addlinespace[3pt]
\multicolumn{10}{@{}l}{\textit{Qwen3.5-9B math, fast trainer on the L40S; against S8TL}} \\
S8TLF & W8A8 & FP8 (ours), TIS & L40S & 3 & 9.3 & 49.6--50.5 & 57.7--60.2 & $-2.0$ [$-5.1$, $+1.1$] & 0/3 \\
\addlinespace[3pt]
\multicolumn{10}{@{}l}{\textit{Qwen3.5-9B math with offload; against S0OL}} \\
S0OL & bf16 & bf16 + FLA & L40S & 7 & 8.4 & 49.6--51.1 & 59.9--64.8 & \textit{reference} & 0/7 \\
S8OL & W8A8 & FP8 (ours), TIS & L40S & 7 & \textbf{7.2} & 49.5--50.7 & 58.1--62.4 & $-1.1$ [$-3.1$, $+0.9$] & 0/7 \\
S4OL & W8A8 & bf16 + FLA, TIS & L40S & 3 & \textbf{7.2} & 50.2--50.8 & 59.2--63.6 & \textbf{$-0.3$} [$-4.6$, $+4.1$] & 0/3 \\
\addlinespace[3pt]
\multicolumn{10}{@{}l}{\textit{Qwen3.5-9B math on the L40S without correction (S8RA: activations rounded alike); against S0OL}} \\
S8RA & W8A8 & FP8 W8A8 (ours) & L40S & 3 & \textbf{8.2} & 49.6--51.3 & 51.1--60.5 & $-6.5$ [$-17.5$, $+4.5$] & 1/3 \\
S8ULF & W8A8 & FP8 (ours) & L40S & 3 & 8.5 & 49.5--50.8 & 52.5--60.1 & \textbf{$-6.3$} [$-15.4$, $+2.8$] & 0/3 \\
\bottomrule
\end{tabular}
\begin{tablenotes}[para,flushleft]\scriptsize
\item[a] Seeds 3--4 ran on A100s; GPU-hours over the H100 seeds.
\end{tablenotes}
\end{threeparttable}
\end{table*}

\begin{table*}[tp]
\caption{\textbf{Every LoRA-RL configuration (continued from \cref{tab:arms}): the 27B, one copy shared by the engine and the trainer, and Qwen3-4B.} Columns as in \cref{tab:arms}; each block's header names its benchmarks and, for the 27B, the accuracy before RL. The 27B is scored on the bf16 engine (on the L40S also on OlympiadBench and Minerva only, as the 9B), except where a header says otherwise. One FP8 copy of the 27B with its KV cache kept ran out of memory on the L40S.}
\label{tab:arms2}
\centering\footnotesize\setlength{\tabcolsep}{1.8pt}
\begin{threeparttable}
\begin{tabular}{@{}llllrrrlr@{}}
\toprule
Code & Engine & Trainer, correction & GPU & Runs & GPU-h & Accuracy (\%) & $\Delta$ (points) $\uparrow$ & $\le$ \\
\cmidrule(lr){7-7}
 & & & & & $\downarrow$ & final & [95\% CI] & pre-RL \\
\midrule
\multicolumn{9}{@{}l}{\textit{Qwen3.8-27B math, MATH-500, OlympiadBench and Minerva (before RL 61.6--62.8\%); against Q27}} \\
Q27 & bf16 & bf16 & 2$\times$A100 & 3 & 15.7 & 70.8--74.6 & \textit{reference} & 0/3 \\
S8T1g & W8A16, Marlin & FP8 (deq.), TIS & A100 & 3 & 15.3 & 71.9--73.3 & $-0.4$ [$-4.4$, $+3.6$] & 0/3 \\
S8T27H & W8A8 & FP8 (ours), TIS & H100 & 2 + 1$^{\mathrm{b}}$ & 4.4 & 72.4--73.5 & \textbf{$+0.2$} [$-3.9$, $+4.2$] & 0/2 \\
S0T27A & bf16, offloaded & bf16 + FLA & A100 & 3 & 7.3 & 71.2--72.0 & $-1.2$ [$-5.6$, $+3.1$]$^{\mathrm{c}}$ & 0/3 \\
S8T27A & W8A16, Marlin, offloaded & FP8 (deq.) + FLA, TIS & A100 & 3 & 6.5 & 70.6--72.0 & $-1.3$ [$-5.3$, $+2.7$]$^{\mathrm{c}}$ & 0/3 \\
S8S27H & W8A8 & FP8 (ours), one copy, TIS & H100 & 5 & \textbf{3.3} & 68.1--72.6 & $-2.5$ [$-6.2$, $+1.2$] & 0/5 \\
\addlinespace[3pt]
\multicolumn{9}{@{}l}{\textit{Qwen3.8-27B math on the H100, the same benchmarks; against S0T27HO}} \\
S0T27HO & bf16, offloaded & bf16 + FLA & H100 & 3 & 4.0 & 70.4--72.6 & \textit{reference} & 0/3 \\
S8T27HO & W8A8, offloaded & FP8 (ours), TIS & H100 & 3 & 3.4 & 70.6--73.7 & \textbf{$+0.7$} [$-2.5$, $+3.9$] & 0/3 \\
S4T27HO & W8A8, offloaded & bf16 + FLA, TIS & H100 & 3 & \textbf{3.2} & 70.7--72.4 & $-0.1$ [$-2.4$, $+2.3$] & 0/3 \\
\addlinespace[3pt]
\multicolumn{9}{@{}l}{\textit{Qwen3.8-27B math on the H100, the same benchmarks, one copy, KV cache released for training; against S8T27HO}} \\
S8K27H & W8A8 & FP8 (ours), one copy, TIS & H100 & 3 & 3.3 & 70.9--73.6 & $-0.1$ [$-3.5$, $+3.3$] & 0/3 \\
\addlinespace[3pt]
\multicolumn{9}{@{}l}{\textit{Qwen3.8-27B math on one L40S, OlympiadBench and Minerva (before RL 49.7--51.1\%); against Q27 on the same benchmarks}} \\
Q27 & bf16 & bf16 & 2$\times$A100 & 3 & 15.7 & 59.7--63.7 & \textit{reference} & 0/3 \\
S8T27L & W8A8, offloaded & FP8 (ours), TIS & L40S & 3 & \textbf{11.5} & 58.7--63.0 & $-0.9$ [$-5.7$, $+3.8$] & 0/3 \\
\addlinespace[3pt]
\multicolumn{9}{@{}l}{\textit{Qwen3.8-27B math on one L40S, the same benchmarks on each run's own FP8 engine, one copy, KV cache released; against S8T27L}} \\
S8K27L & W8A8 & FP8 (ours), one copy, TIS & L40S & 3 & 11.2 & 57.8--61.1 & $+0.0$ [$-3.9$, $+3.9$] & 0/3 \\
\addlinespace[3pt]
\multicolumn{9}{@{}l}{\textit{Qwen3.5-9B math on the L40S, one copy; against S8TLF (two copies)}} \\
S8S9L & W8A8 & FP8 (ours), one copy, TIS & L40S & 3 & 7.1 & 60.4--61.7 & $+2.4$ [$-0.2$, $+5.1$] & 0/3 \\
\addlinespace[3pt]
\multicolumn{9}{@{}l}{\textit{Qwen3.5-9B math on the L40S, one copy, matched engine; against S8ML (two copies)}} \\
S8MS9L & W8A16, matched & FP8 (ours), one copy & L40S & 5 & 7.3 & 56.1--60.1 & $-2.3$ [$-4.7$, $+0.1$] & 0/5 \\
\addlinespace[3pt]
\multicolumn{9}{@{}l}{\textit{Qwen3.5-9B math on the L40S, one copy, KV cache released for training; against S8OL}} \\
S8K9L & W8A8 & FP8 (ours), one copy, TIS & L40S & 3 & 7.1 & 56.6--61.8 & $-0.3$ [$-5.8$, $+5.1$] & 0/3 \\
\addlinespace[3pt]
\multicolumn{9}{@{}l}{\textit{Qwen3-4B math; against S0L}} \\
S0L & bf16 & bf16 & L40S & 3 & 4.5 & 49.7--50.2 & \textit{reference} & 0/3 \\
S8L & W8A8 & FP8 (deq.) & L40S & 3 & \textbf{3.6} & 49.7--51.4 & $+0.7$ [$-1.3$, $+2.7$] & 0/3 \\
\bottomrule
\end{tabular}
\begin{tablenotes}[para,flushleft]\scriptsize
\item[b] A run that stopped early and was not resumed (seed 2: out of memory at step 88).
\item[c] No reference run.
\end{tablenotes}
\end{threeparttable}
\end{table*}

\begin{table*}[tp]
\caption{\textbf{Equivalence of the RL configurations with their references, three ways.} Final accuracy (OlympiadBench and Minerva pooled) of each configuration minus its reference: on each policy's own engine, the Welch 95\% interval over all runs and the $t$ interval of the per-seed differences over the seeds both ran (a seed fixes the problem order); and the Welch interval with every policy re-scored on one common bf16 engine (avg@8). $p_{\ne 0}$: Welch's test of a zero difference; $p_{\pm 5}$: the two one-sided tests of the $\pm$5-point margin; both Holm-corrected over these twelve comparisons (a family chosen after the data). Runs: the configuration's and the reference's.}
\label{tab:paired}
\centering\scriptsize\setlength{\tabcolsep}{2.4pt}
\resizebox{\textwidth}{!}{\begin{tabular}{@{}llrlrrrlrlr@{}}
\toprule
 & & & \multicolumn{3}{c}{Each policy's own engine} & \multicolumn{2}{c}{Paired by seed} & \multicolumn{3}{c}{One bf16 engine, avg@8} \\
\cmidrule(lr){4-6}\cmidrule(lr){7-8}\cmidrule(l){9-11}
Comparison & GPU & Runs & Welch 95\% interval & $p_{\ne 0}$ & $p_{\pm 5}$ & Seeds & interval & Welch 95\% interval & $p_{\ne 0}$ & $p_{\pm 5}$ \\
\midrule
W8A8 + TIS against tuned bf16 & H100 & 9/9 & $-0.3$ [$-2.0$, $+1.3$] & 1.000 & ${<}0.001$ & 9 & $-0.3$ [$-2.6$, $+1.9$] & $-0.3$ [$-1.6$, $+1.1$] & 1.000 & ${<}0.001$ \\
FlashRL's recipe against tuned bf16 & H100 & 9/9 & $+0.6$ [$-0.9$, $+2.0$] & 1.000 & ${<}0.001$ & 9 & $+0.6$ [$-1.3$, $+2.4$] & $+0.5$ [$-0.6$, $+1.6$] & 1.000 & ${<}0.001$ \\
matched engine against tuned bf16 & H100 & 9/9 & $+0.1$ [$-1.6$, $+1.9$] & 1.000 & ${<}0.001$ & 9 & $+0.1$ [$-1.3$, $+1.5$] & $+0.1$ [$-1.4$, $+1.7$] & 1.000 & ${<}0.001$ \\
W8A8 + TIS, offloaded, against bf16 offloaded & L40S & 7/7 & $-1.1$ [$-3.1$, $+0.9$] & 1.000 & 0.005 & 7 & $-1.1$ [$-3.5$, $+1.3$] & $-0.3$ [$-2.0$, $+1.5$] & 1.000 & ${<}0.001$ \\
FlashRL's recipe, offloaded, against bf16 offloaded & L40S & 3/7 & $-0.3$ [$-4.6$, $+4.1$] & 1.000 & 0.106 & 3 & $-1.2$ [$-8.6$, $+6.3$] & $-0.1$ [$-3.8$, $+3.6$] & 1.000 & 0.046 \\
one copy against two copies (W8A8 + TIS) & L40S & 3/3 & $+2.4$ [$-0.2$, $+5.1$] & 0.601 & 0.108 & 3 & $+2.4$ [$+0.3$, $+4.5$] & $+2.1$ [$-0.3$, $+4.4$] & 0.748 & 0.046 \\
one copy against two copies (matched) & L40S & 5/3 & $-2.3$ [$-4.7$, $+0.1$] & 0.601 & 0.106 & 3 & $-2.8$ [$-5.3$, $-0.3$] & $-2.3$ [$-5.3$, $+0.7$] & 0.902 & 0.046 \\
one copy, KV released, against offloading (W8A8 + TIS) & L40S & 3/7 & $-1.1$ [$-6.8$, $+4.6$] & 1.000 & 0.151 & 3 & $-0.3$ [$-4.2$, $+3.6$] & $-1.6$ [$-4.5$, $+1.3$] & 1.000 & 0.046 \\
two copies (W8A8 + TIS) against bf16 offloaded & L40S & 3/7 & $-3.0$ [$-5.5$, $-0.5$] & 0.305 & 0.151 & 3 & $-3.9$ [$-11.3$, $+3.5$] & $-1.6$ [$-3.8$, $+0.6$] & 1.000 & 0.026 \\
one copy (W8A8 + TIS) against bf16 offloaded & L40S & 3/7 & $-0.6$ [$-2.4$, $+1.3$] & 1.000 & 0.002 & 3 & $-1.5$ [$-9.1$, $+6.1$] & $+0.5$ [$-1.1$, $+2.0$] & 1.000 & ${<}0.001$ \\
two copies (matched) against bf16 offloaded & L40S & 3/7 & $-1.1$ [$-3.3$, $+1.2$] & 1.000 & 0.017 & 3 & $-2.0$ [$-6.5$, $+2.5$] & $+0.4$ [$-2.4$, $+3.1$] & 1.000 & 0.026 \\
one copy (matched) against bf16 offloaded & L40S & 5/7 & $-3.4$ [$-5.7$, $-1.1$] & 0.108 & 0.151 & 5 & $-3.6$ [$-5.8$, $-1.5$] & $-1.9$ [$-3.5$, $-0.3$] & 0.304 & 0.006 \\
\bottomrule
\end{tabular}}
\end{table*}

\textbf{The decomposition on a second model} (\cref{tab:decomp4b}). On Qwen3-4B, a dense model, with five of its math RL adapters, the ordering of Figure~\ref{fig:kl} holds: rounding activations leaves 6.3--7.4$\times$ the bf16 pair's KL, most of it the activations' rounding itself (the fake-quant control), and the unrounded engines retain 1.1--1.6$\times$.

\begin{table*}[tp]
\caption{\textbf{On a second, dense model the decomposition's ordering holds.} Qwen3-4B, five math RL adapters (bf16- or W8A8-trained), 128 DAPO-Math prompts $\times$ 4 samples on one H100: each engine's per-token KL to the weight-matched trainer over that of a bf16 engine and trainer; the fake-quant control's share of the W8A8 engine's excess; and a trainer that rounds its activations as the engine does, over the W8A16 trainer.}
\label{tab:decomp4b}
\centering\footnotesize\setlength{\tabcolsep}{3pt}
\begin{tabular}{@{}lrrrrr@{}}
\toprule
 & \multicolumn{3}{c}{KL over the bf16 pair's $\downarrow$} & Fake-quant & Rounded alike \\
\cmidrule(lr){2-4}
Adapter & W8A8 & matched W8A16 & Marlin W8A16 & share & over W16 trainer \\
\midrule
bf16-trained, seed 0 & 6.47$\times$ & 1.35$\times$ & \textbf{1.34$\times$} & 0.76 & 1.35$\times$ \\
bf16-trained, seed 1 & 6.80$\times$ & \textbf{1.20$\times$} & 1.49$\times$ & 0.74 & 1.36$\times$ \\
bf16-trained, seed 2 & 7.41$\times$ & \textbf{1.51$\times$} & 1.60$\times$ & 0.86 & 1.50$\times$ \\
W8A8-trained, seed 0 & 6.31$\times$ & \textbf{1.14$\times$} & 1.30$\times$ & 0.79 & 1.49$\times$ \\
W8A8-trained, seed 1 & 7.16$\times$ & 1.38$\times$ & \textbf{1.30$\times$} & 0.76 & 1.37$\times$ \\
\bottomrule
\end{tabular}
\end{table*}

\textbf{The mismatch over training} (\cref{tab:drift}). Two hundred steps cannot show a collapse that comes later, but the mismatch's trend can: with the matched engine and no correction it stays at the level of a bf16 engine and trainer, while uncorrected W8A8 rollouts more than double theirs.

\begin{table*}[tp]
\caption{\textbf{Without correction, the matched engine's mismatch stays at the bf16 level over training; W8A8 rollouts' grows.} Qwen3.5-9B on the H100, the nine-seed configurations of \cref{fig:rlacc} and uncorrected W8A8 rollouts; mean absolute difference between the trainer's and the engine's log-probabilities of the sampled tokens, averaged over each window of steps and then over runs (descriptive, from the runs' step logs). TIS clipped 0.08\% of the tokens' importance weights in the first ten steps and 0.16\% in the last twenty with FlashRL's recipe, and 0.01\% and 0.06\% with the FP8 trainer (at most 0.31\% in any step).}
\label{tab:drift}
\centering\footnotesize\setlength{\tabcolsep}{3pt}
\begin{tabular}{@{}lrrrrr@{}}
\toprule
 & & \multicolumn{3}{c}{Mismatch, mean $|\Delta\log p|$, at steps ($\downarrow$)} & \\
\cmidrule(lr){3-5}
Configuration & Runs & 1--10 & 91--110 & 181--200 & Late/early $\downarrow$ \\
\midrule
bf16 engine and trainer & 9 & 0.0055 & 0.0061 & 0.0064 & 1.18$\times$ \\
FlashRL's recipe (W8A8, bf16 trainer, TIS) & 9 & 0.0272 & 0.0308 & 0.0348 & 1.28$\times$ \\
W8A8, FP8 trainer, TIS & 9 & 0.0173 & 0.0192 & 0.0220 & 1.27$\times$ \\
matched engine, no correction & 9 & \textbf{0.0070} & \textbf{0.0068} & \textbf{0.0075} & \textbf{1.07$\times$} \\
W8A8, FP8 trainer, no correction & 6 & 0.0174 & 0.0244 & 0.0394 & 2.26$\times$ \\
\bottomrule
\end{tabular}
\end{table*}

\textbf{Where the GPU-hours go on the H100} (\cref{tab:gensplit}). Generation takes most of each step, and the matched engine's extra cost on the H100 is almost all in generation.

\begin{table*}[tp]
\caption{\textbf{On the H100, the matched engine's extra GPU time is almost all in generation.} GPU-hours per 200-step run of the nine-seed H100 arms of \cref{fig:rlacc}, split into generation and the rest of each step (training and the adapter's sync), from the runs' step logs.}
\label{tab:gensplit}
\centering\footnotesize\setlength{\tabcolsep}{3pt}
\begin{tabular}{@{}lrrrrr@{}}
\toprule
Configuration & Runs & Generation $\downarrow$ & Training $\downarrow$ & Total $\downarrow$ & Generation share \\
\midrule
bf16 engine and trainer & 9 & 2.00 & 1.22 & 3.22 & 62\% \\
FlashRL's recipe & 9 & 1.77 & \textbf{1.19} & \textbf{2.96} & 60\% \\
W8A8 with TIS, FP8 trainer & 9 & \textbf{1.74} & 1.36 & 3.10 & 56\% \\
matched engine, FP8 trainer & 9 & 2.81 & 1.39 & 4.20 & 67\% \\
\bottomrule
\end{tabular}
\end{table*}

\textbf{What each part costs and buys} (\cref{tab:rlparts,tab:rlshbf16}; Figure~\ref{fig:memory} shows the memory of each placement).

\begin{table*}[tp]
\caption{\textbf{Colocated LoRA RL on one GPU, one part at a time: FP8 weights let two copies fit, and one shared copy returns memory to the KV cache.} Means over runs: where the trainer's base waits during rollouts, the engine's memory share and KV cache, seconds per step generating and training, GPU-hours per run, and final accuracy minus the block's bf16 run with the Welch 95\% half-width. $^{\mathrm{a}}$The engine releases its KV cache while training; $^{\mathrm{b}}$one copy that keeps its KV cache. The 27B's two-copy (6 runs) and one-copy (7 runs) rows pool every completed run; \cref{tab:arms2} lists the runs.}
\label{tab:rlparts}
\centering\footnotesize\setlength{\tabcolsep}{3pt}
\begin{tabular}{@{}llllrrrrrrrr@{}}
\toprule
 & & & \multicolumn{2}{c}{Trainer's base} & \multicolumn{2}{c}{Engine} & \multicolumn{2}{c}{Seconds per step} & & & \multicolumn{1}{c}{$\Delta$ acc.} \\
\cmidrule(lr){4-5}\cmidrule(lr){6-7}\cmidrule(lr){8-9}
Engine & Trainer & Corr. & Where & GiB $\downarrow$ & Share & KV (GiB) $\uparrow$ & Rollout $\downarrow$ & Train $\downarrow$ & GPU-h $\downarrow$ & Runs & \multicolumn{1}{c}{(points) $\uparrow$} \\
\midrule
\multicolumn{12}{@{}l}{\textit{Qwen3.5-9B on one L40S (48\,GB), 200 steps}} \\
bf16 & bf16 & none & GPU & 16.7 & \multicolumn{7}{l}{\textit{out of memory at every engine share tried}} \\
bf16 & bf16 & none & host & 16.7 & 0.80 & 15.3 & 105 & 46 & 8.4 & 7 & \textit{61.8\,\%} \\
W8A8 & bf16 & TIS & host & 16.7 & 0.80 & \textbf{22.7} & \textbf{84} & 47 & 7.2 & 3 & $\mathbf{-0.3}\pm4.3$ \\
W8A8 & FP8 (ours) & TIS & host & 10.2 & 0.80 & 21.3 & \textbf{84} & 47 & 7.2 & 7 & $-1.1\pm2.0$ \\
W8A8 & FP8 (ours) & TIS & GPU & 10.2 & 0.45 & 5.8 & 123 & 45 & 9.3 & 3 & $-3.0\pm2.5$ \\
W8A8\rlap{$^{\mathrm{b}}$} & FP8 (ours) & TIS & shared & \textbf{0.0} & 0.55 & 10.3 & \textbf{84} & \textbf{44} & \textbf{7.1} & 3 & $-0.6\pm1.8$ \\
W8A8 & FP8 (ours) & TIS & shared\rlap{$^{\mathrm{a}}$} & \textbf{0.0} & 0.80 & 21.4 & \textbf{84} & 45 & \textbf{7.1} & 3 & $-2.2\pm5.5$ \\
W8A16 (ours) & FP8 (ours) & none & GPU & 10.2 & 0.45 & 7.7 & 114 & \textbf{44} & 8.8 & 3 & $-1.1\pm2.3$ \\
W8A16 (ours)\rlap{$^{\mathrm{b}}$} & FP8 (ours) & none & shared & \textbf{0.0} & 0.55 & 12.1 & 86 & 45 & 7.3 & 5 & $-3.4\pm2.3$ \\
\addlinespace[3pt]
\multicolumn{12}{@{}l}{\textit{Qwen3.8-27B on one H100 (80\,GB), 100 steps}} \\
bf16 & bf16 & none & GPU & 50.1 & \multicolumn{7}{l}{\textit{does not fit: 101\,GiB of weights}} \\
bf16 & bf16 & none & host & 50.1 & 0.85 & 12.6 & 110 & \textbf{33} & 4.0 & 3 & \textit{71.4\,\%} \\
W8A8 & FP8 (ours) & TIS & host & 27.4 & 0.85 & 34.8 & 83 & 39 & 3.4 & 3 & $\mathbf{+0.7}\pm3.2$ \\
W8A8 & FP8 (ours) & TIS & GPU & 27.4 & 0.48 & 5.5 & 123 & 35 & 4.4 & 6 & $+0.0\pm2.2$ \\
W8A8\rlap{$^{\mathrm{b}}$} & FP8 (ours) & TIS & shared & \textbf{0.0} & 0.65 & 19.5 & \textbf{82} & 35 & \textbf{3.3} & 7 & $-0.3\pm2.4$ \\
W8A8 & FP8 (ours) & TIS & shared\rlap{$^{\mathrm{a}}$} & \textbf{0.0} & 0.85 & \textbf{35.3} & \textbf{82} & 35 & \textbf{3.3} & 3 & $+0.6\pm2.9$ \\
\bottomrule
\end{tabular}
\end{table*}

\begin{table*}[tp]
\caption{\textbf{With 2{,}048-token rollouts, one bf16 copy that keeps its KV cache trains as fast as one FP8 copy on the 9B; on the 27B the FP8 copy is 1.33$\times$ faster.} 12 steps of the RL recipe with one bf16 copy shared by engine and trainer that keeps its KV cache while training (Unsloth releases it) against the same steps of full runs with one FP8 copy and with bf16 offloading; seconds per step over steps 3--12 and the run's peak host memory.}
\label{tab:rlshbf16}
\centering\footnotesize\setlength{\tabcolsep}{2.5pt}
\begin{tabular}{@{}llrrrrr@{}}
\toprule
 & & & & \multicolumn{2}{c}{Seconds per step} & \\
\cmidrule(lr){5-6}
Model, GPU & Copy & Share & KV (GiB) $\uparrow$ & Rollout $\downarrow$ & All $\downarrow$ & Host peak (GiB) $\downarrow$ \\
\midrule
Qwen3.5-9B, L40S & one bf16 & 0.55 & 5.5 & 187 & 245 & 8.6 \\
Qwen3.5-9B, L40S & one bf16 & 0.70 & 12.2 & \textbf{120} & 174 & 10.1 \\
Qwen3.8-27B, H100 & one bf16 & 0.80 & 9.1 & 136 & 165 & 14.2 \\
\midrule
Qwen3.5-9B, L40S & one FP8 & 0.55 & 10.3 & 122 & \textbf{173} & \textbf{7.4} \\
Qwen3.5-9B, L40S & bf16, offloaded & 0.80 & \textbf{15.3} & 127 & 182 & 48.2 \\
Qwen3.8-27B, H100 & one FP8 & 0.65 & \textbf{19.5} & \textbf{94} & \textbf{124} & \textbf{10.6} \\
Qwen3.8-27B, H100 & bf16, offloaded & 0.85 & 12.6 & 129 & 161 & 159.5 \\
\bottomrule
\end{tabular}
\end{table*}

\begin{table*}[tp]
\caption{\textbf{Of the placements we ran, only FP8 trainers fit the 27B on a 48\,GB GPU, and with long rollouts one FP8 copy runs close to offloading's speed without its host memory.} Seconds per RL step (bold: fastest): the 27B's 100-step runs on a 48\,GB L40S at 2{,}048-token rollouts (mean over three seeds) and 12-step probes in thinking mode with rollouts of up to 8{,}192 or 16{,}384 tokens (steps 3--12). OOM: out of GPU memory (at 16{,}384 tokens the bf16 trainers' runs stopped before their first or, for FlashRL's offloaded recipe, their second step; both FP8 trainers ran all 12 steps, their mean completion length reaching 10.7--10.8k tokens). $^{d}$Rerun with chunked log-probs, the LM head computed in recomputed chunks so that the full logits never exist: FlashRL's offloaded recipe then ran all 12 steps, at 73\,GiB of GPU and 127\,GiB of host memory; its responses trained fewer tokens per step (176k against 265--287k), so its step time is not like for like; bf16 offloading with chunked log-probs still ran out of GPU memory in its third step. All in our trainer, which keeps activation checkpoints and a micro-batch's logits on the GPU. Host, training: the 27B run's median host memory over its steps 2--4 on the H100 (Table~\ref{tab:rlprobe} gives each placement's peak). One bf16 copy runs out of memory at engine share 0.70 as shown; at 0.55 it fits the 9B at 8{,}192 tokens, at 2{,}216\,s per step (Table~\ref{tab:rlprobe}). FlashRL: vLLM's FP8 engine beside a bf16 trainer, with TIS. Unsloth's FP8 standby shares one FP8 copy as the last row does, with a trainer that rounds activations alike; its bf16 standby needs at least bf16 offloading's GPU memory while it trains. $^{a}$Not run: it needs more GPU memory than a placement that ran out of it. $^{b}$From the probe's own trace. $^{c}$Not run: its copies of the base alone exceed the GPU's memory. n/r: not run. Over full 100-step runs at 8{,}192 tokens on an H100 (Table~\ref{tab:rlong27}) the final accuracies (OlympiadBench and Minerva, non-thinking, 2{,}048 tokens) are: one FP8 copy 50.6 and 52.1 (two seeds), FlashRL's offloaded recipe 51.0, FP8 offloading 51.6, bf16 offloading 51.1, against 51.1 for the 27B before RL: these runs change accuracy little at this evaluation, so they compare cost, not learning.}
\label{tab:longroll}
\centering\footnotesize\setlength{\tabcolsep}{5pt}
\begin{tabular}{@{}lrrrrr@{}}
\toprule
 & Host, training & \multicolumn{2}{c}{48\,GB L40S} & \multicolumn{2}{c}{80\,GB H100, the 27B} \\
\cmidrule(lr){3-4}\cmidrule(l){5-6}
Placement of the trainer's base & 27B (GiB) $\downarrow$ & 27B, 2{,}048 $\downarrow$ & 9B, 8{,}192 $\downarrow$ & 8{,}192 $\downarrow$ & 16{,}384 $\downarrow$ \\
\midrule
FlashRL: FP8 engine, bf16 trainer, two copies & -- & $\times^{c}$ & $\times^{a}$ & $\times^{a}$ & $\times^{a}$ \\
FP8 engine and trainer, two copies & 9 & $\times^{c}$ & OOM & OOM & $\times^{a}$ \\
bf16, one shared copy, KV cache kept & \textbf{6}$^{b}$ & $\times^{c}$ & OOM & OOM & OOM \\
bf16, trainer's base offloaded (veRL-style) & 159 & $\times^{c}$ & 1314 & 659 & OOM \\
FlashRL, trainer's base offloaded & 125$^{b}$ & $\times^{c}$ & 1076 & 421 & \textbf{606}$^{d}$ \\
FP8, trainer's base offloaded & 93 & 415 & 1025 & \textbf{418} & 673 \\
FP8, one shared copy & 10 & OOM & 1167 & 476 & n/r \\
FP8, one copy, KV cache released (EpiLoRA) & 8 & \textbf{404} & \textbf{1012} & 422 & 701 \\
\bottomrule
\end{tabular}
\end{table*}

\begin{table*}[tp]
\caption{\textbf{The long-rollout probes behind Table~\ref{tab:rlprobe}.} Per placement: seconds per step and generation seconds per step (steps 3--12; $^{a}$steps 3--11, engine at 55\% of the GPU), mean response length, trained tokens per step (responses in groups with zero reward variance are not trained) and the engine's KV cache. Bold: best within each model, and any within 2\% of it.}
\label{tab:rlprobedetail}
\centering\footnotesize\setlength{\tabcolsep}{3pt}
\begin{tabular}{@{}lrrrrr@{}}
\toprule
Placement & s/step $\downarrow$ & generation (s) $\downarrow$ & response (tokens) & trained tokens & KV (GiB) $\uparrow$ \\
\midrule
\multicolumn{6}{@{}l}{\textit{9B, one L40S}} \\
One bf16 copy, KV kept$^{a}$ & 2{,}216 & 1{,}886 & 7{,}992 & 527k & 4.2 \\
bf16, trainer offloaded & 1{,}314 & 963 & 7{,}902 & 558k & 15.3 \\
FlashRL, trainer offloaded & 1{,}076 & \textbf{732} & 7{,}906 & 549k & \textbf{21.4} \\
EpiLoRA, FP8 trainer offloaded & \textbf{1{,}025} & \textbf{725} & 7{,}888 & 544k & \textbf{21.4} \\
EpiLoRA, one FP8 copy & \textbf{1{,}012} & \textbf{722} & 7{,}795 & 531k & \textbf{21.4} \\
\addlinespace[2pt]
\multicolumn{6}{@{}l}{\textit{27B, one H100}} \\
bf16, trainer offloaded & 659 & 521 & 5{,}575 & 244k & 12.6 \\
FlashRL, trainer offloaded & \textbf{421} & \textbf{286} & 5{,}549 & 248k & \textbf{34.8} \\
EpiLoRA, FP8 trainer offloaded & \textbf{418} & \textbf{291} & 5{,}549 & 206k & \textbf{34.8} \\
EpiLoRA, one FP8 copy & \textbf{422} & \textbf{290} & 5{,}545 & 207k & \textbf{34.8} \\
\bottomrule
\end{tabular}
\end{table*}

\FloatBarrier
\section{Software and hardware}
\label{app:repro}

\textbf{Tuning vLLM's LoRA kernels.} For each model and GPU, we record the shape of every shrink and expand call of a serving run and time, for each shape, rank (64, for some models also 16), slot count (1 and 32; 8 for the 72B) and token count (1 to 8{,}192), vLLM's default configuration and the 12 to 47 others that run (block sizes, split-K, warps) with CUDA graphs and the L2 cache flushed; we keep the fastest whose output matches the default's, a median 1.2--1.4$\times$ faster than the default over 140 to 400 keys per model and GPU. vLLM looks the files up by GPU name: for Qwen3.5-9B, the H200 and GH200 use the H100's, as we did not tune on those GPUs (on Qwen3-8B, tuning on the H200 itself kept the H100's configuration for 112 of 160 keys).
Software: vLLM~0.28.0, PyTorch~2.13.0, Triton~3.7.1, FlashInfer~0.6.16, Transformers~5.16.1, SGLang~0.5.9 and TensorRT-LLM in their own environments; CUDA~12.8 and 13.1.
Hardware: NVIDIA L40S (48\,GB, Ada); H100 SXM (80\,GB), H200 NVL (141\,GB) and GH200 (96\,GB, Hopper); A100 (80\,GB, RL only); a 2g.48gb MIG slice of an RTX PRO 6000 (Blackwell).

\FloatBarrier
\end{document}